\documentclass[12pt]{article}
\pdfoutput=1
\DeclareMathAlphabet{\scr}{U}{rsfs}{m}{n}
\usepackage{latexsym}
\usepackage[mathscr]{eucal}
\usepackage{amsfonts}
\usepackage{amscd}
\usepackage{cite}
\usepackage{array}
\usepackage{amssymb}
\usepackage{colordvi}
\usepackage[centertags]{amsmath}
\usepackage{enumerate}
\usepackage{graphicx}
\usepackage{booktabs}
\usepackage{theorem}
\usepackage[footnotesize]{caption}
\usepackage{multirow}
\usepackage{color}
\usepackage{dsfont}
\usepackage{dsfont}
\usepackage{amsmath}
\usepackage{pifont}
\usepackage{bbm}
\usepackage{bm}

\newcommand{\cleqn}{\setcounter{equation}{0}}
\newcommand{\newc}{\newcommand}
\newc{\be}{\begin{equation}}
\newc{\ee}{\end{equation}}
\newc{\bea}{\begin{eqnarray}}
\newc{\eea}{\end{eqnarray}}
\newc{\ol}{\overline}
\newc{\wt}{\widetilde}
\newc{\bs}{\boldsymbol}
\newc{\m}{\mathcal}

\begin{document}

\title{\hfill ~\\[0mm]
       \textbf{Generalised CP and $A_4$ Family Symmetry }        }
\date{}
\author{\\[1mm]Gui-Jun Ding$^{1\,}$\footnote{E-mail: {\tt
dinggj@ustc.edu.cn}}~,~
Stephen F. King$^{2\,}$\footnote{E-mail: {\tt king@soton.ac.uk}}~,
Alexander J. Stuart$^{2\,}$\footnote{E-mail: {\tt a.stuart@soton.ac.uk}}\\
\\[1mm]
  \it{\small $^1$Department of Modern Physics, University of Science and
    Technology of China,}\\
  \it{\small Hefei, Anhui 230026, China}\\[4mm]
  \it{\small $^2$School of Physics and Astronomy, University of
  Southampton,}\\
  \it{\small Southampton, SO17 1BJ, United Kingdom}\\[4mm]
}

\maketitle

\begin{abstract}
\noindent

We perform a comprehensive study of family symmetry models based on $A_4$ combined with the generalised CP symmetry $H_{\rm{CP}}$. We investigate the lepton mixing parameters which can be obtained from the original symmetry $A_4\rtimes H_{\rm{CP}}$ breaking to different remnant symmetries in the neutrino and charged lepton sectors. We find that only one case is
phenomenologically viable, namely  $G^{\nu}_{\rm{CP}}\cong Z^{S}_2\times
H^{\nu}_{\rm{CP}}$ in the neutrino sector and $G^{l}_{\rm{CP}}\cong
Z^{T}_3\rtimes H^{l}_{\rm{CP}}$ in the charged lepton sector,
leading to the prediction of no CP violation, namely $\delta_{CP}$ and the Majorana phases $\alpha_{21}$ and $\alpha_{31}$ are all equal to either zero or $\pi$. We then propose an effective supersymmetric model based on the symmetry $A_4\rtimes H_{\rm{CP}}$ in which trimaximal lepton mixing is predicted together with either zero CP violation or $\delta_{CP}\simeq\pm \pi/2$ with non-trivial Majorana phases. An ultraviolet completion of the effective model yields a neutrino mass matrix which depends on only three real parameters. As a result of this, all three CP phases and the absolute neutrino mass scale are determined, the atmospheric mixing angle is maximal, and the Dirac CP can either be preserved with $\delta_{CP}=0,\pi$ or maximally broken with $\delta_{CP}=\pm \pi /2$ and sharp predictions for the Majorana phases and neutrinoless double beta decay.

\end{abstract}

\thispagestyle{empty}
\vfill
\newpage
\setcounter{page}{1}

\section{Introduction}
\cleqn

After the measurement of the reactor mixing angle $\theta_{13}$ by the Daya
Bay~\cite{An:2012eh}, RENO~\cite{Ahn:2012nd}, and Double Chooz~\cite{Abe:2011fz} reactor neutrino experiments, all three lepton
mixing angles $\theta_{12}$, $\theta_{23}$, $\theta_{13}$ and both
mass-squared differences $\Delta m^2_{sol}$ and $\Delta m^2_{atm}$ have been
measured to reasonably good accuracy. Yet within the standard framework of three-neutrino oscillations, the Dirac CP phase and neutrino mass ordering still elude measurement so far. Furthermore, if neutrinos are Majorana particles, there exist two more unknown Majorana CP phases which may play a role in neutrinoless double-beta decay searches. Thus, determining the exact neutrino mass ordering and measuring the Dirac and Majorana CP violating phases are the primary goals of future neutrino oscillation experiments.
The CP violation has been firmly established in the quark sector and it is natural to expect that CP violation occurs in the lepton sector as well. It is insightful to note that hints of a nonzero $\delta_{\rm{CP}}$ have begun to show up in global analysis of neutrino oscillation data~\cite{Tortola:2012te,Fogli:2012ua,GonzalezGarcia:2012sz}.

What would we learn from the measurements of the lepton CP violating phases?
What is the underlying physics? These questions are particularly imperative
in view of foreseeable future experimental programs to measure the
CP-violation in the neutrino oscillations sector. In the past years, much effort has been devoted to explaining the structure of the lepton mixing angles through the introduction of family symmetries. In this scheme, one generally assumes a non-abelian discrete flavour group which is broken to different subgroups in the neutrino and charged lepton sectors. The mismatch between these two subgroups leads to particular predictions for the lepton mixing angles. For recent reviews, see Ref.~\cite{Altarelli:2010gt} and
Ref.~\cite{Ishimori:2010au} for the model building and relevant group theory
aspects, respectively. Motivated by this approach one can extend the family symmetry to include a generalised CP symmetry $H_{\rm{CP}}$ ~\cite{Ecker:1981wv} which will allow the prediction of both CP phases and mixing angles.

The possibility of combining a family symmetry with a generalised CP symmetry has already been discussed in the literature. For example, the simple $\mu-\tau$ reflection symmetry, which is a combination of the canonical CP transformation and the $\mu-\tau$ exchange symmetry, has been discussed and successfully implemented in a number of models where both atmospheric mixing angle $\theta_{23}$ and Dirac CP phase $\delta_{\rm{CP}}$ were predicted to be maximal\cite{Harrison:2002kp,Grimus:2003yn,Farzan:2006vj}. Additionally in Ref.~\cite{Feruglio:2012cw}, the phenomenological consequences of imposing both an $S_4$ flavour symmetry and a generalised CP symmetry have been analysed in a model-independent way. They found that all lepton mixing angles and CP phases depend on one free parameter for the symmetry breaking of $S_4\rtimes H_{\rm{CP}}$ to $Z_2\times \rm{CP}$ in the neutrino sector and to some abelian subgroup of $S_4$ in the charged lepton sector. Concrete $S_4$ family models with a generalised CP symmetry have been constructed in Refs.~\cite{Ding:2013hpa,Feruglio:2013hia,Luhn:2013lkn} where the spontaneous breaking of the $S_4\rtimes H_{\rm{CP}}$ down to $Z_2\times \rm{CP}$ in the neutrino sector was implemented. Other models with a family symmetry and a generalised CP symmetry can also be found in Refs.~\cite{Krishnan:2012me,Mohapatra:2012tb,Nishi:2013jqa}. In
addition, there are other theoretical frameworks comprising both family
symmetry and CP violation~\cite{Branco:1983tn,Chen:2009gf,Antusch:2011sx}.

In this work, we study generalised CP symmetry in the context of
the most popular family symmetry $A_4$\footnote{$A_4$ models with spontaneous CP violation are proposed in Refs.~\cite{Antusch:2013kna,Ahn:2013mva}, where a CP symmetry is assumed to exist at a high energy scale.} (please see Ref.~\cite{Barry:2010zk,Ding:2011gt} for a classification of the $A_4$
models on the market). The generalised CP transformation compatible with an $A_4$ family symmetry is clarified, and a model-independent analysis of the lepton mixing matrix is performed by scanning all of the possible remnant subgroups in the neutrino and charged lepton sectors. We construct an effective $A_4\rtimes H_{\rm{CP}}$ model, where non-renormalisable operators are involved. The lepton mixing is predicted to be trimaximal pattern in the model, and the Dirac phase is trivial or nearly maximal. Furthermore, this effective model is promoted to a renormalisable one in which the higher order operators are under control.

The remainder of this paper is organised as follows. In Section~\ref{sec:cp}, we present the general CP transformations consistent with the $A_4$ family symmetry. In Section~\ref{sec:general}, we perform a thorough scan of leptonic mixing parameters which can be obtained from the remnant symmetries of the underlying combined symmetry group $A_4\rtimes H_{\rm{CP}}$. We find that only one case out of all possibilities is phenomenologically viable.  This case predicts both Dirac and Majorana phases to be trivial. In Section~\ref{sec:model_effective} we specify the structure of the model at leading order, and the required vacuum alignment is justified. In subsection \ref{sec:NLO}, we analyse the subleading Next-to-Leading-Order (NLO) corrections induced by higher dimensional operators and phenomenological predictions of the model are presented. In Section~\ref{sec:UV_completion}, we address the ultraviolet completion of the model which significantly increases the predictability of the theory such that all the mixing angles, CP phases and the absolute neutrino mass scale are fixed. We conclude in Section~\ref{sec:conclusion}. The details of the group theory
of $A_4$ are collected in Appendix~\ref{sec:appendix_A} and Appendices \ref{K4}-\ref{GlK4} contain the implications of preserving other subgroups of $A_4$ different than $G_{\nu}=Z_2$ and $G_l=Z_3$.  Finally, Appendix~\ref{sec:appendix_B} describes the diagonalisation of a general $2\times2$ symmetric complex matrix.

\section{\label{sec:cp}Generalised CP transformations with family symmetry}
\cleqn

\subsection{General family symmetry group}

In general, it is nontrivial to combine the family symmetry $G_f$ and the generalised CP symmetry together because the definition of the generalised CP transformations must be compatible with the family symmetry.  Thus, the generalised CP transformations are subject to certain consistency conditions~\cite{Grimus:1995zi,Feruglio:2012cw,Holthausen:2012dk}. Namely, for a set of fields $\varphi$ in a generic irreducible representation $\mathbf{r}$ of $G_f$, it transforms under the action of $G_f$ as
\begin{equation}
\varphi(x)\stackrel{G_f}{\longrightarrow} \rho_{\mathbf{r}}(g) \varphi(x),
\qquad g \in G_f\,,
\end{equation}
where $\rho_{\mathbf{r}}(g)$ denotes the representation matrix for the
element $g$ in the irreducible representation $\mathbf{r}$, the generalised
CP transformation is of the form
\begin{equation}
\varphi(x)\stackrel{CP}{\longrightarrow}X_{\mathbf{r}}\,\varphi^{*}(x')\,,
\end{equation}
where $x'=(t,-\mathbf{x})$ and the obvious action of CP on the spinor
indices is omitted for the case of $\varphi$ being spinor. Here we are considering the ``minimal'' theory in which the generalised CP transforms the field $\varphi\sim\mathbf{r}$ into its complex conjugate $\varphi^{*}\sim\mathbf{r}^{*}$, and the transformation into another field $\varphi'^{*}\sim\mathbf{r}'^{*}$ with $\mathbf{r}'\neq\mathbf{r}$ is beyond the present scope since both $\varphi$ and $\varphi'$ would be required to be present in pair and correlated with each other in that case. Notice that $X_{\mathbf{r}}$ should be a unitary matrix to keep the kinetic term invariant. Now if we first perform a CP transformation, then apply a family symmetry transformation, and finally an inverse CP transformation is followed, i.e.
\begin{equation}
\varphi(x)\stackrel{CP}{\longrightarrow}X_{\mathbf{r}}\,\varphi^{*}(x')\stackrel{G_f}{\longrightarrow}X_{\mathbf{r}}\rho^{*}_{\mathbf{r}}(g)
\varphi^{*}(x')\stackrel{CP^{-1}}{\longrightarrow}X_{\mathbf{r}}\rho^{*}_{\mathbf{r}}(g)X^{-1}_{\mathbf{r}}\varphi(x)\,,
\end{equation}
the theory should still be invariant since it is invariant under each
transformation individually. To make the theory consistent the resulting net transformation should be
equivalent to a family symmetry transformation $\rho_{\mathbf{r}}(g')$ of
some family group element $g'$, i.e.
\begin{equation}
\label{eq:consistency}X_{\mathbf{r}}\rho^{*}_{\mathbf{r}}(g)X^{-1}_{\mathbf{r}}=\rho_{\mathbf{r}}(g'),
\qquad g' \in G_f\,,
\end{equation}
where the elements $g$ and $g'$ must be the same for all irreducible
representations of $G_f$. Eq. (\ref{eq:consistency}) is the important \textit{consistency condition} which has to be fulfilled in order to impose both generalised CP and family symmetry invariance simultaneously. It also implies that the generalised CP transformation $X_{\mathbf{r}}$ maps the group element $g$ into $g'$ and that the family group structure is preserved under this mapping. Therefore Eq. (\ref{eq:consistency}) defines a homomorphism of the family symmetry group $G_f$. Notice that in the case where $\rho_{\mathbf{r}}$ is a faithful representation, the elements $g$ and $g'$ have the same order, the mapping defined in Eq. (\ref{eq:consistency}) is bijective, and thus the associated CP transformation becomes an automorphism \cite{Holthausen:2012dk}. It is notable that both $e^{i\theta}X_{\mathbf{r}}$ and $\rho_{\mathbf{r}}(h)X_{\mathbf{r}}$ also satisfy the consistency equation of Eq.~\eqref{eq:consistency} for a generalised CP transformation $X_{\mathbf{r}}$, where $\theta$ is real and $h$ is any element of $G_f$. Therefore the possible form of the CP transformation $X_{\mathbf{r}}$ is only determined by the consistency equation up to an overall arbitrary phase and family symmetry transformation $\rho_{\mathbf{r}}(h)$ for a given
irreducible representation $\mathbf{r}$. In the following, we investigate the generalised CP transformations consistent with an $A_4$ family symmetry for different irreducible representations, i.e. $G_f=A_4$.

\subsection{$A_4$ family symmetry}

The $A_4$ group can be generated by two generators $S$ and
$T$, which are of orders two and three, respectively (see Appendix~\ref{sec:appendix_A} for the details of the group theory of $A_4$). To include a generalised CP symmetry consistent with an $A_4$ family symmetry, it is sufficient to only impose the consistency condition in Eq.~\eqref{eq:consistency} on the group generators:
\begin{equation}
X_{\mathbf{r}}\rho^{*}_{\mathbf{r}}(S)X^{-1}_{\mathbf{r}}=\rho_{\mathbf{r}}(S'),\qquad
X_{\mathbf{r}}\rho^{*}_{\mathbf{r}}(T)X^{-1}_{\mathbf{r}}=\rho_{\mathbf{r}}(T')\,.
\end{equation}
To do this, we start with the faithful triplet representation $\mathbf{3}$.  Then the order of $S'$ and $T'$ will be 2 and 3, respectively. Therefore $S'$ and $T'$ can only belong to certain conjugacy classes of $A_4$. Namely,
\begin{equation}
S'\in 3 C_2,\qquad  T'\in 4 C_3\cup 4 C_3^2
\end{equation}
It is remarkable that the consistency condition of Eq.~\eqref{eq:consistency} must hold for all representations $\mathbf{r}$ simultaneously. However, because of the models constructed in later sections, we assume that our theory contains only one of the nontrivial singlet irreducible representations (either $\mathbf{1}'$ or $\mathbf{1}''$) in the flavon sector and further restrict ourselves to a minimal case where there exists only one flavon transforming under that nontrivial singlet irreducible representation (in addition to other flavons transforming under the $\mathbf{1}$ and $\mathbf{3}$ representations).  However, in these models there does exist a
$\mathbf{1}'$ and $\mathbf{1}''$ in the matter sector.  Yet, additional symmetry forbids the interchanging of these fields under the generalised CP symmetry. Therefore we have chosen to define a generalised CP symmetry without the interchanging of fields transforming under conjugate representations, e.g.  fields transforming under $\mathbf{1}'$ and $\mathbf{1}''$ representations.
Then, the element $T'$ can further be constrained by these nontrivial singlet representations $\mathbf{1}'$ and $\mathbf{1}''$, where the corresponding generalised CP transformations $X_{\mathbf{1}',\mathbf{1}''}$ are numbers with absolute value equal to 1, and then we have
\begin{equation}
\rho_{\mathbf{1}',\mathbf{1}''}(T')=X_{\mathbf{1}',\mathbf{1}''}\rho^{*}_{\mathbf{1}',\mathbf{1}''}(T)X^{-1}_{\mathbf{1}',\mathbf{1}''}
=\rho^{*}_{\mathbf{1}',\mathbf{1}''}(T)=\omega^{\mp2}
\label{T}
\end{equation}
Consequently, the element $T'$ can only be in the conjugacy class $4C_3^2$.  In summary, the consistency equation applied to our ``minimal'' case restricts $S'$ and $T'$ to
\begin{equation}
S'\in 3 C_2,\qquad  T'\in 4 C_3^2.
\end{equation}
For the simple case of $S'=S$ and $T'=T^2$ in the $\mathbf{3}$-dimensional representation, the associated CP transformation satisfying Eq.~\eqref{eq:consistency} can be found straightforwardly:
\begin{equation}
X_{\mathbf{0}}=\begin{pmatrix} 1&0&0\\
0&1&0 \\0&0&1
\end{pmatrix}\equiv\mathbbm{1}_3\,,
\end{equation}
which is the canonical CP transformation. The remaining eleven possible
choices for $S'$ and $T'$ lead to different solutions for $X_{\mathbf{3}}$.  These solutions are listed in Table~\ref{tab:CP_transformation} and can be neatly summarised in a compact way:
\begin{equation}
\label{eq:GCP_3}X_{\mathbf{3}}=\rho_{\mathbf{3}}(g),\quad g\in A_4\,.
\end{equation}
For the singlet representations $\mathbf{1}$, $\mathbf{1}'$ and
$\mathbf{1}''$, we take
\begin{equation}
X_{\mathbf{1},\mathbf{1}',\mathbf{1}''}=\rho_{\mathbf{1},\mathbf{1}',\mathbf{1}''}(g),\quad
g\in A_4\,.
\end{equation}
Therefore the generalised CP transformation consistent with an $A_4$ family
symmetry is of the same form as the family group transformation, i.e.
\begin{equation}
\label{eq:CP_A4}X_{\mathbf{r}}=\rho_{\mathbf{r}}(g),\quad g\in A_4\,.
\end{equation}
\begin{table}[t!]
\begin{center}
\resizebox{\textwidth}{!}{
\begin{tabular}{|c|c|c||c|c|c|}\hline\hline
  $X_{\mathbf{3}}$    &    $S\rightarrow S'$   &     $T\rightarrow T'$  &
  $X_{\mathbf{3}}$    &    $S\rightarrow S'$   &     $T\rightarrow T'$
  \\ \hline

  &   &    &     &   &   \\ [-0.15in]

$X_0=\left(\begin{array}{ccc}
1&0&0  \\
0&1&0  \\
0&0&1
\end{array}\right)$      &  \multirow{10}{*}{$S$ }    &    $T^2$    &

$\rho_{\mathbf{3}}(T^2)=\left(
\begin{array}{ccc}
 1 & 0 & 0 \\
 0 & \omega & 0 \\
 0 & 0 & \omega^2
\end{array}
\right)$    &    \multirow{10}{*}{$T^2ST$ }    &    $T^2$     \\  [0.25in]
\cline{1-1}  \cline{3-3}  \cline{4-4}  \cline{6-6}

  &   &    &     &   &   \\ [-0.15in]

$\rho_{\mathbf{3}}(T^2ST)=\frac{1}{3}\left(
\begin{array}{ccc}
 -1 & 2 \omega^2  & 2 \omega \\
 2 \omega & -1 & 2 \omega^2  \\
 2 \omega^2  & 2 \omega & -1
\end{array}
\right)$        &      &      $ST^2$   &
$\rho_{\mathbf{3}}(T^2S)=\frac{1}{3}\left(
\begin{array}{ccc}
 -1 & 2 & 2 \\
 2 \omega & -\omega & 2 \omega \\
 2 \omega^2  & 2 \omega^2  & -\omega^2
\end{array}
\right)$    &   &  $ST^2$   \\ [0.25in] \cline{1-1}  \cline{3-3} \cline{4-4}
\cline{6-6}

  &   &    &     &   &   \\ [-0.15in]

$\rho_{\mathbf{3}}(TST^2)=\frac{1}{3}\left(
\begin{array}{ccc}
 -1 & 2 \omega & 2 \omega^2  \\
 2 \omega^2  & -1 & 2 \omega \\
 2 \omega & 2 \omega^2  & -1
\end{array}
\right)$    &      &    $T^2S$   &
$\rho_{\mathbf{3}}(ST^2S)=\frac{1}{3}\left(
\begin{array}{ccc}
 -1 & 2 \omega^2  & 2 \omega \\
 2 \omega^2  & -\omega & 2 \\
 2 \omega & 2 & -\omega^2
\end{array}
\right)$   &    &  $T^2S$  \\  [0.25in] \cline{1-1}  \cline{3-3} \cline{4-4}
\cline{6-6}

  &   &    &     &   &   \\ [-0.15in]

$\rho_{\mathbf{3}}(S)=\frac{1}{3}\left(
\begin{array}{ccc}
 -1 & 2 & 2 \\
 2 & -1 & 2 \\
 2 & 2 & -1
\end{array}
\right)$  &    &   $ST^2S$  &   $\rho_{\mathbf{3}}(ST^2)=\frac{1}{3}\left(
\begin{array}{ccc}
 -1 & 2 \omega & 2 \omega^2  \\
 2 & -\omega & 2 \omega^2  \\
 2 & 2 \omega & -\omega^2
\end{array}
\right)$   &    &   $ST^2S$    \\  [0.25in] \hline

  &   &    &     &   &   \\ [-0.15in]

$\rho_{\mathbf{3}}(T)=\left(
\begin{array}{ccc}
 1 & 0 & 0 \\
 0 & \omega^2  & 0 \\
 0 & 0 & \omega
\end{array}
\right)$  &   \multirow{5}{*}{$TST^2$ }    &    $T^2$   &
$\rho_{\mathbf{3}}(TS)=\frac{1}{3}\left(
\begin{array}{ccc}
 -1 & 2 & 2 \\
 2 \omega^2  & -\omega^2  & 2 \omega^2  \\
 2 \omega & 2 \omega & -\omega
\end{array}
\right)$   &   \multirow{5}{*}{$TST^2$ }   &   $T^2S$
\\  [0.25in] \cline{1-1}  \cline{3-3} \cline{4-4}  \cline{6-6}

  &   &    &     &   &   \\ [-0.15in]

$\rho_{\mathbf{3}}(STS)=\frac{1}{3}\left(
\begin{array}{ccc}
 -1 & 2 \omega & 2 \omega^2  \\
 2 \omega & -\omega^2  & 2 \\
 2 \omega^2  & 2 & -\omega
\end{array}
\right)$  &   &    $ST^2$   &    $\rho_{\mathbf{3}}(ST)=\frac{1}{3}\left(
\begin{array}{ccc}
 -1 & 2 \omega^2  & 2 \omega \\
 2 & -\omega^2  & 2 \omega \\
 2 & 2 \omega^2  & -\omega
\end{array}
\right)$   &   &   $ST^2S$

\\  [0.25in] \hline\hline

\end{tabular}
}
\caption{\label{tab:CP_transformation} The 12 non-trivial generalised CP transformations consistent with an $A_4$ family symmetry for the triplet representation $\mathbf{3}$ in the chosen basis determined by the consistency equation $X_{\mathbf{3}}\rho_{\mathbf{3}}^{*}(g)X^{-1}_{\mathbf{3}}=\rho_{\mathbf{3}}(g')$.   These CP transformations realise non-trivial outer automorphisms which \textit{change} the conjugacy class of $T$ from $4C_3$ to $4C_3^2$. Notice that even though they are outer automorphisms they are represented by $A_4$ group elements, e.g. the mapping $(S,T)\rightarrow (S',T')=(S,T^2)$ is acheived via the $A_4$ identity element $X_0$ by  $X_0\rho_{\mathbf{3}}^{*}(S)X_0^{-1}=\rho_{\mathbf{3}}(S)$ and $X_{0}\rho_{\mathbf{3}}^{*}(T)X^{-1}_{0}=\rho_{\mathbf{3}}(T^2)$.}
\end{center}
\end{table}
Now that we have found all generalised CP transformations consistent with the $A_4$ family symmetry,\footnote{Had we allowed the flavons to transform under all nontrivial $A_4$ irreducible representations (call them e.g. $\phi_{\mathbf{1}'}$,  $\phi_{\mathbf{1}''}$ and $\phi_{\mathbf{3}}$) then the transformation
\begin{equation}
\label{eq:CP_referee}\phi_{\mathbf{1}'}\rightarrow\phi^{*}_{\mathbf{1}''}, \quad  \phi_{\mathbf{1}''}\rightarrow\phi^{*}_{\mathbf{1}'},  \quad \phi_{\mathbf{3}}\rightarrow\left(\begin{array}{ccc}
1  &  0  &  0 \\
0  &  0  &  1 \\
0  &  1  &  0
\end{array}\right)\phi^{*}_{\mathbf{3}}
\end{equation} could generate an alternate set of 12 other generalised CP transformations.  We see that this kind of CP transformation can only be realised if both $\phi_{\mathbf{1}'}$ and $\phi_{\mathbf{1}''}$ are present and are interchanged under the CP transformation.} we proceed by investigating their implications on lepton masses and mixings.

\section{\label{sec:general}General analysis of lepton mixing from preserved
family and CP symmetries }
\cleqn

\subsection{General family symmetry}

To obtain definite predictions for both the lepton mixing angles and CP violating phases from symmetry, we impose the family symmetry $G_f$ and the generalised CP symmetry $H_{\rm{CP}}$ simultaneously at high energies. Then the family symmetry is spontaneously broken to the $G_{\nu}$ and $G_l$ subgroups in the neutrino and the charged lepton sector respectively, and the remnant CP symmetries from the breaking of $H_{\rm{CP}}$ are $H^{\nu}_{\rm{CP}}$ and $H^{l}_{\rm{CP}}$,
respectively. The mismatch between the remnant symmetry groups $G_{\nu}\rtimes H^{\nu}_{\rm{CP}}$ and $G_l\rtimes H^{l}_{\rm{CP}}$ gives rise to particular values for both mixing angles and CP phases. As usual, the three generations of the left-handed (LH) lepton
doublets are unified into a three-dimensional representation
$\rho_{\mathbf{3}}$ of $G_f$. The invariance under the residual family
symmetries $G_{\nu}$ and $G_l$ implies that the neutrino mass matrix
$m_{\nu}$ and the charged lepton mass matrix $m_{l}$ satisfy
\begin{eqnarray}
\nonumber&&\rho^{T}_{\mathbf{3}}(g_{\nu_i})m_{\nu}\rho_{\mathbf{3}}(g_{\nu_
i})=m_{\nu},\quad g_{\nu_i}\in G_{\nu},\\
\label{eq:inv_family}&&\rho^{\dagger}_{\mathbf{3}}(g_{l_i})m_{l}m^{\dagger}_{l}\rho_{\mathbf{3}}(g_{l_i})=m_{l}m^{\dagger}_{l},\quad
g_{l_i}\in G_l\,.
\end{eqnarray}
where the charged lepton mass matrix $m_l$ is given in the convention in which the left-handed (right-handed) fields are on the left-hand (right-hand) side of $m_l$. Moreover, the neutrino and the charged lepton mass matrices are constrained by the residual CP symmetry via
\begin{eqnarray}
\nonumber&&X^{T}_{\mathbf{3}\nu}m_{\nu}X_{\mathbf{3}\nu}=m^{*}_{\nu}, \qquad
\quad X_{\mathbf{3}\nu}\in H^{\nu}_{\rm{CP}},\\
\label{eq:inv_CP}&&X^{\dagger}_{\mathbf{3}l}m_{l}m^{\dagger}_{l}X_{\mathbf{3}l}=(m_{l}m^{\dagger}_{l})^{*},\quad
X_{\mathbf{3}l}\in H^{l}_{\rm{CP}}\,.
\end{eqnarray}
Since there are both remnant family and CP symmetries, the corresponding
consistency equation similar to Eq.~\eqref{eq:consistency} has to be
satisfied. Namely, the elements $X_{\mathbf{r}\nu}$ of $H^{\nu}_{\rm{CP}}$ and
$X_{\mathbf{r}l}$ of $H^{l}_{\rm{CP}}$ should satisfy
\begin{eqnarray}
\nonumber&&X_{\mathbf{r}\nu}\rho^{*}_{\mathbf{r}}(g_{\nu_i})X^{-1}_{\mathbf{r}\nu}=\rho_{\mathbf{r}}(g_{\nu_j}),\qquad
g_{\nu_i},g_{\nu_j}\in G_{\nu},\\
\label{eq:consistency_remnant}&&X_{\mathbf{r}l}\rho^{*}_{\mathbf{r}}(g_{l_i})X^{-1}_{\mathbf{r}l}=\rho_{\mathbf{r}}(g_{l_j}),\qquad
g_{l_i},g_{l_j}\in G_{l}.
\end{eqnarray}
Given a set of solutions $X_{\mathbf{r}\nu}$ and $X_{\mathbf{r}l}$, we can
straightforwardly check that $\rho_{\mathbf{r}}(g_{\nu_i})X_{\mathbf{r}\nu}$
and $\rho_{\mathbf{r}}(g_{l_i})X_{\mathbf{r}l}$ are solutions as well. The
invariance conditions of Eqs.~(\ref{eq:inv_family})-(\ref{eq:inv_CP}) allow us to reconstruct the mass matrices $m_{\nu}$ and
$m_lm^{\dagger}_l$, and eventually determine the lepton mixing matrix
$U_{PMNS}$. Furthermore, if two other residual family symmetries $G'_{\nu}$ and $G'_{l}$ are conjugate to $G_{\nu}$ and $G_{l}$ under the element $h\in G_f$, i.e.
\begin{equation}
 G'_{\nu}=hG_{\nu}h^{-1},\qquad G'_{l}=hG_{l}h^{-1}\,,
\end{equation}
then the associated residual CP symmetries $H^{\nu'}_{\rm{CP}}$ and $H^{l'}_{\rm{CP}}$
are related to $H^{\nu}_{\rm{CP}}$ and $H^{l}_{\rm{CP}}$ as
\begin{equation}
\label{eq:CP_conju}H^{\nu'}_{\rm{CP}}=\rho_{\mathbf{r}}(h)H^{\nu}_{\rm{CP}}\rho^{T}_{\mathbf{r}}(h),\qquad
H^{l'}_{\rm{CP}}=\rho_{\mathbf{r}}(h)H^{l}_{\rm{CP}}\rho^{T}_{\mathbf{r}}(h)\,,
\end{equation}
and the corresponding neutrino and charged lepton mass matrices are of the form
\begin{equation}
\label{eq:mass_matr_conju}m'_{\nu}=\rho^{*}_{\mathbf{3}}(h)m_{\nu}\rho^{\dagger}_{\mathbf{3}}(h),\qquad
m'_{l}m'^{\dagger}_{l}=\rho_{\mathbf{3}}(h)m_{l}m^{\dagger}_{l}\rho^{\dagger}_{\mathbf{3}}(h).
\end{equation}
Therefore, the remnant subgroups $G'_{\nu}$ and $G'_{l}$ lead to the same
mixing matrix $U_{PMNS}$ as $G_{\nu}$ and $G_{l}$ do.

Having completed a general discussion of the implementation of a generalised CP symmetry with a family symmetry, we now concentrate on the case of interest in which the family symmetry $G_f=A_4$ and a generalised CP symmetry $H_{\rm{CP}}$ consistent with $A_4$ is imposed.  Thus, the theory respects the full symmetry $A_4\rtimes H_{\rm{CP}}$. In the following, we perform a model independent study of the constraints that these symmetries impose on the neutrino mass matrix, the charged lepton mass matrix and the PMNS matrix by scanning all the possible remnant symmetries $G^{\nu}_{\rm{CP}}\cong G_{\nu}\rtimes H^{\nu}_{\rm{CP}}$ and $G^{l}_{\rm{CP}}\cong G_{l}\rtimes H^{l}_{\rm{CP}}$.  We begin this study with an analysis of the neutrino sector.

\subsection{Neutrino sector from a subgroup of $A_4\rtimes H_{\rm{CP}}$}

As shown in Appendix~\ref{K4}, the case $G_{\nu}=K_4\cong Z_2\times Z_2$ is not phenomenologically viable.
To resolve this issue, we assume that the underlying symmetry $A_4\rtimes
H_{\rm{CP}}$ is broken into $G^{\nu}_{\rm{CP}}\cong Z_2\times H^{\nu}_{\rm{CP}}$\footnote{As has been shown in previous work~\cite{Ding:2013hpa}, if the remnant family symmetry is $Z_2=\left\{1,Z\right\}$ with $Z^2=1$, a
consistent CP transformation $X_{\mathbf{r}}$ should satisfy
$X_{\mathbf{r}}\rho^{*}_{\mathbf{r}}(Z)X^{-1}_{\mathbf{r}}=\rho_{\mathbf{r}}(Z'),~
Z'\in Z_2$. For the faithful triplet representation
$\mathbf{r}=\mathbf{3}$, $Z'$ will be of the same order as $Z$. Consequently $Z'$ can only be equal to $Z$ exactly.  Thus the consistency
equation is uniquely fixed to be
$X_{\mathbf{r}}\rho^{*}_{\mathbf{r}}(Z)X^{-1}_{\mathbf{r}}=\rho_{\mathbf{r}}(Z)$.
This means that the generalised CP transformation
will commute with $Z_2$, and the semidirect product will reduce to the direct product.} in the neutrino sector~\cite{Feruglio:2012cw}.
Since the three $Z_2$ subgroups in Eq.~\eqref{eq:Z2_subg} are related by
conjugation as $Z^{(2)}=T^{2}Z^{S}_2(T^{2})^{-1}$ and
$Z^{TST^2}_2=TZ^{S}_2T^{-1}$, it is sufficient to only consider $G^{\nu}_{\rm{CP}}\cong Z^{S}_2\times
H^{\nu}_{\rm{CP}}$, where the element $X_{\mathbf{r}\nu}$ of $H^{\nu}_{\rm{CP}}$
should satisfy
\begin{equation}
X_{\mathbf{r}\nu}\rho^{*}_{\mathbf{r}}(S)X^{-1}_{\mathbf{r}\nu}=\rho_{\mathbf{r}}(S)\,.
\end{equation}
It is found that only 4 of the 12 non-trivial CP transformations are
acceptable\footnote{In Ref.~\cite{Feruglio:2012cw}, the authors chose a different basis and  proposed that three cases are admissible for $G_{\nu}=Z^{S}_2$ in $A_4$. Case II of Ref.~\cite{Feruglio:2012cw} exactly corresponds to
$X_{\mathbf{r}\nu}=\{\rho_{\mathbf{r}}(1), \rho_{\mathbf{r}}(S)\}$ of the
present work. However, the CP transformations for their Cases I and III map $\left(S,T\right)$ to $\left(S,T\right)$ and $\left(S,TS\right)$ respectively. They belong to another 12 CP transformations defined in Eq. (\ref{eq:CP_referee}). Thererefore, both $\phi_{\mathbf{1}'}$ and $\phi_{\mathbf{1}''}$ should be present in the Lagrangian to define these CP transformations. Furthermore, the scenario of $X_{\mathbf{r}\nu}=\rho_{\mathbf{r}}(T^2ST), \rho_{\mathbf{r}}(TST^2)$ found in our work was omitted in Ref.~\cite{Feruglio:2012cw} because the authors required that the CP transformation should be both unitary and symmetric.  Although it only needs to be unitary (not necessarily symmetric). However, they claimed that non-symmetric CP transformations consistent with the remnant $Z_2$ flavour symmetry generally implies a partially degenerate neutrino mass spectrum.},
\begin{equation}
\label{eq:CP_Z2}H^{\nu}_{\rm{CP}}=\left\{\rho_{\mathbf{r}}(1),\rho_{\mathbf{r}}(S),\rho_{\mathbf{r}}(T^2ST),\rho_{\mathbf{r}}(TST^2)\right\}\,.
\end{equation}
Thus, the neutrino mass matrix is constrained by
\begin{eqnarray}
\label{eq:family_inv_Z2}&&\rho^{T}_{\mathbf{3}}(S)m_{\nu}\rho_{\mathbf{3}}(S)=m_{\nu},\\
\label{eq:cp_inv_Z2}&&X^{T}_{\mathbf{3}\nu}m_{\nu}X_{\mathbf{3}\nu}=m^{*}_{\nu}\,,
\end{eqnarray}
where Eq.~\eqref{eq:family_inv_Z2} is the invariance condition under
$Z^{S}_2$, and it implies that the neutrino mass matrix is of the form
\begin{equation}
\label{eq:nu_mass_Z2}m_{\nu}=\alpha\left(\begin{array}{ccc}
2 & -1 & -1 \\
-1 & 2 & -1 \\
-1 & -1 &  2
\end{array}\right)+\beta\left(\begin{array}{ccc}
1 &  0 & 0 \\
0 & 0  &  1  \\
0 &  1 & 0
\end{array}\right)+\gamma\left(\begin{array}{ccc}
0 &  1 & 1 \\
1 &  1 & 0  \\
1 &  0 &  1
\end{array}\right)+\epsilon\left(\begin{array}{ccc}
0 & 1 & -1  \\
1 & -1 &  0 \\
-1 & 0 & 1
\end{array}\right)\,,
\end{equation}
where $\alpha$, $\beta$, $\gamma$ and $\epsilon$ are complex parameters, and they are further constrained by the remnant CP symmetry shown in Eq.~\eqref{eq:cp_inv_Z2}. In order to diagonalise the neutrino mass matrix
$m_{\nu}$ in Eq. (\ref{eq:nu_mass_Z2}), we first apply the tri-bimaximal transformation $U_{TB}$ to yield
\begin{equation}
m'_{\nu}=U^{T}_{TB}m_{\nu}U_{TB}=\left(\begin{array}{ccc}
3\alpha+\beta-\gamma   &   0   & -\sqrt{3}\;\epsilon  \\
0   &   \beta+2\gamma   &  0   \\
-\sqrt{3}\;\epsilon  &   0   &  3\alpha-\beta+\gamma
\end{array}\right)\,,
\end{equation}
where
\begin{equation}
U_{TB}=\left(
\begin{array}{ccc}
 \sqrt{\frac{2}{3}} & \frac{1}{\sqrt{3}}
   & 0 \\
 -\frac{1}{\sqrt{6}} & \frac{1}{\sqrt{3}}
   & -\frac{1}{\sqrt{2}} \\
 -\frac{1}{\sqrt{6}} & \frac{1}{\sqrt{3}}
   & \frac{1}{\sqrt{2}}
\end{array}
\right)\,.
\label{TB}
\end{equation}
Now we return to the investigation of the residual CP symmetry constraint of Eq.~\eqref{eq:cp_inv_Z2}. Two distinct phenomenological predictions arise for the different choices of $X_{\mathbf{r}\nu}$:
\begin{itemize}
  \item {$X_{\mathbf{r}\nu}=\rho_{\mathbf{r}}(1), \rho_{\mathbf{r}}(S)$}

   For this case, we see that we can straightforwardly solve Eq.~\eqref{eq:cp_inv_Z2} and find that
   all four parameters $\alpha$, $\beta$, $\gamma$ and $\epsilon$ are real. Then $m'_{\nu}$ can be further diagonalised by
\begin{equation}
U'^{T}_{\nu}m^{\prime}_{\nu}U'_{\nu}=\text{diag}(m_1,m_2,m_3),\qquad
U'_{\nu}=R(\theta)P\,,
\end{equation}
where $P$ is a unitary diagonal matrix with entries $\pm1$ or
$\pm i$ which renders the light neutrino masses $m_{1,2,3}$ positive, and
\begin{equation}
R(\theta)=\left(\begin{array}{ccc}
\cos\theta  &    0   &  \sin\theta  \\
0  &  1  &    0  \\
-\sin\theta   &   0   &   \cos\theta
\end{array}\right)\,
\label{R}
\end{equation}
is a rotation matrix with
\begin{equation}
\label{eq:tan_theta}\tan2\theta=\frac{\sqrt{3}\;\epsilon}{\beta-\gamma}\,.
\end{equation}
This diagonalisation reveals that the light neutrino masses $m_{1,2,3}$ are given by
\begin{eqnarray}
\nonumber&&m_1=\left|3\alpha+\text{sign}\left((\beta-\gamma)\cos2\theta\right)\sqrt{(\beta-\gamma)^2+3\epsilon^2}\right|,\\
\nonumber&&m_2=\left|\beta+2\gamma\right|,\\
&&m_3=\left|3\alpha-\text{sign}\left((\beta-\gamma)\cos2\theta\right)\sqrt{(\beta-\gamma)^2+3\epsilon^2}\right|.
\end{eqnarray}
We conclude that this case is acceptable.

  \item {$X_{\mathbf{r}\nu}=\rho_{\mathbf{r}}(T^2ST),
      \rho_{\mathbf{r}}(TST^2)$}

  In this case, it can be seen that the $\alpha$ of Eq.~\eqref{eq:nu_mass_Z2} is purely imaginary, and the remaining parameters $\beta$, $\gamma$ and $\epsilon$ are real. Then the hermitian combination $m'^{\dagger}_{\nu}m'_{\nu}$ turns out to be of the form:
\begin{equation}
m'^{\dagger}_{\nu}m'_{\nu}=\text{diag}\left(-9\alpha^2+(\beta-\gamma)^2+3\epsilon^2,(\beta+2\gamma)^2,-9\alpha^2+(\beta-\gamma)^2+3\epsilon^2\right)\,,
\end{equation}
which implies $m_1=m_3$.  Clearly, this is not consistent with the experimental observation that the three light neutrinos have different masses. Note that the generalised CP transformations $X_{\mathbf{r}\nu}=\rho_{\mathbf{r}}(T^2ST), \rho_{\mathbf{r}}(TST^2)$ are not symmetric in the chosen basis, and hence we confirm the argument of Ref.~\cite{Feruglio:2012cw} that non-symmetric CP transformations consistent with the remnant $Z_2$ family symmetry in the neutrino sector lead to partially degenerate neutrino masses.

\end{itemize}
Since the remaining choices $G_{\nu}=Z^{T^2ST}_2$ or $G_{\nu}=Z^{TST^2}_2$ are related to the discussed case $G_{\nu}=Z^{S}_2$ by conjugation, the corresponding remnant CP symmetry is $\rho_{\mathbf{r}}(T^2)H^{\nu}_{\rm{CP}}\rho^{T}_{\mathbf{r}}(T^2)$ or
$\rho_{\mathbf{r}}(T)H^{\nu}_{\rm{CP}}\rho^{T}_{\mathbf{r}}(T)$, respectively, where
$H^{\nu}_{\rm{CP}}$ is given by Eq.~\eqref{eq:CP_Z2}. Then their corresponding neutrino mass matrices are of the form
$\rho^{*}_{\mathbf{3}}(T^2)m_{\nu}\rho^{\dagger}_{\mathbf{3}}(T^2)$ or
$\rho^{*}_{\mathbf{3}}(T)m_{\nu}\rho^{\dagger}_{\mathbf{3}}(T)$, respectively, with $m_{\nu}$ given in Eq.~\eqref{eq:nu_mass_Z2}. Now that we have finished a systematic discussion of the effects of the residual flavour and CP symmetries on the neutrino mass matrix, we turn to analyse their effects on the charged lepton mass matrix.

\subsection{Charged lepton sector  from a subgroup of $A_4\rtimes H_{\rm{CP}}$\label{sec:chargedlepton}}

In Appendices~\ref{GlZ2} and \ref{GlK4} we consider the cases $G_{l}=Z_2$ and $K_4$ and show that they are not phenomenologically viable. Here we consider the successful case that $G_{l}$ is one of the $Z_3$ subgroups shown in Eq.~\eqref{eq:Z3_subg}. Since the four $Z_3$ subgroups are conjugate to each other, i.e.
\begin{eqnarray}
\nonumber&&\hskip-0.5in
(TST^2)Z^{T}_3(TST^2)^{-1}=Z^{ST}_3,\quad
(T^2ST)Z^{T}_3(T^2ST)^{-1}=Z^{TS}_3,\quad
SZ^{T}_3S=Z^{STS}_3,\\
&&\hskip-0.5in SZ^{ST}_3S=Z^{TS}_S,\quad
(T^2ST)Z^{ST}_3(T^2ST)^{-1}=Z^{STS}_3,\quad
(TST^2)Z^{TS}_3(TST^2)^{-1}=Z^{STS}_3\,,
\end{eqnarray}
we choose $G_{l}=Z^{T}_3$ for demonstration.  Then the combined
symmetry group $A_4\rtimes H_{\rm{CP}}$ is broken to $G^{l}_{\rm{CP}}\cong
Z^{T}_3\rtimes H^{l}_{\rm{CP}}$ in the charged lepton sector. The element
$X_{\mathbf{r}l}$ of $H^{l}_{\rm{CP}}$ should satisfy the consistency
equation\footnote{The alternative
$X_{\mathbf{r}l}\rho^{*}_{\mathbf{r}}(T)X^{-1}_{\mathbf{r}l}=\rho_{\mathbf{r}}(T)$
is ruled out by the singlet representations $\mathbf{1}'$ and
$\mathbf{1}''$ as discussed below Eq.~\eqref{T}.}
\begin{equation}
X_{\mathbf{r}l}\rho^{*}_{\mathbf{r}}(T)X^{-1}_{\mathbf{r}l}=\rho_{\mathbf{r}}(T^2)\,.
\end{equation}
It is found that the remnant CP transformation $H^{l}_{\rm{CP}}$ can be
\begin{equation}
\label{eq:CP_Z3_cc}H^{l}_{\rm{CP}}=\left\{\rho_{\mathbf{r}}(1),
\rho_{\mathbf{r}}(T), \rho_{\mathbf{r}}(T^2)\right\}\,.
\end{equation}
Similar to the neutrino mass matrix, the charged lepton mass matrix $m_l$ must respect both the residual family symmetry $Z^{T}_3$ and the generalised CP symmetry $H^{l}_{\rm{CP}}$, i.e.
\begin{eqnarray}
\nonumber&&\rho^{\dagger}_{\mathbf{3}}(T)m_{l}m^{\dagger}_{l}\rho_{\mathbf{3}}(T)=m_{l}m^{\dagger}_{l},\\
\label{eq:cl_cons_remn}&&\rho^{\dagger}_{\mathbf{3}}(1)m_{l}m^{\dagger}_{l}\rho_{\mathbf{3}}(1)=(m_{l}m^{\dagger}_{l})^{*},
\end{eqnarray}
where $X_{\mathbf{r}l}=\rho_{\mathbf{r}}(1)$ from Eq. (\ref{eq:CP_Z3_cc}) has been taken. For the value
$X_{\mathbf{r}l}=\rho_{\mathbf{r}}(T)$ or $X_{\mathbf{r}l}=\rho_{\mathbf{r}}(T^2)$, the resulting constraint is
equivalent to Eq.~\eqref{eq:cl_cons_remn}. One can easily see that
$m_{l}m^{\dagger}_{l}$ is diagonal in this case,
\begin{equation}
\label{eq:cc_mm_Z3}m_{l}m^{\dagger}_{l}=\text{diag}(m^2_e,m^2_{\mu},m^2_{\tau})\,,
\end{equation}
where $m_e$, $m_{\mu}$ and $m_{\tau}$ are the electron, muon and tau masses,
respectively. For the other choices $G_{l}=Z^{ST}_3, Z^{TS}_3$ and
$Z^{STS}_3$, the corresponding residual CP symmetry and the mass matrix
$m_{l}m^{\dagger}_{l}$ follow from the general relations
Eq.~\eqref{eq:CP_conju} and Eq.~\eqref{eq:mass_matr_conju} immediately with
$h=TST^2, T^2ST$ and $S$, respectively.

\subsection{Lepton mixing from $A_4\rtimes H_{\rm{CP}}$ broken to $G^{\nu}_{\rm{CP}}\cong Z^{S}_2\times
H^{\nu}_{\rm{CP}}$ and $G^{l}_{\rm{CP}}\cong
Z^{T}_3\rtimes H^{l}_{\rm{CP}}$}

In the context of family symmetry and its extension of including generalised CP symmetry, a specific lepton mixing pattern arises from the mismatch
between the symmetry breaking in the neutrino and the charged lepton
sectors. In this section, we perform a comprehensive analysis of all possible lepton mixing matrices obtainable from the implementation of an $A_4$ family symmetry and its corresponding generalised CP symmetry by considering all possible residual symmetries $G^{\nu}_{\rm{CP}}$ and $G^{l}_{\rm{CP}}$ discussed in previous sections.

Immediately we can disregard the cases predicting partially degenerate lepton masses.  Therefore, breaking to the subgroups $G^{\nu}_{\rm{CP}}\cong K_4\rtimes H^{\nu}_{\rm{CP}}$ or $G^{l}_{\rm{CP}}\cong K_4\rtimes H^{l}_{\rm{CP}}$ will be neglected in the following. Furthermore, in order that the elements of $G_{\nu}$ and $G_{l}$ give rise to the entire family symmetry group $A_4$, we take $G_{l}$ to be one of the $Z_3$ subgroups shown in Eq.~\eqref{eq:Z3_subg}. Then, there are $3\times4=12$ combinations for $G_{\nu}=Z_2$ and $G_{l}=Z_3$. However, we find that all of these are conjugate to each other\footnote{For example, the choice $G'_{\nu}=Z^{T^2ST}_2, G'_{l}=Z^{TS}_3$ is conjugate to $G_{\nu}=Z^{S}_2, G_{l}=Z^{T}_3$ via $G'_{\nu}=(T^2S)G_{\nu}(T^2S)^{-1}$ and
$G'_{l}=(T^2S)G_{l}(T^2S)^{-1}$.}. As a result, all possible symmetry
breaking chains of this kind lead to the same lepton mixing matrix $U_{PMNS}$. This important point is further confirmed by straightforward calculations which are lengthy and tedious.

Without loss of generality, it is sufficient to consider the representative values $G_{\nu}=Z^{S}_2=\left\{1,S\right\}$ and $G_{l}=Z^{T}_3=\left\{1,T,T^2\right\}$, and the original symmetry $A_4\rtimes H_{\rm{CP}}$ is broken to $Z^{S}_2\times H^{\nu}_{\rm{CP}}$ in the neutrino sector and $Z^{T}_3\rtimes H^{l}_{\rm{CP}}$ in the charged lepton sector, where $H^{\nu}_{\rm{CP}}=\left\{\rho_{\mathbf{r}}(1),\rho_{\mathbf{r}}(S)\right\}$\footnote{$X_{\mathbf{r}\nu}=\left\{\rho_{\mathbf{r}}(T^2ST), \rho_{\mathbf{r}}(TST^2)\right\}$ leads to degenerate light neutrino masses, and it is ignored here.} and $H^{l}_{\rm{CP}}=\left\{\rho_{\mathbf{r}}(1),\rho_{\mathbf{r}}(T),\rho_{\mathbf{r}}(T^2)\right\}$.
In this case, $m_{l}m^{\dagger}_{l}$ is diagonal as shown in
Eq.~\eqref{eq:cc_mm_Z3}. Therefore, no rotation of the charged lepton fields is needed to get to the mass eigenstate basis, and the lepton mixing comes completely from the neutrino sector. In the PDG convention~\cite{pdg}, the PMNS matrix is cast in the form
\begin{equation}
U_{PMNS}=V\,\text{diag}(1,e^{i\frac{\alpha_{21}}{2}},e^{i\frac{\alpha_{31}}{2}}),
\label{eq:pmns_pdg}
\end{equation}
with
\begin{equation}
V=\left(\begin{array}{ccc}
c_{12}c_{13}  &   s_{12}c_{13}   &   s_{13}e^{-i\delta_{CP}}  \\
-s_{12}c_{23}-c_{12}s_{23}s_{13}e^{i\delta_{CP}}   &  c_{12}c_{23}-s_{12}s_{23}s_{13}e^{i\delta_{CP}}  &  s_{23}c_{13}  \\
s_{12}s_{23}-c_{12}c_{23}s_{13}e^{i\delta_{CP}}   & -c_{12}s_{23}-s_{12}c_{23}s_{13}e^{i\delta_{CP}}  &  c_{23}c_{13}
\end{array}\right).
\end{equation}
where we use the shorthand notation $c_{ij}=\cos\theta_{ij}$ and $s_{ij}=\sin\theta_{ij}$, $\delta_{CP}$ is the Dirac CP phase, $\alpha_{21}$ and $\alpha_{31}$ are the Majorana CP phases. Using this PDG convention we find that the resulting PMNS matrix is:
\begin{equation}
\label{eq:pmns_Z2_Z3}U_{PMNS}=U_{TB}R(\theta)P=\left(\begin{array}{ccc}
\frac{2}{\sqrt{6}}\cos\theta  &   \frac{1}{\sqrt{3}}  &
\frac{2}{\sqrt{6}}\sin\theta \\
-\frac{1}{\sqrt{6}}\cos\theta+\frac{1}{\sqrt{2}}\sin\theta  &
\frac{1}{\sqrt{3}}  &
-\frac{1}{\sqrt{6}}\sin\theta-\frac{1}{\sqrt{2}}\cos\theta  \\
-\frac{1}{\sqrt{6}}\cos\theta-\frac{1}{\sqrt{2}}\sin\theta &
\frac{1}{\sqrt{3}}  &
-\frac{1}{\sqrt{6}}\sin\theta+\frac{1}{\sqrt{2}}\cos\theta
\end{array}\right)P \,,
\end{equation}
where as shown previously $P$ is a unitary diagonal matrix with entries $\pm1$ or $\pm i$ and $R(\theta)$ and $U_{TB}$ are given in Eq.~\eqref{TB} and Eq.~\eqref{R}. Hence, the lepton mixing angles and CP phases are
\begin{eqnarray}
\nonumber&\hskip-0.2in\sin\delta_{\rm{CP}}=\sin\alpha_{21}=\sin\alpha_{31}=0,\\
\label{eq:mixing_parameters_Z2_Z3}&\hskip-0.2in\sin^2\theta_{13}=\frac{2}{3}\sin^2\theta,\quad
\sin^2\theta_{12}=\frac{1}{2+\cos2\theta}=\frac{1}{3\cos^2\theta_{13}},\quad
\sin^2\theta_{23}=\frac{1}{2}\left[1+\frac{\sqrt{3}\;\sin2\theta}{2+\cos2\theta}\right]\,,
\end{eqnarray}
which implies the three CP phases $\delta_{CP}$, $\alpha_{21}$, $\alpha_{31}=0,\pi$, and therefore there is no CP violation in this case. Note that the same results are found in Ref.~\cite{Feruglio:2012cw}.

To summarise the arguments of the preceding section, if one imposes the symmetry $A_4\rtimes H_{\rm{CP}}$, which is spontaneously broken to certain residual family and CP symmetries in order to obtain definite predictions for mixing angles and CP phases, then only the symmetry breaking of $A_4\rtimes H_{\rm{CP}}$ to $G^{\nu}_{\rm{CP}}\cong Z_2\times H^{\nu}_{\rm{CP}}$ in the neutrino sector and $G^{l}_{\rm{CP}}\cong Z_3\rtimes H^{l}_{\rm{CP}}$ in the charged lepton sector can lead to lepton mixing angles in the experimentally preferred range. However, there is no CP violation in this case.  This is consistent with the result found for $S_4\rtimes H_{\rm{CP}}$
for the case where $G^{\nu}_{\rm{CP}}\cong Z^{S}_2\times
H^{\nu}_{\rm{CP}}$ with $X_{\mathbf{r}\nu}=\left\{\rho_{\mathbf{r}}(1), \rho_{\mathbf{r}}(S)\right\}$
\cite{Ding:2013hpa}. For $S_4\rtimes H_{\rm{CP}}$ it was possible to achieve maximal CP violation
for the case $G^{\nu}_{\rm{CP}}\cong Z^{S}_2\times
H^{\nu}_{\rm{CP}}$ with $X_{\mathbf{r}\nu}=\left\{\rho_{\mathbf{r}}(U), \rho_{\mathbf{r}}(SU)\right\}$.
This case is not directly accessible for $A_4\rtimes H_{\rm{CP}}$
since the $U$ generator is absent, although it is accidentally present at LO in the models that we now discuss.

\section{\label{sec:model_effective}Model with $A_4$ and generalised CP
symmetries}
\cleqn

Guided by the general analysis of previous sections, we construct an effective model in this section. The predictions of Eq.~\eqref{eq:mixing_parameters_Z2_Z3} are realised
if the remnant CP is preserved otherwise the Dirac CP phase is approximately maximal. The model is based on $A_4\rtimes H_{\rm{CP}}$, which is supplemented by the extra symmetries $Z_4\times Z_6\times U(1)_R$. The auxiliary symmetry $Z_4\times Z_6$ separates the neutrino sector from the charged lepton sector, eliminates unwanted dangerous operators and it is also helpful to produce the mass hierarchy among the charged leptons. As usual both left-handed (LH) lepton doublets $l$ and the right-handed (RH) neutrinos $\nu^{c}$ are embedded into triplet representation $\mathbf{3}$, while the RH charged leptons $e^c$, $\mu^c$ and $\tau^c$ transform as the $A_4$ singlets $\mathbf{1}$, $\mathbf{1}''$ and $\mathbf{1}'$, respectively. All the fields of the model together with their assignments under the symmetry groups are listed in Table~\ref{tab:transformation_effective}.

It will be seen that in the ensuing model, the $A_4\rtimes H_{\rm{CP}}$ symmetry is broken to $Z^{S}_2\times H^{\nu}_{\rm{CP}}$ in the neutrino sector and $Z^{T}_3\rtimes H^{l}_{\rm{CP}}$ in the charged lepton sector at leading order. An accidental $Z_2^U$ symmetry, which is the $\mu-\tau$ exchange symmetry, arises due to the absence of flavons transforming as $\mathbf{1}'$ or $\mathbf{1}''$. As a result, the leading order (LO) lepton mixing is tri-bimaximal. The next-to-leading order (NLO) corrections will subsequently correct the mixing pattern, bringing it into agreement with experiment. In the following, we begin by analysing vacuum alignment and Yukawa operators of the model at LO, then turn to the NLO analysis.

\begin{table} [t!]
\begin{center}
{\footnotesize
\begin{tabular}{|c||c|c|c|c|c|c||c|c|c|c|c|c||c|c|c|c|c|}
\hline\hline

&   &     &    &    &    &    &  &    &   &    &    &     &     &    &    &
&      \\ [-0.15in]

{\tt Field}& $l$ &  $\nu^c$ &  $e^c$ & $\mu^c$ & $\tau^c$ & $h_{u,d}$ &
$\varphi_T$ &  $\zeta$ & $\varphi_S$ & $\xi(\tilde{\xi})$ &  $\chi$ &  $\rho$
& $\varphi^0_T$  & $\varphi^0_S$ & $\xi^0$  &  $\chi^{0}$   &   $\rho^{0}$
\\
\hline
$A_4$ & $\mathbf{3}$ & $\mathbf{3}$ & $\mathbf{1}$ & $\mathbf{1}''$ &
$\mathbf{1}'$ & $\mathbf{1}$  &  $\mathbf{3}$ & $\mathbf{1}$ & $\mathbf{3}$
& $\mathbf{1}$ &  $\mathbf{1}''$ & $\mathbf{1}$ &  $\mathbf{3}$ &
$\mathbf{3}$ & $\mathbf{1}$  &   $\mathbf{1}''$  &  $\mathbf{1}$ \\
\hline

$Z_4$ & $-1$ & $-1$ & $-i$ & 1 & $i$ & 1 & $i$ & $i$ & $1$ & $1$ & $1$ &
$1$ & $-1$  &  $1$ & $1$  & $1$ &  $1$  \\
\hline

$Z_6$  &  $\omega^4_6$  &  $\omega^2_6$  &  $\omega^2_6$  &  $\omega^2_6$  &
$\omega^2_6$  &  1  & 1 &  1 & $\omega^2_6$  &  $\omega^2_6$  &
$\omega^5_6$  &  $\omega^3_6$  &   $1$  &  $\omega^2_6$  &   $\omega^2_6$  &
$\omega^2_6$  & $1$  \\ \hline

$U(1)_R$ & $1$& $1$ & $1$ & $1$ & $1$ &  $0$ & $0$& $0$  & $0$ & $0$ & $0$ &
$0$ & $2$ & $2$ & $2$ & $2$ & $2$  \\
\hline \hline
\end{tabular}}
\caption{\label{tab:transformation_effective}Field content and their
transformation rules under the family symmetry $A_4 \times Z_4\times Z_6$
and $U(1)_R$, where $\omega_6=e^{2\pi i/6}$.}
\end{center}
\end{table}

\subsection{\label{subsec:vacuum_alignment}Vacuum alignment}

The vacuum alignment problem can be solved by the supersymmetric driving
field method introduced in Ref.~\cite{Altarelli:2005yx}. This approach
utilises a global  $U(1)_R$ continuous symmetry which contains the discrete
$R$-parity as a subgroup. The flavon and Higgs fields are uncharged
under $U(1)_R$, the matter fields have $R$ charge equal to $+1$ and the
so-called driving fields $\varphi^0_T$, $\varphi^0_S$, $\xi^0$, $\chi^0$ and
$\rho^0$ carry two units of $R$ charge. The most general driving
superpotential $w_d$ invariant under the family symmetry $A_4\times
Z_4\times Z_6$ can be written as
\begin{equation}
\label{eq:driving}w_d=w^{l}_d+w^{\nu}_d\,,
\end{equation}
where $w^{l}_d$ is the superpotential for the flavons entering the charged
lepton sector at leading order (LO), i.e.
\begin{equation}
w^{l}_d=f_1\left(\varphi^0_T\varphi_T\right)\zeta+f_2\left(\varphi^0_T\varphi_T\varphi_T\right)\,
\end{equation}
and $w^{\nu}_d$ is the superpotential involving the flavon fields of the
neutrino sector, i.e.
\begin{eqnarray}
\nonumber&&w^{\nu}_d=g_1\tilde{\xi}\left(\varphi^{0}_S\varphi_S\right)+g_2\left(\varphi^{0}_S\varphi_S\varphi_S\right)+g_3\xi^0\left(\varphi_S\varphi_S\right)+g_4\xi^0\xi^2+g_5\xi^0\xi\tilde{\xi}+g_6\xi^0\tilde{\xi}^2
\\
\label{eq:wd_nu_LO}&&\qquad+g_7\chi^0\left(\varphi_S\varphi_S\right)'+g_8\chi^0\chi^2+M^2_{\rho}\rho^0+g_9\rho^0\rho^2\,,
\end{eqnarray}
where the fields $\xi$ and $\tilde{\xi}$ are defined in such a way that only
the latter couples to the combination $\left(\varphi^{0}_S\varphi_S\right)$.
Notice that  $(\ldots)$ indicate a contraction to the singlet $\mathbf{1}$,  $(\ldots)'$ a contraction to the singlet $\mathbf{1}'$ and $(\ldots)''$ a contraction to the singlet $\mathbf{1}''$. Moreover, all couplings in $w_d$ are real, since we have imposed the generalised CP $H_{\rm{CP}}$ as a symmetry of the model. In the SUSY limit, the vacuum alignment is determined by the vanishing of the derivative of the driving superpotential $w_d$ with respect to each component of the driving field, i.e. the $F-$ terms of the driving fields must vanish. Therefore, the vacuum in the charged lepton sector is determined by
\begin{eqnarray}
\nonumber&&\frac{\partial
w_d}{\partial\varphi^0_{T_1}}=f_1\varphi_{T_1}\zeta+\frac{2}{3}f_2\left(\varphi^2_{T_1}-\varphi_{T_2}\varphi_{T_3}\right)=0\,,\\
\nonumber&&\frac{\partial
w_d}{\partial\varphi^0_{T_2}}=f_1\varphi_{T_3}\zeta+\frac{2}{3}f_2\left(\varphi^2_{T_2}-\varphi_{T_1}\varphi_{T_3}\right)=0\,,\\
&&\frac{\partial
w_d}{\partial\varphi^0_{T_3}}=f_1\varphi_{T_2}\zeta+\frac{2}{3}f_2\left(\varphi^2_{T_3}-\varphi_{T_1}\varphi_{T_2}\right)=0\,.
\end{eqnarray}
This set of equations admit two inequivalent solutions. The first solution is
\begin{equation}
\langle\zeta\rangle=0,\qquad\langle\varphi_T\rangle=v_T\left(\begin{array}{c}
1\\
1\\
1
\end{array}\right)\,,
\end{equation}
where $v_T$ is undetermined, and the second solution is
\begin{equation}
\label{eq:vev_charged}\langle\zeta\rangle=v_{\zeta},\qquad\langle\varphi_T\rangle=\left(\begin{array}{c}
v_T\\0\\
0
\end{array}\right)\quad \text{with} \quad v_T=-\frac{3f_1}{2f_2}v_{\zeta}\,.
\end{equation}
Note that the phase of $v_{\zeta}$ can be absorbed into the lepton fields.  Therefore we can take $v_{\zeta}$ to be real without loss of generality, and then the VEV $v_T$ is real as well. Since the couplings $f_1$ and $f_2$ naturally have absolute values of $\mathcal{O}(1)$, the vacuum expectation values (VEVs) $v_{\zeta}$ and $v_T$ are expected to be of the same order of magnitude. In the present work, we choose this solution and shall show that the mass hierarchies among the charged lepton masses can be naturally produced for
\begin{equation}
\frac{v_T}{\Lambda}\sim\frac{v_{\zeta}}{\Lambda}\sim\mathcal{O}(\lambda^2)\,,
\end{equation}
where $\lambda$ is of the order of Cabibbo angle $\theta_{c}\simeq0.23$.
Similarly the $F-$ term conditions for the flavon fields $\xi$,
$\tilde{\xi}$, $\varphi_S$ and $\chi$ are
\begin{eqnarray}
\nonumber&&\frac{\partial
w_d}{\partial\varphi^0_{S_1}}=g_1\tilde{\xi}\varphi_{S_1}+\frac{2}{3}g_2\left(\varphi^2_{S_1}-\varphi_{S_2}\varphi_{S_3}\right)=0\,,\\
\nonumber&&\frac{\partial
w_d}{\partial\varphi^0_{S_2}}=g_1\tilde{\xi}\varphi_{S_3}+\frac{2}{3}g_2\left(\varphi^2_{S_2}-\varphi_{S_1}\varphi_{S_3}\right)=0\,,\\
\nonumber&&\frac{\partial
w_d}{\partial\varphi^0_{S_3}}=g_1\tilde{\xi}\varphi_{S_2}+\frac{2}{3}g_2\left(\varphi^2_{S_3}-\varphi_{S_1}\varphi_{S_2}\right)=0\,,\\
\nonumber&&\frac{\partial
w_d}{\partial\xi^0}=g_3\left(\varphi^2_{S_1}+2\varphi_{S_2}\varphi_{S_3}\right)+g_4\xi^2+g_5\xi\tilde{\xi}+g_6\tilde{\xi}^2=0\,,\\
&&\frac{\partial
w_d}{\partial\chi^0}=g_7\left(\varphi^2_{S_3}+2\varphi_{S_1}\varphi_{S_2}\right)+g_8\chi^2=0\,.
\end{eqnarray}
Disregarding the ambiguity caused by  $A_4$ family symmetry
transformations, we find the solution
\begin{equation}
\label{eq:vev_neutrino}\langle\xi\rangle=v_{\xi},\quad
\langle\tilde{\xi}\rangle=0,\quad
\langle\varphi_S\rangle=v_S\left(\begin{array}{c}
1\\
1\\
1
\end{array}\right),\quad \langle\chi\rangle=v_{\chi}\,,
\end{equation}
where the VEVs $v_{\xi}$, $v_S$ and $v_{\chi}$ are related by
\begin{equation}
\label{eq:vev_relation}v^2_{S}=-\frac{g_4}{3g_3}v^2_{\xi},\qquad
v^2_{\chi}=\frac{g_4g_7}{g_3g_8}v^2_{\xi}\,,
\end{equation}
where $v_{\xi}$ is undetermined and generally complex. Consequently the
VEVs $v_S$ and $v_{\chi}$ are complex as well. Since all couplings are real due to the invariance under the generalised CP symmetry $H_{\rm{CP}}$,
the three VEVs $v_{\xi}$, $v_S$ and $v_{\chi}$ share the same
phase, up to the phase difference $0$, $\pi$ or $\pm\pi/2$ determined by the
sign of $g_3g_4$ and $g_7g_8$.\footnote{Note that it is possible to obtain more complicated phase differences by coupling more flavons together in the flavon potential~\cite{Antusch:2011sx}. Consequently the corresponding driving superpotential becomes non-renormalisable. For example, if Eq. (\ref{eq:vev_relation}) instead appeared schematically as $v^3_{S}\sim v^3_{\xi}$, then phase differences of $\frac{2k\pi}{3}$ and $\frac{\left(2k-1\right)\pi}{3}$ with $k=1,2,3$ could be obtained. More generally, if one obtains a relation like $v^p_{S}\sim v^p_{\xi}$, then phase differences of $\frac{2k\pi}{p}$ and $\frac{\left(2k-1\right)\pi}{p}$ with $k=1,2,\dots, p$ could be realised.}

Finally, the minimisation equation for the vacuum of $\rho$ is
\begin{equation}
\frac{\partial w_d}{\partial\rho^0}=M^2_{\rho}+g_9\rho^2=0\,,
\end{equation}
which leads to
\begin{equation}
\label{eq:vev_rho}\langle\rho\rangle=v_{\rho},\quad \text{with}\quad
v^2_{\rho}=-M^2_{\rho}/g_{9}\,.
\end{equation}
Obviously the VEV $v_{\rho}$ can only be real or purely imaginary depending
on the coupling $g_9$ being negative or positive, respectively. As we shall
see, agreement with the experimental data (in particular
the measured sizeable $\theta_{13}$) can be achieved if
\begin{equation}
\frac{v_{\xi}}{\Lambda}\sim\frac{v_{S}}{\Lambda}\sim\frac{v_{\chi}}{\Lambda}\sim\frac{v_{\rho}}{\Lambda}\sim\mathcal{O}(\lambda)\,.
\end{equation}
Thus, there is a moderate hierarchy of order $\lambda$ between the VEVs of
the flavon fields in the neutrino and the charged lepton sectors. This
hierarchy can be accommodated since the two sets of VEVs are determined by
different minimisation conditions. Now that we have studied the vacuum alignments possible in this model, we proceed by constructing the explicit charged lepton and neutrino mass matrices.

\subsection{The model at leading order}

From Table \ref{tab:transformation_effective}, it is seen that the effective superpotential for the charged lepton masses is given by
\begin{eqnarray}
\nonumber&&\hskip-0.35in
w_l=\frac{y_{\tau}}{\Lambda}\left(l\varphi_T\right)''\tau^ch_d+\frac{y_{\mu_1}}{\Lambda^2}\left(l\varphi_T\varphi_T\right)'\mu^ch_d+\frac{y_{\mu_2}}{\Lambda^2}\left(l\varphi_T\right)'\zeta\mu^ch_d
+\frac{y_{e_1}}{\Lambda^3}\left(l\varphi_T\right)\left(\varphi_T\varphi_T\right)e^ch_d\\
\nonumber&&\hskip-0.35in\quad+\frac{y_{e_2}}{\Lambda^3}\left(l\varphi_T\right)'\left(\varphi_T\varphi_T\right)''e^ch_d+\frac{y_{e_3}}{\Lambda^3}\left(l\varphi_T\right)''\left(\varphi_T\varphi_T\right)'e^ch_d
+\frac{y_{e_4}}{\Lambda^3}\left(\left(l\varphi_T\right)_{3_S}\left(\varphi_T\varphi_T\right)_{3_S}\right)e^ch_d\\
\label{eq:wl_LO}&&\hskip-0.35in\quad+\frac{y_{e_5}}{\Lambda^3}\left(\left(l\varphi_T\right)_{3_A}\left(\varphi_T\varphi_T\right)_{3_S}\right)e^ch_d+\frac{y_{e_6}}{\Lambda^3}\left(l\varphi_T\varphi_T\right)\zeta e^ch_d+\frac{y_{e_7}}{\Lambda^3}\left(l\varphi_T\right)\zeta^2e^ch_d+\ldots\,,
\end{eqnarray}
where dots represent the higher dimensional operators which will be
discussed later, and all coupling constants are constrained to be real
by the generalised CP symmetry. Due to the auxiliary $Z_4$ symmetry, the
relevant electron, muon and tau mass terms involve one flavon, two flavons
and three flavons, respectively. Substituting the VEVs of $\varphi_T$ and
$\zeta$ in Eq.~\eqref{eq:vev_charged}, a diagonal charged lepton mass matrix is generated with
\begin{eqnarray}
\nonumber&&m_e=\left(y_{e_1}+\frac{4}{9}y_{e_4}+\frac{2}{3}y_{e_6}\frac{v_{\zeta}}{v_T}+y_{e_7}\frac{v^2_{\zeta}}{v^2_T}\right)\frac{v^3_T}{\Lambda^3}v_d\,,\\
&&m_{\mu}=\left(\frac{2}{3}y_{\mu_1}+y_{\mu_2}\frac{v_{\zeta}}{v_T}\right)\frac{v^2_T}{\Lambda^2}v_d\,,\qquad
m_{\tau}=y_{\tau}\frac{v_T}{\Lambda}v_d\,,
\end{eqnarray}
where $v_d=\langle h_d\rangle$. The VEVs of the flavons $\varphi_T$ and $\zeta$ are responsible for the spontaneous breaking of both
family symmetry and generalised CP symmetry here. Furthermore, it is obvious that the $A_4$ family symmetry is broken to the $Z^{T}_3$ subgroup in the charged lepton sector. As was pointed out in the vacuum alignment of Section~\ref{subsec:vacuum_alignment}, both $v_{T}$ and $v_{\zeta}$ can be set to be real. Therefore the generalised CP symmetry is broken to $H^{l}_{\rm{CP}}=\left\{\rho _{\mathbf{r}}(1),\rho _{\mathbf{r}}(T),\rho _{\mathbf{r}}(T^2)\right\}$ in the charged lepton sector. It is remarkable that the observed charged lepton mass hierarchies are naturally reproduced for $v_T/\Lambda\sim v_{\zeta}/\Lambda\sim\lambda^2$. In the following, we turn to discuss the neutrino sector. Neutrino masses are generated by the seesaw mechanism~\cite{seesaw}, and the LO superpotential for the neutrino masses, which is invariant under the imposed family symmetry $A_4\times Z_4\times Z_6$, is of the form
\begin{equation}
\label{eq:neutrino_LO}w_{\nu}=y\left(l\nu^c\right)h_u+y_1\left(\nu^c\nu^c\right)\xi+\tilde{y}_1\left(\nu^c\nu^c\right)\tilde{\xi}+y_3\left(\nu^c\nu^c\varphi_S\right)\,,
\end{equation}
where all couplings are real because of invariance under the
generalised CP transformations defined in Section~\ref{sec:cp}. We
can straightforwardly read out the Dirac neutrino mass matrix,
\begin{equation}
m_D=y\left(\begin{array}{ccc}
1  &  0   &  0  \\
0  &  0   &  1  \\
0  &  1   &  0
\end{array}\right)v_u\,,
\end{equation}
where $v_u=\langle h_u\rangle$ is the VEV of the Higgs field $h_u$. Given
the vacuum configuration of Eq.~\eqref{eq:vev_neutrino}, which breaks the
$A_4$ family symmetry to $G_{\nu}=Z^{S}=\{1,S\}$, the Majorana neutrino
mass matrix $m_M$ for the heavy RH neutrinos is
\begin{equation}
\label{eq:mM}m_M=\left(\begin{array}{ccc}
y_1v_{\xi}+2y_3v_{S}/3  &  -y_3v_{S}/3  &  -y_3v_{S}/3 \\
-y_3v_{S}/3   & 2y_3v_{S}/3   &  y_1v_{\xi}-y_3v_{S}/3\\
-y_3v_{S}/3  &  y_1v_{\xi}-y_3v_{S}/3  &  2y_3v_{S}/3
\end{array}\right)\,.
\end{equation}
Notice that this mass matrix also has an accidental $Z_2^U$ symmetry, which is the $\mu-\tau$ exchange symmetry, arising due to the absence of flavons transforming as $\mathbf{1}'$ or $\mathbf{1}''$.
It is exactly diagonalised by the tri-bimaximal mixing matrix $U_{TB}$, i.e.
\begin{equation}
U^{T}_{TB}m_MU_{TB}=\text{diag}(y_1v_{\xi}+y_3v_S,y_1v_{\xi},-y_1v_{\xi}+y_{3}v_S)\,.
\end{equation}
Then, the light neutrino mass matrix follows from the seesaw formula
\begin{equation}
m_{\nu}=-m_Dm^{-1}_Mm^{T}_D=U_{TB}\text{diag}(m_1,m_2,m_3)U^{T}_{TB}\,,
\end{equation}
where
\begin{equation}
\label{eq:neutrino_mass_LO}m_1=-\frac{y^2v^2_u}{y_1v_{\xi}+y_3v_S},\quad
m_2=-\frac{y^2v^2_u}{y_1v_{\xi}},\quad
m_3=\frac{y^2v^2_{u}}{y_1v_{\xi}-y_3v_S}
\end{equation}
Note that these masses obey the mass sum rule
\begin{equation}
\label{eq:sumrule}
\frac{1}{m_1}-\frac{1}{m_3}=\frac{2}{m_2}.
\end{equation}
However the sum rule will be violated by NLO corrections.

Recalling that the charged lepton mass matrix is diagonal, therefore lepton
flavour mixing is predicted to be of the tri-bimaximal form at LO. Since the
common phase of $v_{\xi}$ and $v_S$ can always be absorbed by a redefinition
of the fields, we can take the product $y_1v_{\xi}$ to be real without loss
of generality. Then $y_3v_S$ will be either real or purely imaginary
depending on $g_3g_4$ being negative or positive, as shown in
Eq.~\eqref{eq:vev_relation}. For the case that $y_3v_S$ is imaginary, we can
easily check that the remnant CP symmetry in the neutrino sector is
$H^{\nu}_{\rm{CP}}=\{\rho_{\mathbf{r}}(T^2ST),\rho_{\mathbf{r}}(TST^2)\}$, and we have $|m_1|=|m_3|$ from Eq.~\eqref{eq:neutrino_mass_LO}, which implies the light neutrino masses are degenerate. Therefore this case is not phenomenologically viable, and it will be disregarded in the following.

\begin{table} [t!]
\begin{center}
\begin{tabular}{|c|c|c|c|c|c|c|c|}
\hline\hline
$x$ &  $~\alpha_{21}~$  &  $~\alpha_{31}~$ &  $|m_1|\text{(meV)}$ &
 $|m_2|\text{(meV)}$ & $|m_3|\text{(meV)}$ & $|m_{\beta\beta}|\text{(meV)}$
 &
   \text{mass order} \\ \hline
0.79  & 0 & $\pi$ & 5.83 & 10.44 & 50.07  &  7.36 & \text{NO} \\ \hline
1.19 & 0  &  0  &  4.43 & 9.73 & 49.93  &  6.20 & \text{NO} \\ \hline
$-2.01$ &  $\pi$  & 0  &  51.50 & 52.22 & 17.33 & 16.93 & \text{IO}  \\
 \hline \hline
\end{tabular}
\caption{\label{tab:LO_predictions}The LO predictions for the Majorana
phases $\alpha_{21}$ and $\alpha_{31}$, the light neutrino masses $|m_i|(i=1,2,3)$ and the effective mass
$|m_{\beta\beta}|$ of the neutrinoless double-beta decay, where
$x=y_3v_S/(y_1v_{\xi})$. Note that $\delta_{CP}$ is undetermined due to vanishing $\theta_{13}$ at LO.}
\end{center}
\end{table}

Hence we are left with the case that $v_{\xi}$ and $v_{S}$ are of the same phase up to relative sign, and then the generalised CP symmetry is broken to
$H^{\nu}_{\rm{CP}}=\{\rho_{\mathbf{r}}(1),\rho_{\mathbf{r}}(S)\}$ at LO. The neutrino mass-squared differences are given by
\begin{eqnarray}
\nonumber&&\Delta m^2_{sol}\equiv
|m_2|^2-|m_1|^2=\left(\frac{y^2v^2_u}{y_1v_{\xi}}\right)^2\frac{x^2+2x}{\left(1+x\right)^2}\,,\\
\nonumber&&\Delta
m^2_{atm}\equiv|m_3|^2-|m_1|^2=\left(\frac{y^2v^2_u}{y_1v_{\xi}}\right)^2\frac{4x}{\left(1-x^2\right)^2},\quad\text{for
NO}\,,\\
&&\Delta
m^2_{atm}\equiv|m_2|^2-|m_3|^2=\left(\frac{y^2v^2_u}{y_1v_{\xi}}\right)^2\frac{x^2-2x}{\left(1-x\right)^2},\quad\text{for
IO}\,,
\end{eqnarray}
where $x=y_{3}v_S/(y_1v_{\xi})$ is real. Furthermore, the effective mass
parameter $|m_{\beta\beta}|$ for the neutrinoless double-beta decay is given by
\begin{equation}
|m_{\beta\beta}|=\left|\frac{y^2v^2_u}{y_1v_{\xi}}\right|\left|\frac{3+x}{3\left(1+x\right)}\right|.
\end{equation}
Since the solar neutrino mass squared difference $\Delta m^2_{sol}$ is
positive, we need $x>0$ or $x<-2$. The neutrino spectrum is normal ordering
(NO) for $x>0$ and inverted ordering (IO) for $x<-2$. Imposing the best fit
values for the mass splittings $\Delta
m^2_{sol}=7.50\times10^{-5}\text{eV}^2$ and $\Delta
m^2_{atm}=2.473(2.427)\times10^{-3}\text{eV}^2$ for NO (IO)
spectrum~\cite{GonzalezGarcia:2012sz}, we find three possible values for the
ratio $x$:
\begin{equation}
x\simeq0.792, 1.195, -2.014\,,
\end{equation}
where the first two correspond to NO, while the last one corresponds to IO
spectrum. The corresponding predictions for Majorana phases, the light neutrino masses and $|m_{\beta\beta}|$ are listed in Table~\ref{tab:LO_predictions}. Note that the Dirac phase can not be fixed uniquely in this case because of the vanishing $\theta_{13}$.

Recall that for $S_4\rtimes H_{\rm{CP}}$ it was possible to achieve $\delta_{CP}=\pm \pi /2$
for the case $G^{\nu}_{\rm{CP}}\cong Z^{S}_2\times
H^{\nu}_{\rm{CP}}$ with $X_{\mathbf{r}\nu}=\left\{\rho_{\mathbf{r}}(U), \rho_{\mathbf{r}}(SU)\right\}$ \cite{Ding:2013hpa}.
Although this case is not directly accessible for $A_4\rtimes H_{\rm{CP}}$
since the $U$ generator is absent, we note that at LO the neutrino mass matrix in Eq.~\eqref{eq:mM} has an accidental $X_{\mathbf{r}\nu}=\left\{\rho_{\mathbf{r}}(U), \rho_{\mathbf{r}}(SU)\right\}$ CP symmetry.
This leads to the same prediction for Majorana phases
$\alpha_{21}=0,\pi$ and $\alpha_{31}=0,\pi$ as in the $S_4\rtimes H_{\rm{CP}}$ model.

\subsection{\label{sec:NLO}Next-to-Leading-Order corrections}

In the following, we study the subleading NLO corrections to the previous
superpotentials, which are essential to bring the model into agreement with
data. As will be seen, these corrections will produce a non-zero reactor angle $\theta_{13}$ whose relative smallness with respect to $\theta_{12}$ and $\theta_{23}$ is naturally explained by its generation at NLO. The subleading corrections are indicated by higher dimensional operators which are compatible with all symmetries of the model. The NLO contribution to the driving superpotential $w^{\nu}_d$ is suppressed by one power of $1/\Lambda$ with respect to the LO terms in Eq.~\eqref{eq:wd_nu_LO}, and it is of the form
\begin{eqnarray}
\nonumber&&\delta
w^{\nu}_d=\frac{s}{\Lambda}\left(\varphi^0_S\varphi_S\right)'\chi\rho+\frac{r_1}{\Lambda}\rho^{0}\left(\varphi_S\varphi_S\varphi_S\right)
+\frac{r_2}{\Lambda}\rho^0\left(\varphi_S\varphi_S\right)\xi+\frac{r_3}{\Lambda}\rho^0\left(\varphi_S\varphi_S\right)\tilde{\xi}+\frac{r_4}{\Lambda}\rho^0\xi^3\\
\label{eq:NLO_wdnu}&&\qquad+\frac{r_5}{\Lambda}\rho^0\xi^2\tilde{\xi}+\frac{r_6}{\Lambda}\rho^0\xi\tilde{\xi}^2+\frac{r_7}{\Lambda}\rho^0\tilde{\xi}^3\,,
\end{eqnarray}
where the coupling $s$ and $r_i(i=1\ldots7)$ are real due to the generalised
CP symmetry. The LO vacuum configuration is modified to
\begin{equation}
\langle\xi\rangle=v_{\xi},\quad \langle\tilde{\xi}\rangle=\delta
v_{\tilde{\xi}},\quad \langle\varphi_S\rangle=\left(\begin{array}{c}
v_{S}+\delta v_{S_1} \\
v_{S}+\delta v_{S_2} \\
v_{S}+\delta v_{S_3}
\end{array}\right),\quad \langle\chi\rangle=v_{\chi}+\delta v_{\chi},\quad
\langle\rho\rangle=v_{\rho}+\delta v_{\rho}\,,
\end{equation}
where the VEV of $\xi$ remains undetermined. The new vacuum configuration is
determined by the vanishing of the first derivative of $w^{\nu}_d+\delta
w^{\nu}_d$ with respect to the driving fields $\varphi^{0}_S$, $\xi^0$,
$\chi^{0}$ and $\rho^{0}$. Keeping only the terms linear in the shift
$\delta v$ and neglecting the term $\delta v/\Lambda$, we find
\begin{eqnarray}
\nonumber&\delta v_{S_1}=\delta v_{S_2}=\delta
v_{S_3}=\frac{sg_5}{6g_1g_3}\frac{v_{\xi}}{v_S}\frac{v_{\chi}v_{\rho}}{\Lambda}\equiv\delta
v_S,\\
\label{eq:vev_neutrino_NLO}&\delta
v_{\tilde{\xi}}=-\frac{s}{g_1}\frac{v_{\chi}v_{\rho}}{\Lambda},\quad \delta
v_{\chi}=-\frac{sg_5g_7}{2g_1g_3g_8}\frac{v_{\xi}v_{\rho}}{\Lambda},\quad
\delta v_{\rho}=\frac{g_4r_2-g_3r_4}{2g_3g_9}\frac{v^3_{\xi}}{\Lambda
v_{\rho}}\,.
\end{eqnarray}
We see that the three components of $\varphi_S$ are shifted by the same
amount.  This implies that the vacuum alignment of $\varphi_S$ is not changed. The reason for this is
that only the neutrino flavon fields $\varphi_S$, $\xi$, $\tilde{\xi}$,
$\chi$ and $\rho$ instead of $\varphi_T$ enter into the NLO operators of
Eq.~\eqref{eq:NLO_wdnu}. Hence, the remnant family symmetry $Z^{S}_2=\{1,S\}$ in the neutrino sector is still preserved. This implies $\langle\varphi_{S}\rangle\propto(1,1,1)$. Furthermore, Eq.~\eqref{eq:vev_neutrino_NLO} indicates that $\delta v_{S}$, $\delta v_{\tilde{\xi}}$, $\delta v_{\chi}$ and $\delta v_{\rho}$ are of
order $\lambda^2\Lambda$, i.e. the shifts of the flavon fields in the
neutrino sector are of relative order $\lambda$ with respect to the LO VEVs.
For the driving superpotential $w^{l}_d$, the nontrivial subleading operators, whose contributions can not be absorbed via a redefinition of the LO parameters, are of the form:
\begin{eqnarray}
\nonumber&\left(\varphi^{0}_T\varphi^2_T\varphi^3_{\nu}\right)/\Lambda^3,\qquad
\left(\varphi^0_T\varphi_T\varphi^3_{\nu}\right)\zeta/\Lambda^3,\qquad
\left(\varphi^0_T\varphi^3_{\nu}\right)\zeta^2/\Lambda^3, \\
&\left(\varphi^0_T\varphi^2_T\varphi_{\nu}\right)''\chi^2/\Lambda^3,\qquad
\left(\varphi^0_T\varphi_T\varphi_{\nu}\right)''\chi^2\zeta/\Lambda^3,\qquad
\left(\varphi^0_T\varphi_{\nu}\right)''\chi^2\zeta^2/\Lambda^3 \,,
\end{eqnarray}
where $\varphi_{\nu}=\{\varphi_S,\xi,\tilde{\xi}\}$ denotes the flavon
involved in the neutrino sector at LO. Therefore subleading
contributions to the $F-$terms of the driving field $\varphi^0_T$ are suppressed by $\langle\varphi_{\nu}\rangle^3/\Lambda^3\sim\lambda^3$
with respect to the LO renormalisable terms in $w^{l}_d$. As a result, the vacuum alignment of $\varphi_T$ acquires corrections of order $\lambda^3$:
\begin{equation}
\langle\varphi_T\rangle=v_T\left(\begin{array}{c}
1+\epsilon_1\lambda^3 \\
\epsilon_2\lambda^3 \\
\epsilon_3\lambda^3
\end{array}\right)\,,
\end{equation}
where $\epsilon_i (i=1,2,3)$ are complex numbers with absolute value of $\mathcal{O}(1)$. Inserting this modified vacuum of $\varphi_T$ into the LO expression of $w_l$ in Eq.~\eqref{eq:wl_LO}, the off-diagonal elements of the charged lepton mass matrix become non-zero and are all suppressed by $\lambda^3$ with respect to the diagonal entries. Consequently, the corrected charged lepton mass matrix has the following structure:
\begin{equation}
\label{eq:ml_NLO}m_{l}=\left(\begin{array}{ccc}
m_e  &  \lambda^3 m_{\mu}  &  \lambda^3 m_{\tau}  \\
\lambda^3 m_e  &  m_{\mu}   &  \lambda^3 m_{\tau} \\
\lambda^3 m_e  & \lambda^3 m_{\mu}  &  m_{\tau}
\end{array}\right)\,,
\end{equation}
where only the order of magnitude of each non-diagonal entry is reported.
Therefore the lepton mixing angles receive corrections of order $\lambda^3$
from the charged lepton sector. These can be safely neglected. Another source of correction to the charged lepton mass matrix comes from adding the
product $\varphi^3_{\nu}$ or $\varphi_{\nu}\chi^2$ in all possible ways to
each term of $w_l$. However, the introduction of these additional terms
changes the charged lepton mass matrix in exactly the same way as the
corrections induced by the VEV shifts of $\varphi_T$. Therefore, the general structure of $m_{l}$ shown in Eq.~\eqref{eq:ml_NLO} remains.

Now we turn to study the corrections to the neutrino sector. The higher
order corrections to the neutrino Dirac mass are given by\footnote{The
operator $\left(l\nu^c\right)\rho^2h_u/\Lambda^2$ is omitted here, since its
contribution can be absorbed by redefining the LO parameter $y$ of
Eq.~\eqref{eq:neutrino_LO}.}
\begin{equation}
\left(l\nu^{c}\varphi^3_{\nu}\right)h_u/\Lambda^3+\left(l\nu^c\varphi_{\nu}\right)''\chi^2h_u/\Lambda^3\,,
\end{equation}
where all possible $A_4$ contractions should be considered, and we have
suppressed all real coupling constants. The resulting contributions are
of relative order $\lambda^3$ with respect to the LO term
$y\left(l\nu^c\right)h_u$ in Eq.~\eqref{eq:neutrino_LO} and therefore
negligible. The NLO corrections to the RH Majorana neutrino mass are
\begin{equation}
\label{eq:wnu_NLO}\delta
w_{\nu}=\tilde{y}_1\left(\nu^c\nu^c\right)\delta\tilde{\xi}+y_3\left(\nu^c\nu^c\delta\varphi_S\right)+y_4\left(\nu^c\nu^c\right)'\chi\rho/\Lambda\,,
\end{equation}
where $\delta\tilde{\xi}$ and $\delta\varphi_S$ indicate the shifted vacua
of the flavons $\tilde{\xi}$ and $\varphi_S$. They lead to additional
contributions to $m_M$ as follows:
\begin{equation}
\delta m_M=\left(\begin{array}{ccc}
\tilde{y}_1\delta v_{\tilde{\xi}}+2y_3\delta v_{S}/3  &  -y_3\delta
v_{S}/3+y_4v_{\chi} v_{\rho}/\Lambda  &   -y_3\delta v_{S}/3 \\
-y_3\delta v_{S}/3+y_4v_{\chi}v_{\rho}/\Lambda  & 2y_3\delta v_{S}/3  &
\tilde{y}_1\delta v_{\tilde{\xi}} -y_3\delta v_{S}/3 \\
-y_3\delta v_{S}/3   &   \tilde{y}_1\delta v_{\tilde{\xi}}-y_3\delta v_{S}/3
&  2y_3\delta v_{S}/3+y_4v_{\chi}v_{\rho}/\Lambda
\end{array}\right)\,.
\end{equation}

Notice that this mass matrix breaks the accidental $Z_2^U$ symmetry, which is the $\mu-\tau$ exchange symmetry, arising due to the presence of the $\chi$ flavon transforming as $\mathbf{1}''$, allowing a non-zero reactor angle. It also breaks the accidental $X_{\mathbf{r}\nu}=\left\{\rho_{\mathbf{r}}(U), \rho_{\mathbf{r}}(SU)\right\}$ CP symmetry.
In fact, since we have fewer parameters in the neutrino mass matrix than in the $S_4$ case, we cannot preserve an accidental  $X_{\mathbf{r}\nu}=\left\{\rho_{\mathbf{r}}(U), \rho_{\mathbf{r}}(SU)\right\}$ CP symmetry
whilst breaking the accidental $Z_2^U$ family symmetry. It will therefore lead to different predictions for Majorana phases $\alpha_{21}\neq 0,\pi$ and $\alpha_{31}\neq 0,\pi$ compared to the $S_4\rtimes H_{\rm{CP}}$ model,
however with $\theta_{13}\neq 0$ we will allow the possibility that $\delta_{CP}=\pm \pi/2$ which can be understood from the discussion below Eq.~\eqref{eq:vev_relation}.

As shown in Eq.~\eqref{eq:vev_rho}, the VEV $v_{\rho}$ is real for $g_9<0$
and imaginary for $g_9>0$. Eq.~\eqref{eq:vev_relation} implies that the
phase difference between $v_{\chi}$ and $v_{\xi}$ is $0$, $\pi$ or
$\pm\pi/2$ for the product $g_3g_4g_7g_8>0$ or $g_3g_4g_7g_8<0$,
respectively. Hence the combination $v_{\chi}v_{\rho}$ is real or purely
imaginary once the phase of $v_{\xi}$ is absorbed by redefining the fields.

First, we consider the case that $v_{\chi}v_{\rho}$ is real\footnote{We
could choose $g_9<0$ and $g_3g_4g_7g_8>0$ such that $v_{\chi}$ and $v_{\xi}$
have a common phase up to relative sign and $v_{\rho}$ is real. Consequently
the symmetry $A_4\rtimes H_{\rm{CP}}$ is broken down to
$G^{\nu}_{\rm{CP}}=Z^{S}_2\times H^{\nu}_{\rm{CP}}$ in the neutrino sector with $H^{\nu}_{\rm{CP}}=\left\{\rho_{\mathbf{r}}(1),\rho_{\mathbf{r}}(S)\right\}$. On the other hand, this case can also be realised by taking $g_9>0$ and
$g_3g_4g_7g_8<0$ such that $v_{\rho}$ is imaginary and the phase difference
of $v_{\chi}$ and $v_{\xi}$ is $\pm\pi/2$.}, i.e. the phase difference
between $v_{\chi}v_{\rho}$ and $v_{\xi}$ is 0 or $\pi$, and then both $\delta v_{\tilde{\xi}}$ and $\delta v_S$ will be also real from Eq.~\eqref{eq:vev_neutrino_NLO}. Further recalling that $v_{\xi}$ and $v_S$ should have a common phase to avoid degenerate light neutrino masses, the NLO contributions carry the same phase (up to relative sign) as the LO contribution from Eq.~\eqref{eq:mM} in this case. The corrections due to shifted vacuum of $\tilde{\xi}$ and $\varphi_S$ can be absorbed by a redefinition of the couplings $y_1$ and $y_3$ respectively. Thus the RH neutrino mass matrix $m_M$ including NLO contributions can be parametrised as
\begin{equation}
m_M=\left(\begin{array}{ccc}
\hat{y}_1v_{\xi}+2\hat{y}_3v_S/3 &
-\hat{y}_3v_S/3+y_4v_{\chi}v_{\rho}/\Lambda  & -\hat{y}_3v_S/3 \\
-\hat{y}_3v_S/3+y_4v_{\chi}v_{\rho}/\Lambda  &  2\hat{y}_3v_S/3  &
\hat{y}_1v_{\xi}-\hat{y}_3v_S/3 \\
-\hat{y}_3v_S/3  &  \hat{y}_1v_{\xi}-\hat{y}_3v_S/3   &  2\hat{y}_3v_S/3
+y_4v_{\chi}v_{\rho}/\Lambda
\end{array}\right)\,,
\end{equation}
where $\hat{y}_1=y_1+\tilde{y}_1\delta v_{\tilde{\xi}}/v_{\xi}$ and
$\hat{y}_3=y_3(1+\delta v_S/v_S)$ are real. The light neutrino mass matrix
is given by the seesaw relation
\begin{eqnarray}
\nonumber m_{\nu}&=&-m_{D}m^{-1}_{M}m^{T}_D\\
\label{eq:neutrino_NLO}&=&\alpha\left(\begin{array}{ccc}
2  &  -1  &  -1  \\
-1 &  2   &  -1  \\
-1 &   -1 & 2
\end{array}\right)+\beta\left(\begin{array}{ccc}
1 &  0  &  0  \\
0 &  0  &  1  \\
0 &  1  & 0
\end{array}\right)+\gamma\left(\begin{array}{ccc}
0 & 1  & 1  \\
1 & 1  & 0  \\
1 & 0  & 1
\end{array}\right)+\epsilon\left(\begin{array}{ccc}
0 &  1 &  -1  \\
1 &  -1 & 0   \\
-1 &  0 &  1
\end{array}\right)\,.
\end{eqnarray}
It is the most general neutrino mass matrix consistent with the residual
family symmetry $G_{\nu}=Z^{S}_2=\{1,S\}$, as is shown in
Eq.~\eqref{eq:nu_mass_Z2}. The parameters $\alpha$, $\beta$, $\gamma$ and
$\epsilon$ can be regarded as real and are given by
\begin{eqnarray}
\label{1}
\nonumber&&\alpha=\frac{\hat{y}_3v_S}{3\left(\hat{y}^2_1v^2_{\xi}-\hat{y}^2_3v^2_S-\hat{y}_1y_4v_{\xi}v_{\chi}v_{\rho}/\Lambda+y^2_4v^2_{\chi}v^2_{\rho}/\Lambda^2\right)},\\
\nonumber&&\beta=\frac{-3\hat{y}^2_1v^2_{\xi}+\hat{y}^2_3v^2_S}{3\left(\hat{y}^3_1v^3_{\xi}-\hat{y}_1\hat{y}^2_3v_{\xi}v^2_S-\hat{y}^2_3y_4v^2_Sv_{\chi}v_{\rho}/\Lambda+y^3_4v^3_{\chi}v^3_{\rho}/\Lambda^3\right)},\\
\nonumber&&\gamma=\frac{2\hat{y}^2_3v^2_S+3\hat{y}_1y_4v_{\xi}v_{\chi}v_{\rho}/\Lambda-3y^2_4v^2_{\chi}v^2_{\rho}/\Lambda^2}{6\left(\hat{y}^3_1v^3_{\xi}-\hat{y}_1\hat{y}^2_3v_{\xi}v^2_S-\hat{y}^2_3y_4v^2_Sv_{\chi}v_{\rho}/\Lambda+y^3_4v^3_{\chi}v^3_{\rho}/\Lambda^3\right)},\\
&&\epsilon=\frac{y_4v_{\chi}v_{\rho}/\Lambda}{2\left(\hat{y}^2_1v^2_{\xi}-\hat{y}^2_3v^2_S-\hat{y}_1y_4v_{\xi}v_{\chi}v_{\rho}/\Lambda+y^2_4v^2_{\chi}v^2_{\rho}/\Lambda^2\right)}\,,
\end{eqnarray}
where the overall factor $y^2v^2_u$ has been omitted. We note that the
$\epsilon$ term in Eq.~\eqref{eq:neutrino_NLO}, which is induced by the last
term of the NLO corrections in Eq.~\eqref{eq:wnu_NLO}, is responsible for
the non-zero reactor angle $\theta_{13}$.  It is suppressed by $\lambda$
with respect to the tri-bimaximal mixing preserving contributions $\alpha$,
$\beta$ and $\gamma$ terms. Neglecting the small contributions from the
charged lepton sector, the PMNS matrix is of the form shown in Eq.~\eqref{eq:pmns_Z2_Z3}, and the predictions for lepton mixing angles and CP phases are given in Eq.~\eqref{eq:mixing_parameters_Z2_Z3}. Notice that in this case both Dirac and Majorana CP phases are trivial, and there is no CP violation because the neutrino mass matrix is real except for an overall phase.

In this case, the parameters $\alpha$, $\beta$ and $\gamma$ are real, and $\epsilon$ is also real instead of imaginary, as would be required in order to have an accidental  $X_{\mathbf{r}\nu}=\left\{\rho_{\mathbf{r}}(U), \rho_{\mathbf{r}}(SU)\right\}$ CP symmetry, therefore it leads to different predictions from the $S_4\rtimes H_{\rm{CP}}$ model where $X_{\mathbf{r}\nu}=\left\{\rho_{\mathbf{r}}(U), \rho_{\mathbf{r}}(SU)\right\}$ CP symmetry was preserved~\cite{Ding:2013hpa}.

The lepton mixing is predicted to be the so-called trimaximal mixing pattern. All the three mixing angles depend on one parameter $\theta$ which is of order $\lambda$ and related to the model parameters via
Eq.~\eqref{eq:tan_theta}. Consequently, the reactor angle $\theta_{13}$ is
of order $\lambda$ as well in the present model. For the best fit value $\sin^2\theta_{13}=0.0227$~\cite{GonzalezGarcia:2012sz}, the rotation angle
$\theta$ is determined to be $\theta\simeq\pm0.186$. Consequently we have the solar mixing angle $\sin^2\theta_{12}\simeq0.341$ and the atmospheric mixing angle $\sin^2\theta_{23}\simeq0.393$ or $\sin^2\theta_{23}\simeq0.607$, which are in the experimentally preferred regions.

For the remaining case in which the phase difference between $v_{\chi}v_{\rho}$ and $v_{\xi}$ is $\pm\pi/2$\footnote{This scenario could be realised by taking $g_9<0$, $g_3g_4g_7g_8<0$ or $g_9>0$, $g_3g_4g_7g_8>0$. In this case, the LO residual CP symmetry
$H^{\nu}_{\rm{CP}}=\left\{\rho_{\mathbf{r}}(1),\rho_{\mathbf{r}}(S)\right\}$ is broken completely by the VEVs $v_{\chi}$ and $v_{\rho}$, although the residual family symmetry $G_{\nu}=Z^{S}_2$ is still preserved.},
Eq.~\eqref{eq:vev_neutrino_NLO} implies that the shifts $\delta
v_{\tilde{\xi}}$ and $\delta v_S$ will be imaginary after extracting the overall phase carried by $v_{\xi}$. Then, the RH neutrino mass matrix $m_M$ can be parametrised as
\begin{equation}
\label{eq:mM_NLO}m_M=y_1v_{\xi}\Big[(1+ia\lambda)\left(\begin{array}{ccc}
1 & 0  &0 \\
0 & 0  & 1  \\
0 & 1  & 0
\end{array}\right)+(x+ib\lambda)\left(\begin{array}{ccc}
2/3  &  -1/3  &  -1/3 \\
-1/3 &  2/3  &  -1/3  \\
-1/3 &  -1/3  &  2/3
\end{array}\right)+ic\lambda\left(\begin{array}{ccc}
0  &   1   &   0  \\
1  &  0  &  0  \\
0  &  0  &  1
\end{array}
\right)\Big]\,,
\end{equation}
with
\begin{eqnarray}
a=-\frac{i}{\lambda}\frac{\tilde{y}_1\delta
v_{\tilde{\xi}}}{y_1v_{\xi}},\quad x=\frac{y_3v_S}{y_1v_{\xi}},\quad
b=-\frac{i}{\lambda}\frac{y_3\delta v_S}{y_1v_{\xi}},\quad
c=-\frac{i}{\lambda}\frac{y_4v_{\chi}}{y_1v_{\xi}}\frac{v_{\rho}}{\Lambda}\,,
\end{eqnarray}
where $x$, $a$, $b$ and $c$ are $\mathcal{O}(1)$ real parameters. To first order in $\lambda$, the light neutrino mass matrix followed by a tri-bimaximal transformation is of the form
\begin{equation}
m'_{\nu}=U^{T}_{TB}m_{\nu}U_{TB}=-\frac{y^2v^2_u}{y_1v_{\xi}}\left(\begin{array}{ccc}
\frac{2+2x-i\left(2a+2b-c\right)\lambda}{2(1+x)^2}  &   0   &
\frac{i\sqrt{3}\;c\lambda}{2\left(1-x^2\right)}   \\
0    &    1-i\left(a+c\right)\lambda    &   0  \\
\frac{i\sqrt{3}\;c\lambda}{2(1-x^2)}  &   0   &
\frac{-2+2x+i\left(2a-2b-c\right)\lambda}{2(1-x)^2}
\end{array}\right)\,.
\end{equation}
Following the procedure presented in Appendix \ref{sec:appendix_B}, this matrix $m'_{\nu}$ can
be diagonalized. After lengthy and tedious calculations, we find that the lepton mixing parameters are modified to
\begin{eqnarray}
\nonumber&\hskip-0.2in\sin\theta_{13}\simeq\left|\frac{c}{2\sqrt{2}\;x}\right|\lambda,\quad
\sin^2\theta_{12}=\frac{1}{3}+\mathcal{O}(\lambda^2),\quad
\sin^2\theta_{23}=\frac{1}{2}+\mathcal{O}(\lambda^2)\,,\\
\label{eq:lepton_parameters_NLO}&\hskip-0.2in\left|\sin\delta_{\rm{CP}}\right|=1+\mathcal{O}(\lambda^2),\quad
\left|\sin\alpha_{21}\right|\simeq\left|\frac{3c-2b+2x(a+c)}{2(1+x)}\right|\lambda,\quad
\left|\sin\alpha'_{31}\right|\simeq\left|\frac{x\left(2a-c-2xb\right)}{1-x^2}\right|\lambda\,,
\end{eqnarray}
where $\alpha'_{31}=\alpha_{31}-2\delta_{\rm{CP}}$, and the parameter $\alpha'_{31}$ has been redefined to include the Dirac CP phase $\delta_{\rm{CP}}$. This parametrisation turns out to be very useful and convenient for the analysis of neutrinoless double-beta decay and leptonic CP violation~\cite{Branco:2011zb}. We note that the higher order contributions to both $\theta_{12}$ and $\theta_{23}$ are suppressed such that they are rather close to the tri-bimaximal values. The reactor angle $\theta_{13}$ is predicted to be of order $\lambda$, and thus experimentally preferred value can be achieved. In particular, the Dirac CP violation is approximately maximal with $\delta_{\rm{CP}}\simeq\pm\frac{\pi}{2}$.

\begin{figure}[hptb!]
\begin{center}
\begin{tabular}{ccc}
\hskip-5mm\includegraphics[width=0.38\textwidth]{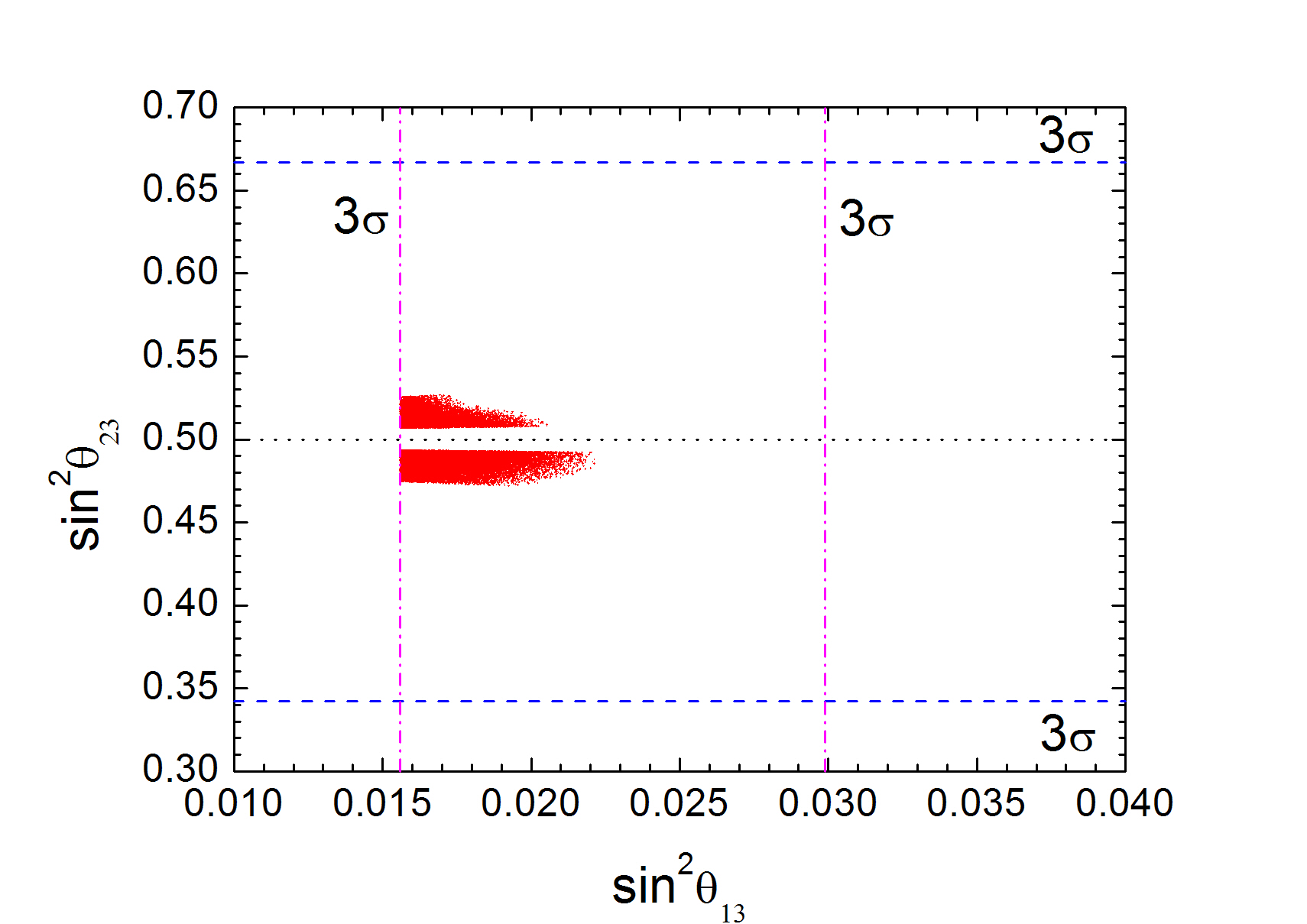} &
\hskip-10mm\includegraphics[width=0.38\textwidth]{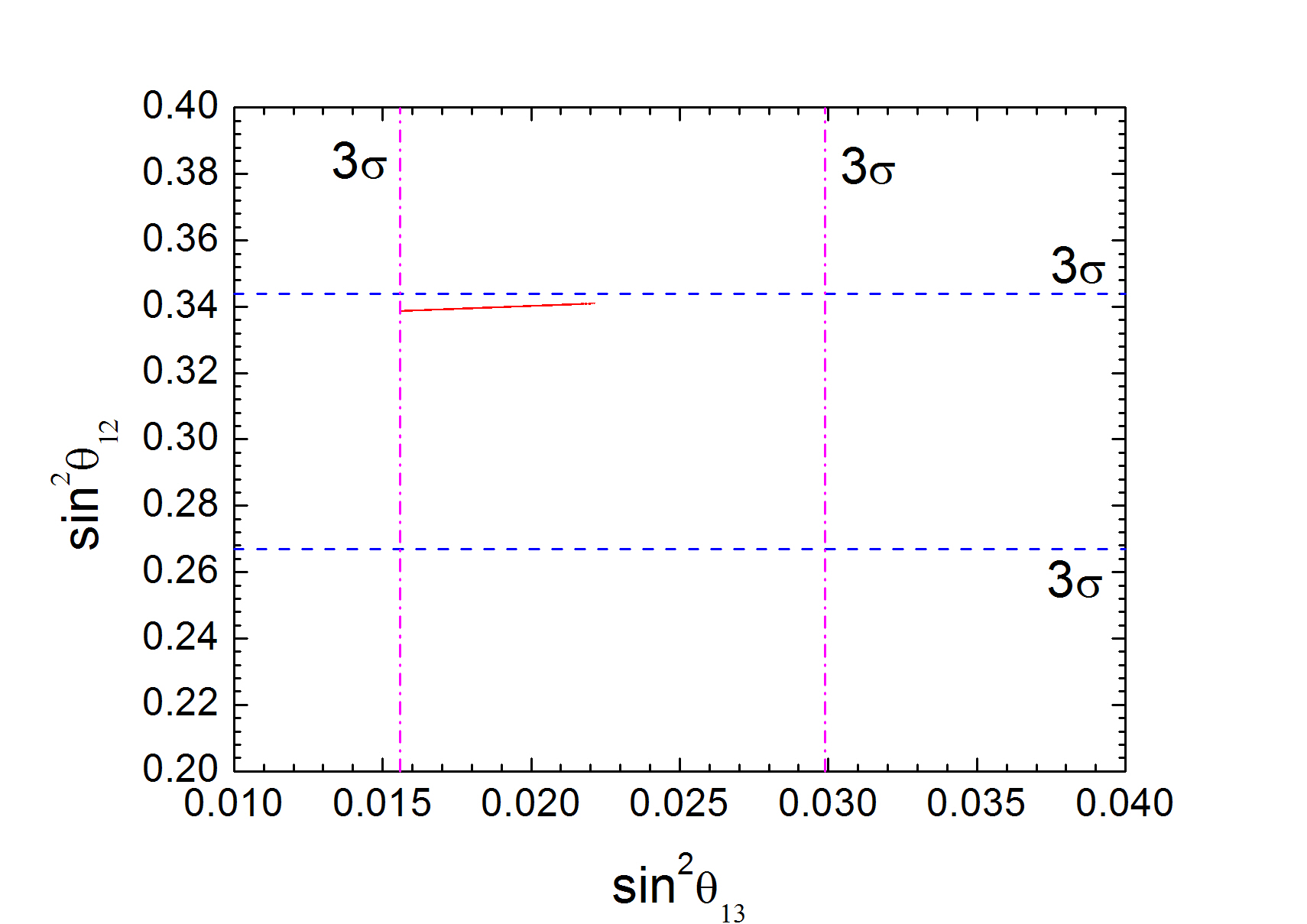}   &
\hskip-10mm\includegraphics[width=0.38\textwidth]{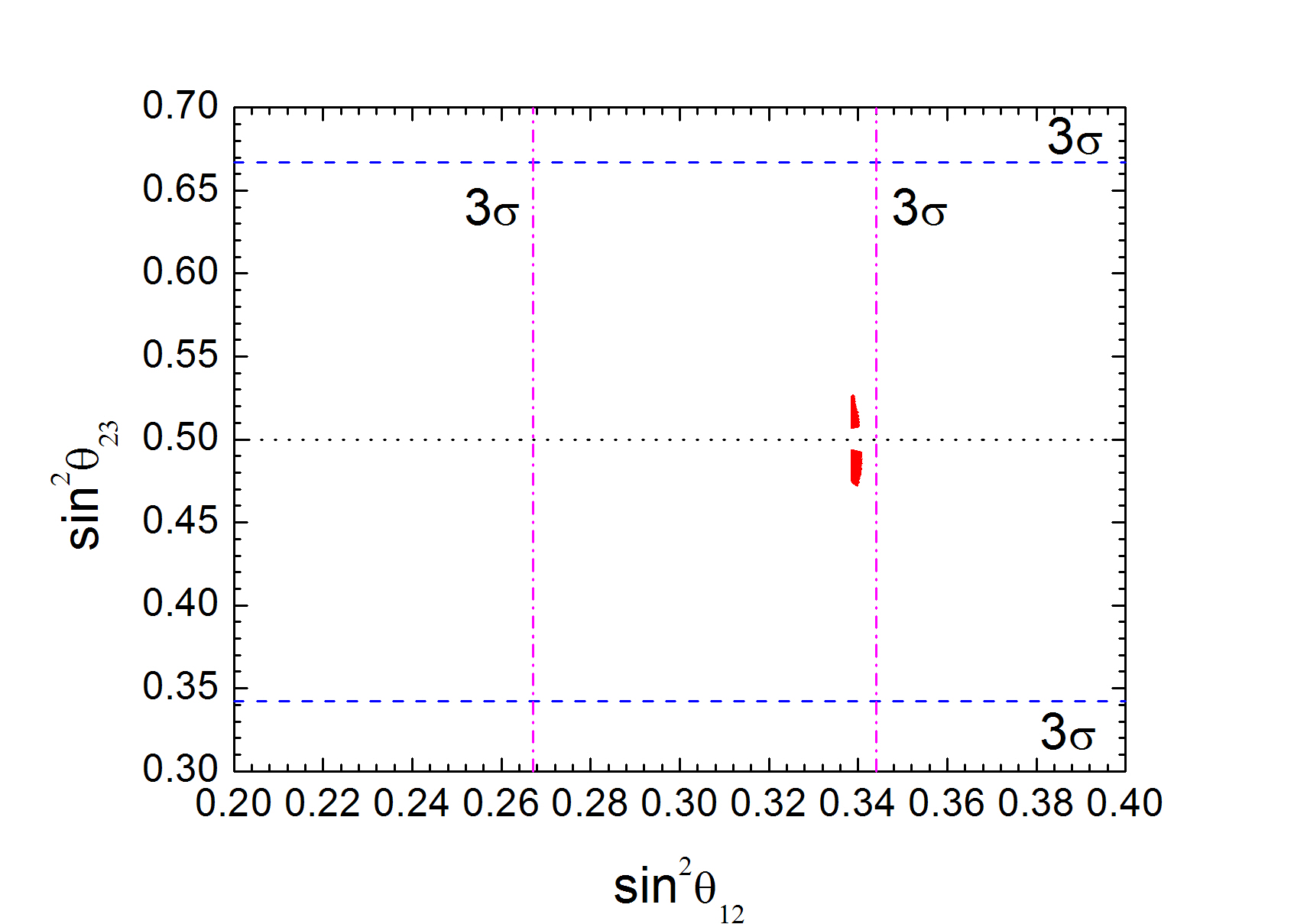}  \\ [-0.05in]
\hskip-5mm\includegraphics[width=0.38\textwidth]{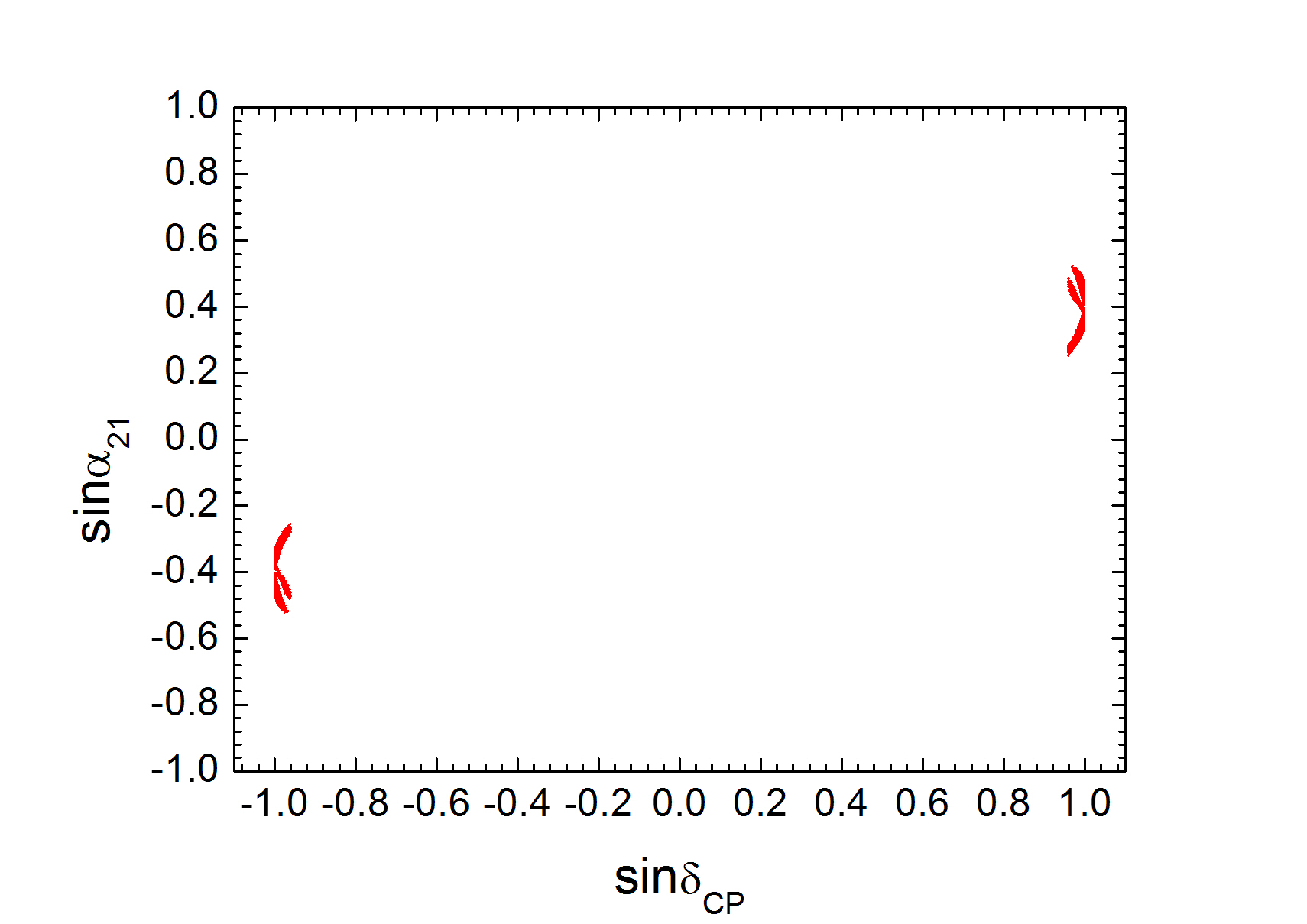}   &
\hskip-10mm\includegraphics[width=0.38\textwidth]{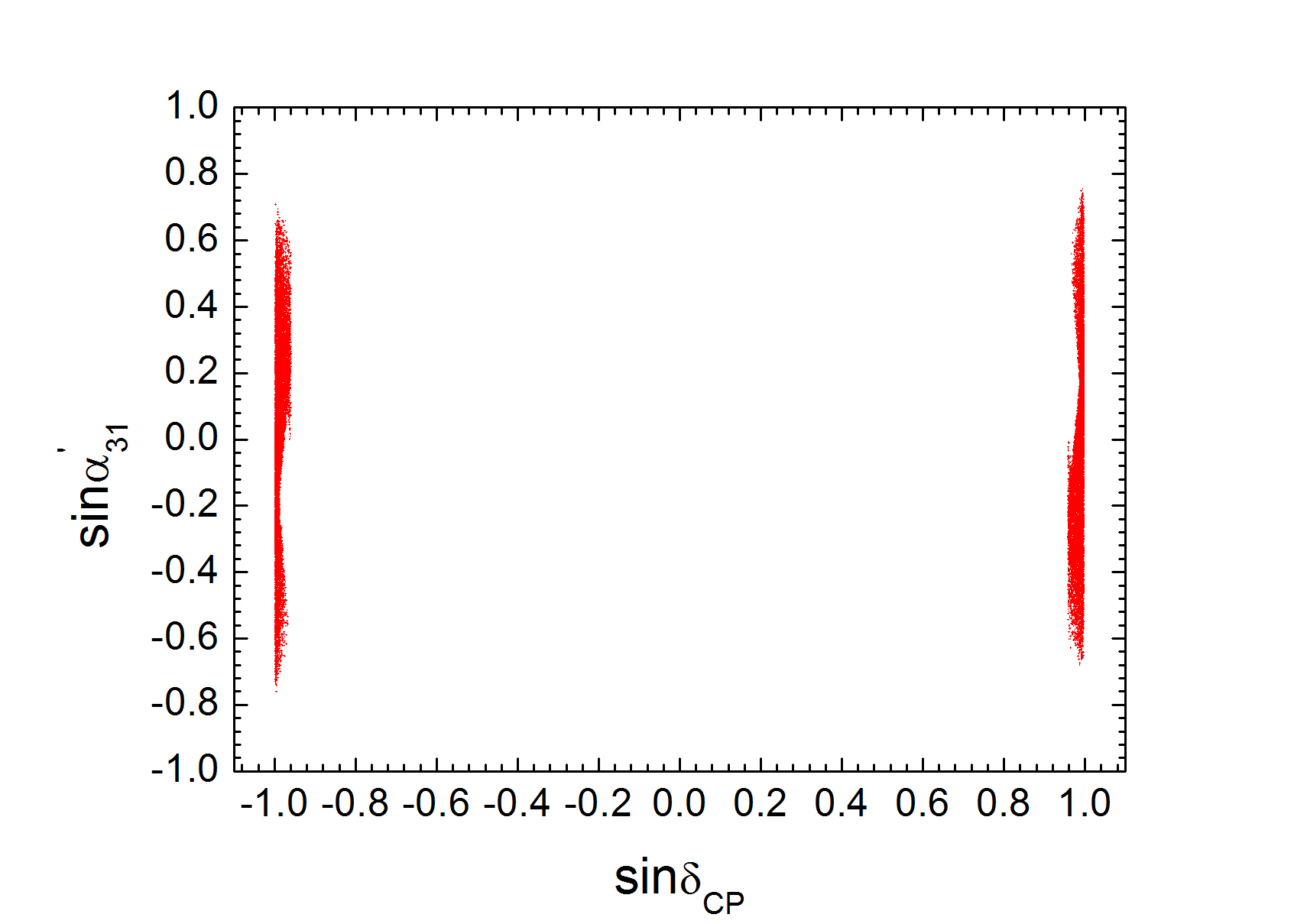}   &
\hskip-10mm\includegraphics[width=0.38\textwidth]{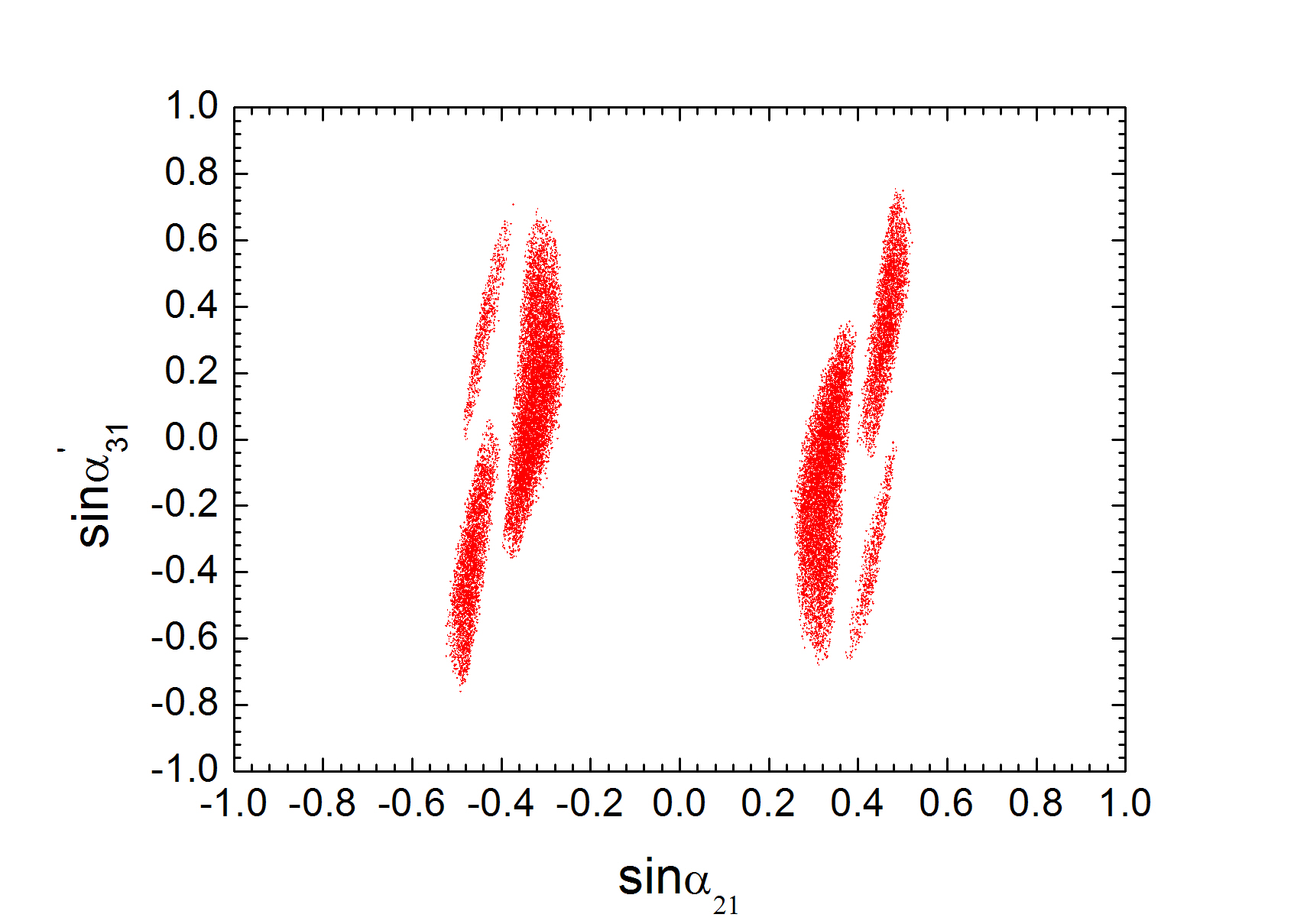}  \\ [-0.05in]
\hskip-5mm\includegraphics[width=0.38\textwidth]{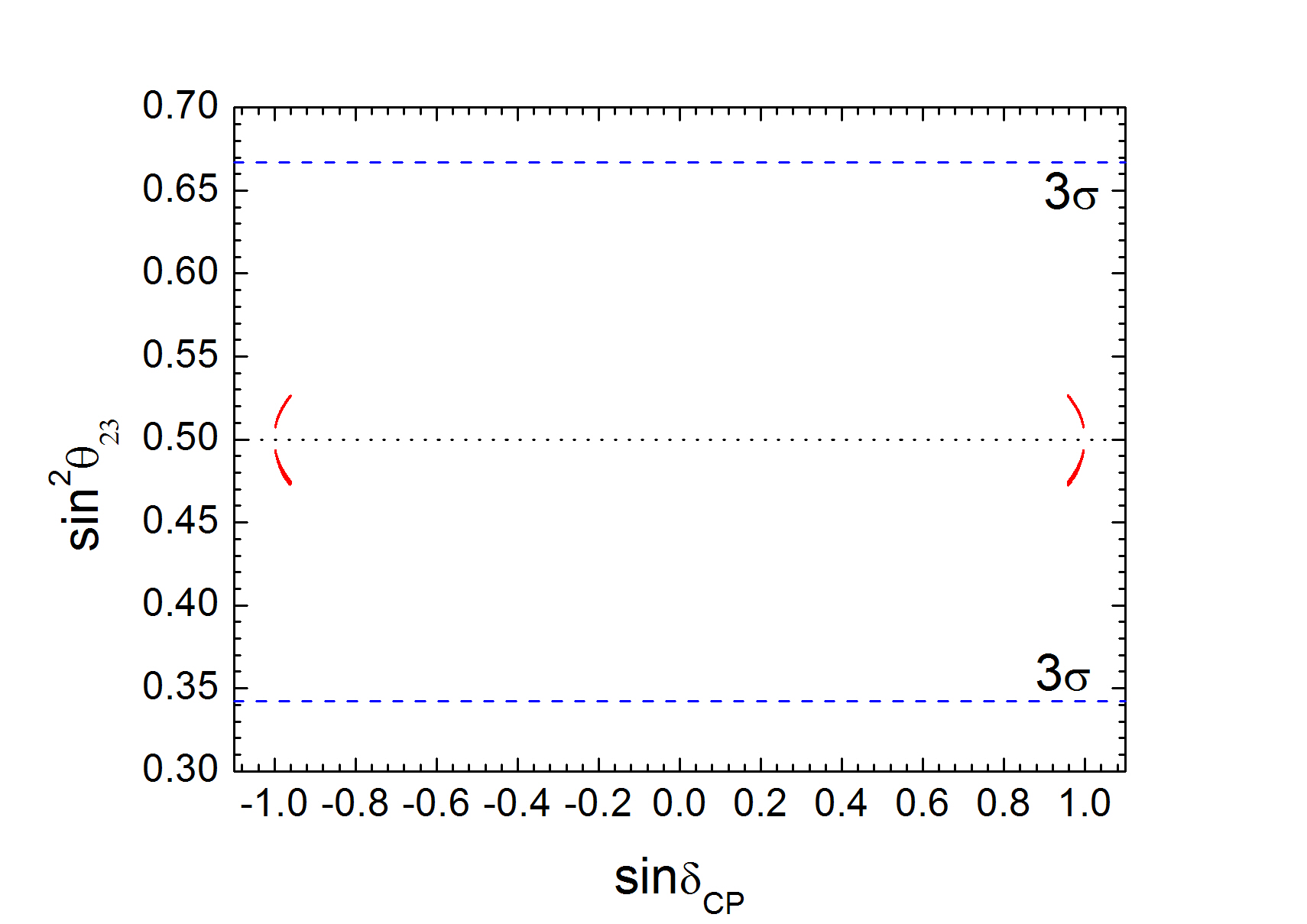}   &
\hskip-10mm\includegraphics[width=0.38\textwidth]{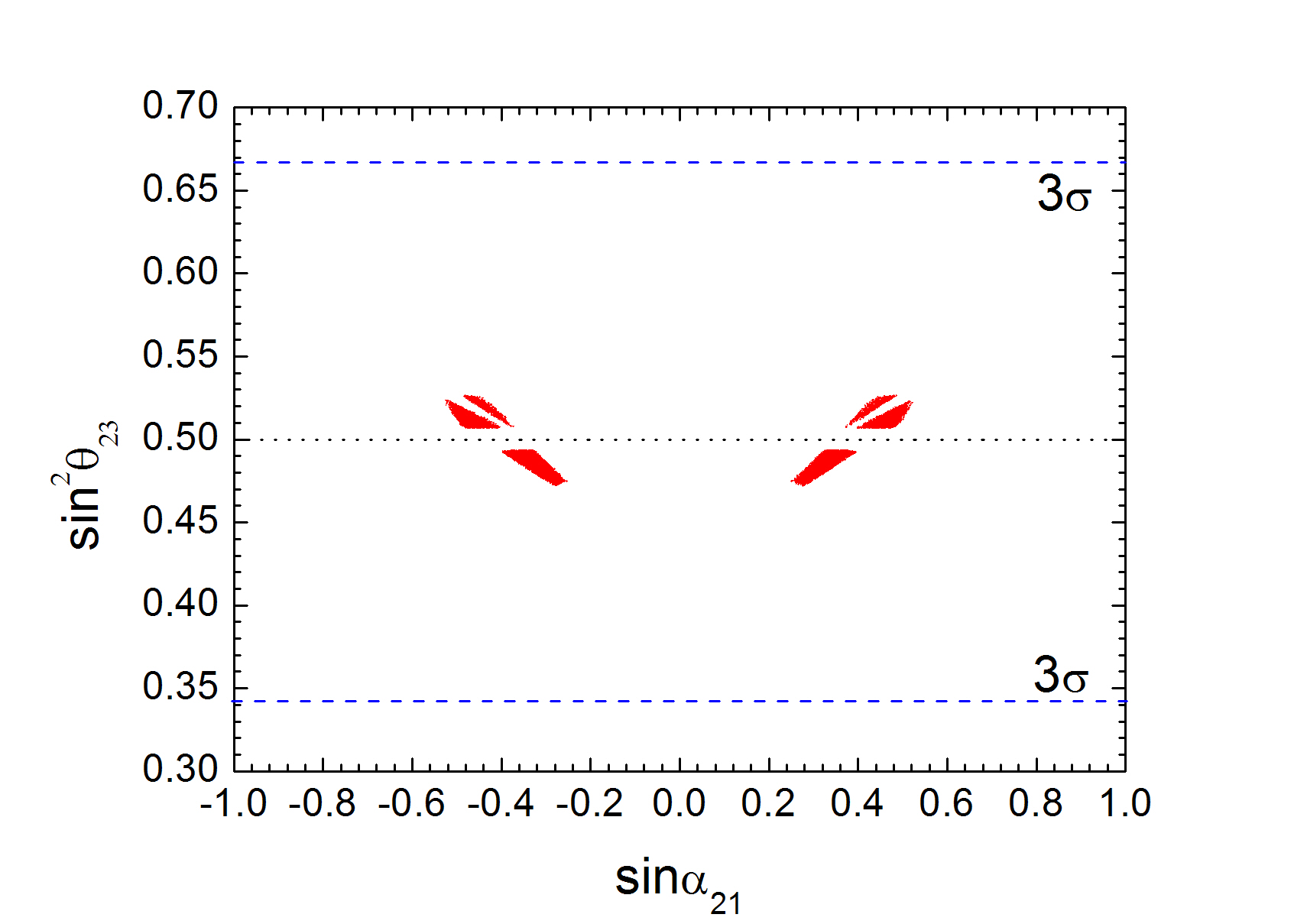}   &
\hskip-10mm\includegraphics[width=0.38\textwidth]{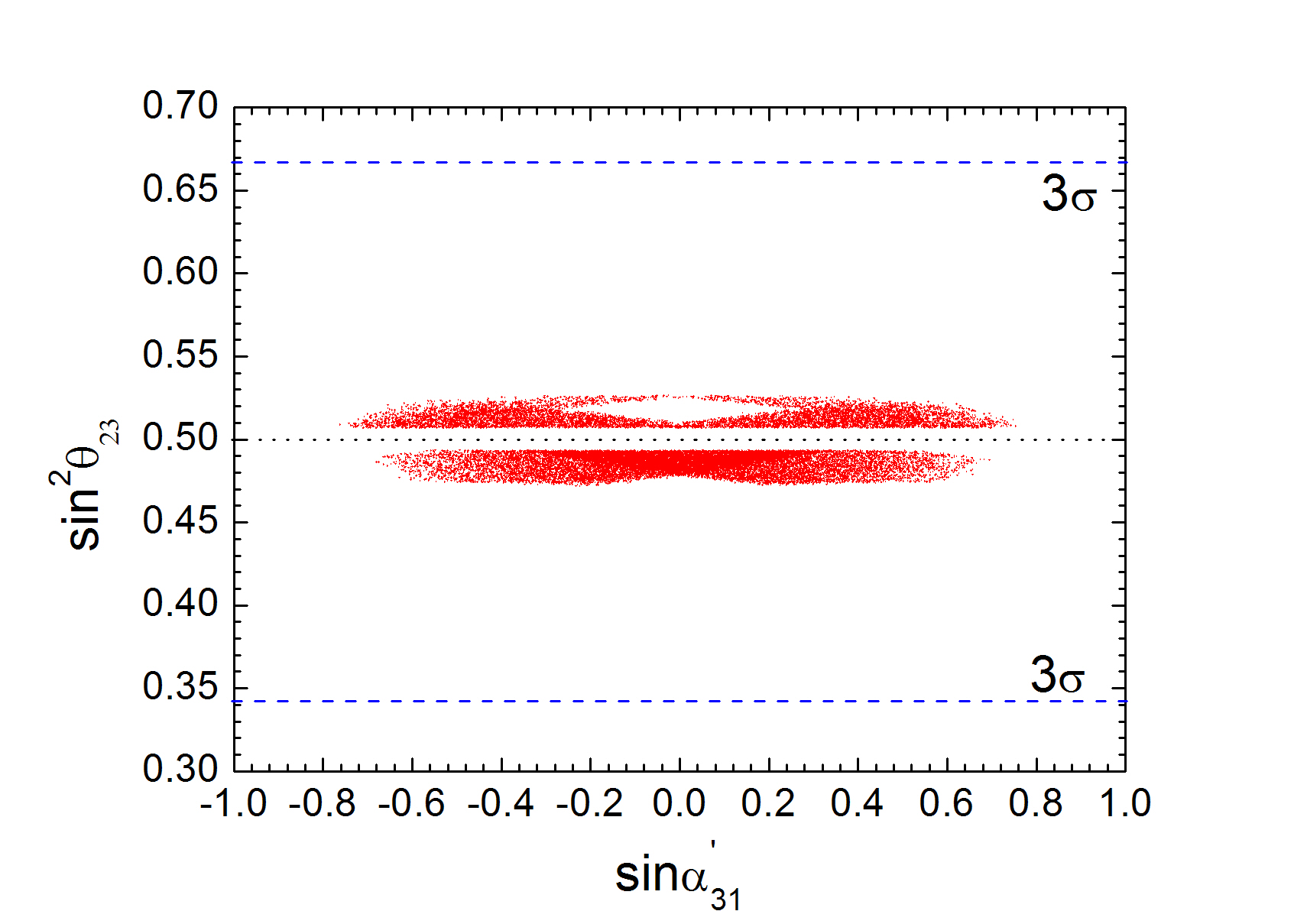}  \\  [-0.05in]

\hskip-5mm\includegraphics[width=0.38\textwidth]{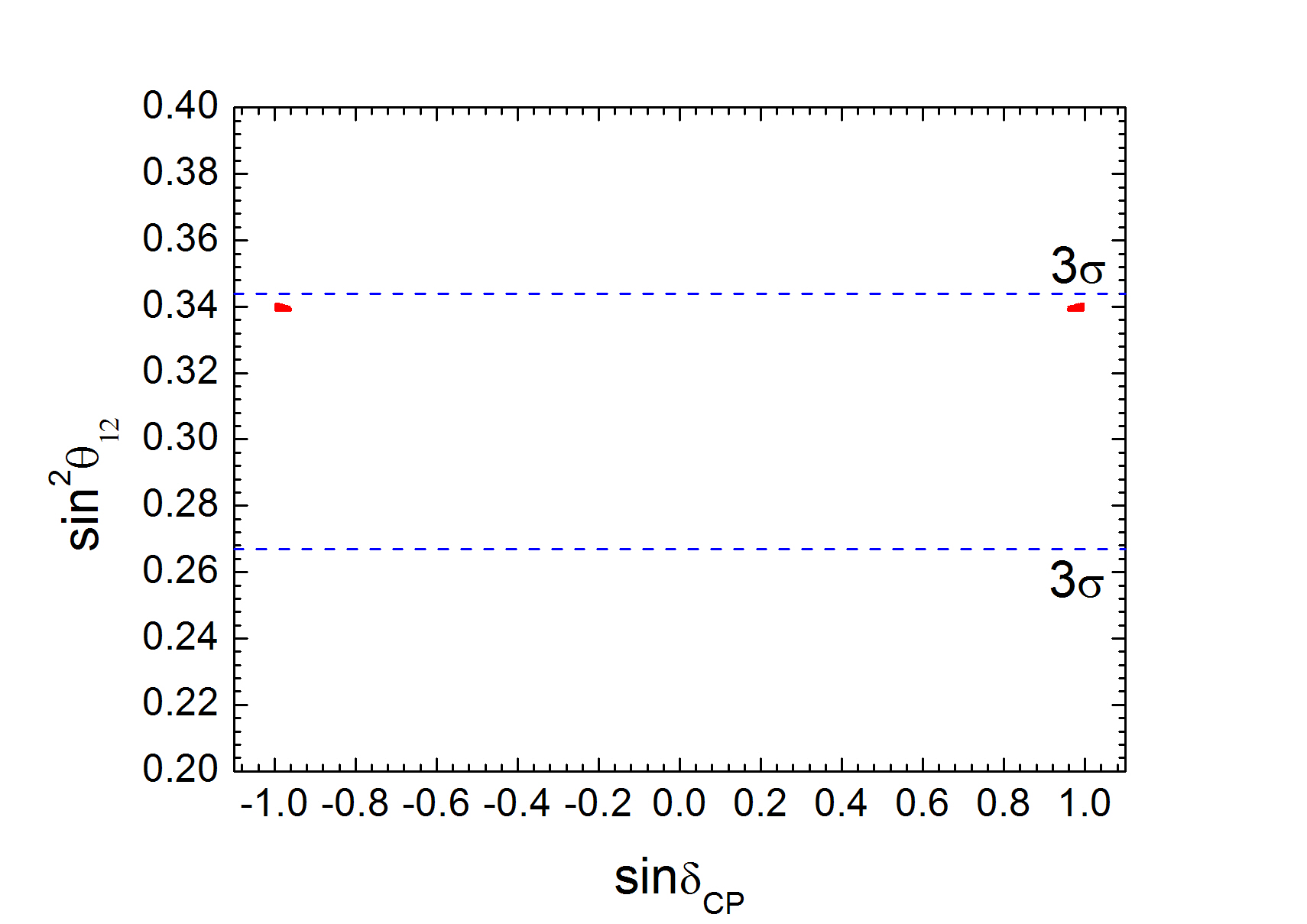}   &
\hskip-10mm\includegraphics[width=0.38\textwidth]{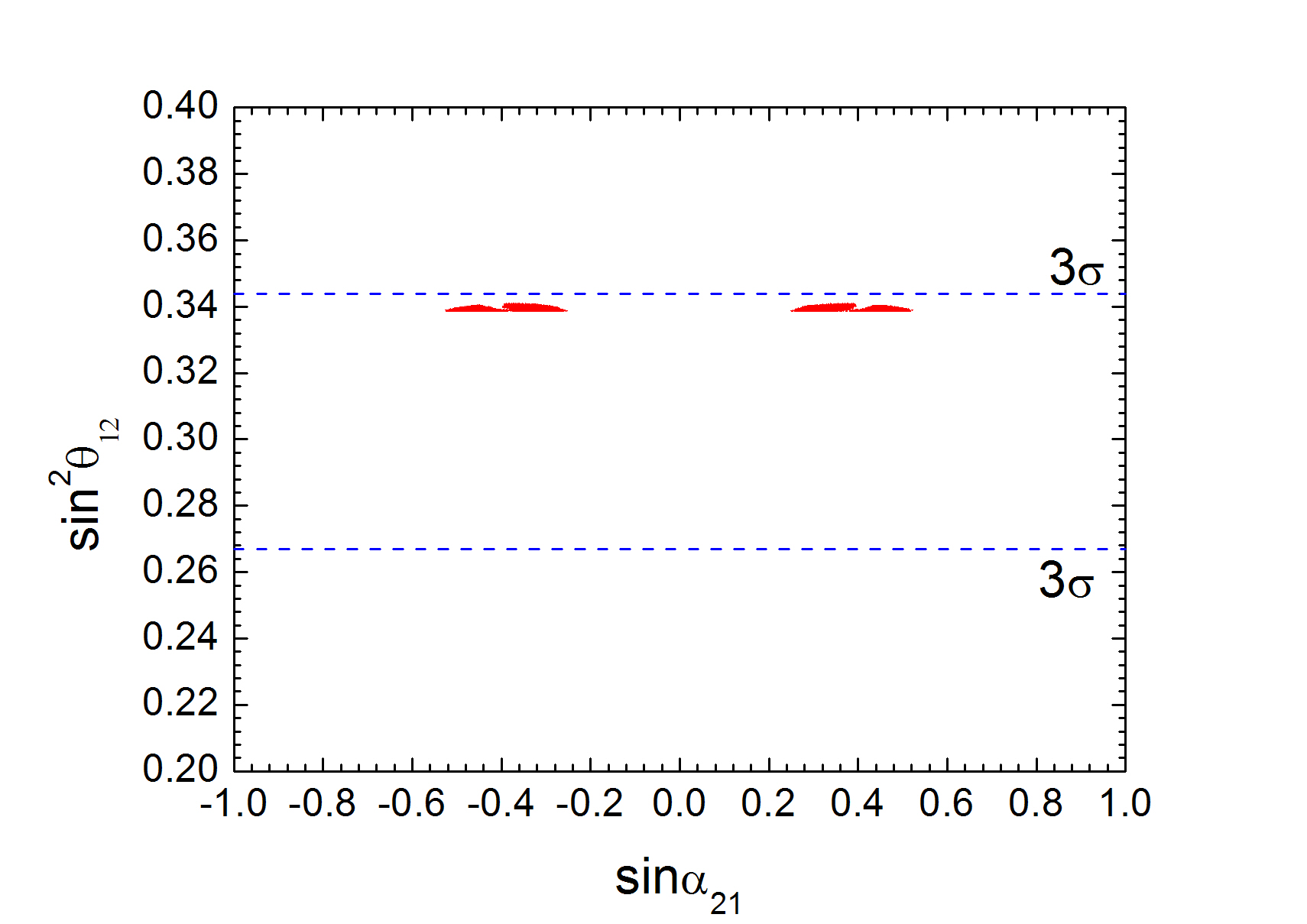}   &
\hskip-10mm\includegraphics[width=0.38\textwidth]{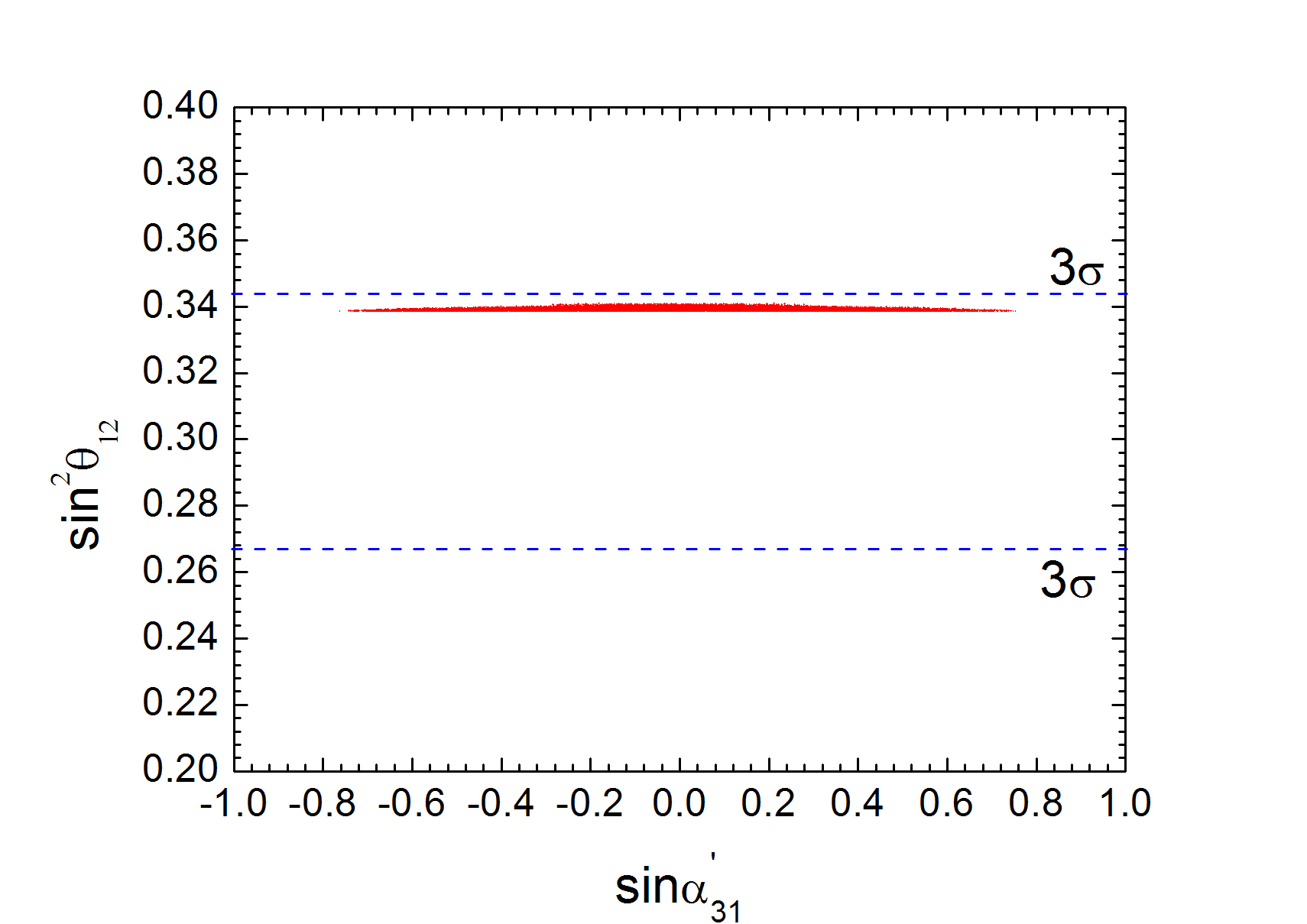}  \\ [-0.05in]

\hskip-5mm\includegraphics[width=0.38\textwidth]{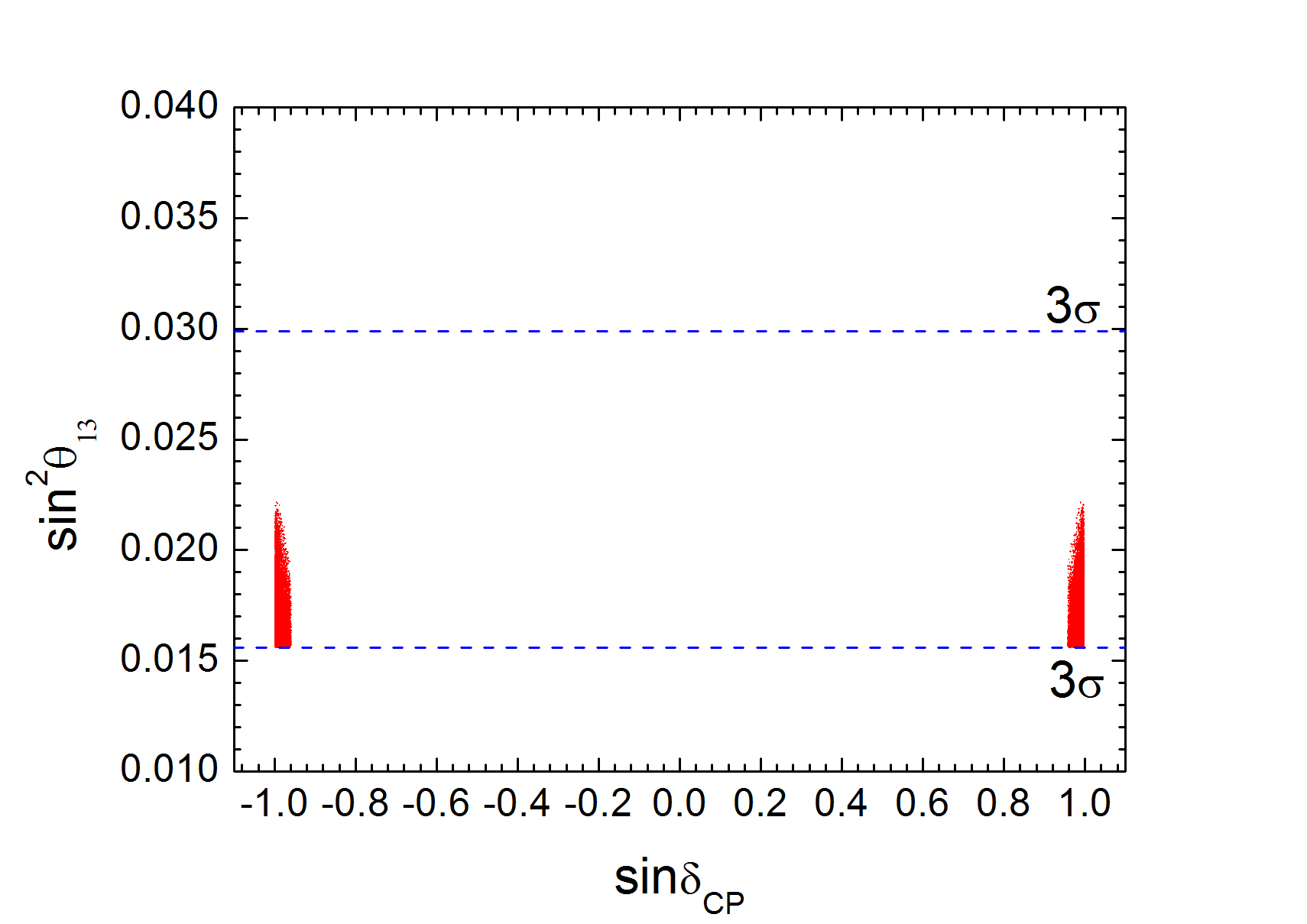}   &
\hskip-10mm\includegraphics[width=0.38\textwidth]{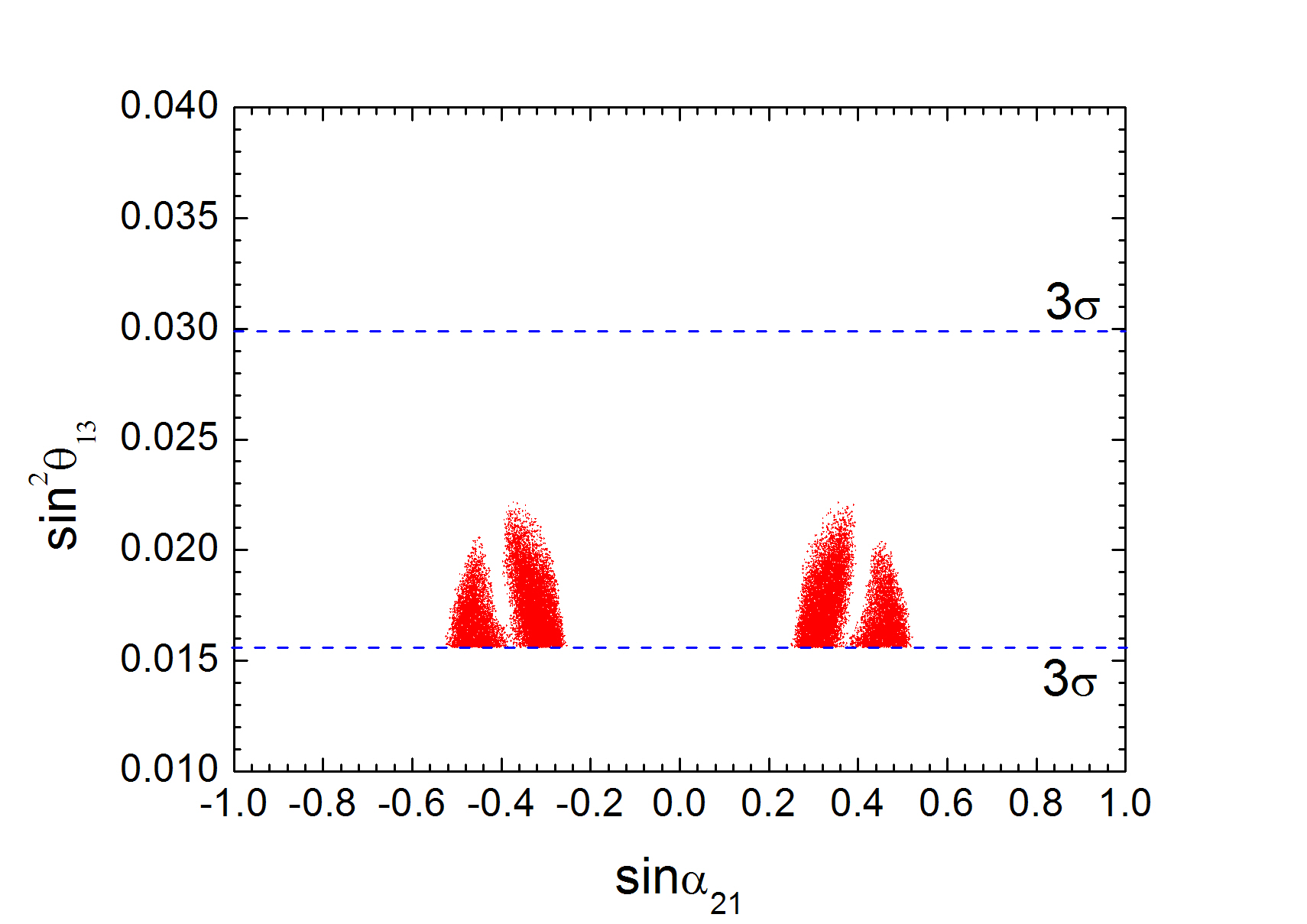}   &
\hskip-10mm\includegraphics[width=0.38\textwidth]{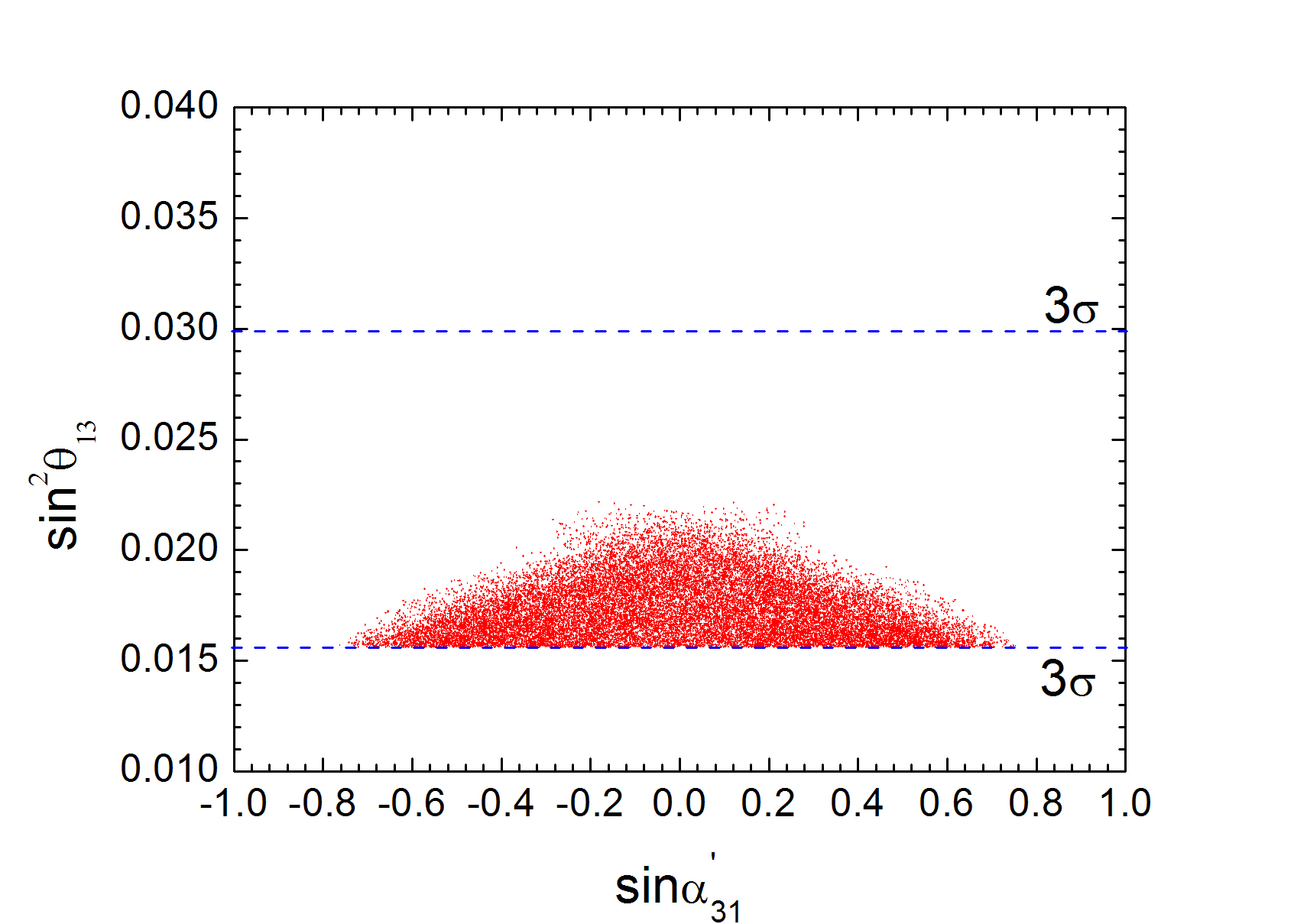}
\end{tabular}
\caption{\label{fig:correlation}The correlations of different flavour mixing parameters, where the horizontal lines and the vertical ones correspond to the $3\sigma$ bound for the mixing angles, which are taken from~\cite{GonzalezGarcia:2012sz}.}
\end{center}
\end{figure}

In order to see more clearly the predictions for the lepton mixing
parameters, we perform a numerical analysis. The expansion
parameter $\lambda$ is fixed at the indicative value 0.15, and the
parameters $x$, $a$, $b$ and $c$ are treated as random real numbers of
absolute value between $1/2$ and $2$. The resulting lepton mixing angles and
the mass-squared differences $\Delta m^2_{sol}$ and $\Delta m^2_{atm}$ are
required to lie in their $3\sigma$ ranges~\cite{GonzalezGarcia:2012sz}.
Correlations among the lepton mixing angles and the CP phases are plotted in
Fig.~\ref{fig:correlation}. Obviously we have almost maximal Dirac CP
phase $\delta_{CP}$, and the numerical results are consistent with the analytical estimates of Eq.~\eqref{eq:lepton_parameters_NLO}.

\section{\label{sec:UV_completion}Ultraviolet completion of the effective model}
\cleqn

\begin{table}[t!]
\begin{center}
\begin{tabular}{|c|c|c|c|c|c|c|c|c|c|c|}\hline\hline
\texttt{Field}  &   $\Omega_1$   &   $\Omega_2$   &   $\Omega_3$   &
$\Omega_4$   &    $\Omega^c_1$   &    $\Omega^c_2$   &    $\Omega^c_3$   &
$\Omega^c_4$   &  $\Sigma$  &   $\Sigma^c$  \\  \hline

$A_4$  & $\mathbf{3} $  &    $\mathbf{1}''$   &   $\mathbf{1}$   &
$\mathbf{1}$  & $\mathbf{3} $  &    $\mathbf{1}'$   &   $\mathbf{1}$   &
$\mathbf{1}$  &  $\mathbf{3}$   &    $\mathbf{3}$ \\  \hline

$Z_4$  &  $-1$   &  $i$   &  $i$  &   1 &  $-1$   &  $-i$   &  $-i$  &   1
&  $-1$   &   $-1$ \\  \hline

$Z_6$  &  $\omega^2_6$   &   $\omega^2_6$     &   $\omega^2_6$   &
$\omega^2_6$   &  $\omega^4_6$   &   $\omega^4_6$     &   $\omega^4_6$   &
$\omega^4_6$   &   $\omega^5_6$   &   $\omega_6$ \\  \hline

$U(1)_R$  &   1    &  1   &  1   &  1    &   1   &   1   &  1   &  1  &  1
&  1

\\\hline\hline

\end{tabular}
\caption{\label{tab:UV_transformation} The transformation rules of the
messenger fields under the family symmetry $A_4\times Z_4\times Z_6$ and
$U(1)_R$.}
\end{center}
\end{table}

In the previously discussed effective model, non-renormalisable terms allowed by the symmetries are included in the superpotential $w_{l}$ of Eq.~\eqref{eq:wl_LO} and the subleading correction terms. It is generally believed that these effective terms arise from a fundamental renormalisable theory at high energies by integrating out the heavy degree of freedom. In this section, we present a ultraviolet (UV) completion of the effective model, which in general has the advantage of improving the predictability of the effective model. In such UV completed models, the non-renormalisable terms of the previously discussed effective model arise from integrating out heavy messenger fields, and some terms included at the effective level will be eliminated if no messenger field exists to mediate them. It is well-known that the UV completion of a low energy effective theory is generally not unique. In this section, we shall present the ``minimal" completion of the above effective model in the sense of having the least number of extra messenger fields and the least number of associated (renormalisable) couplings.

\begin{figure}[t!]
\begin{center}
\includegraphics[width=0.9\textwidth]{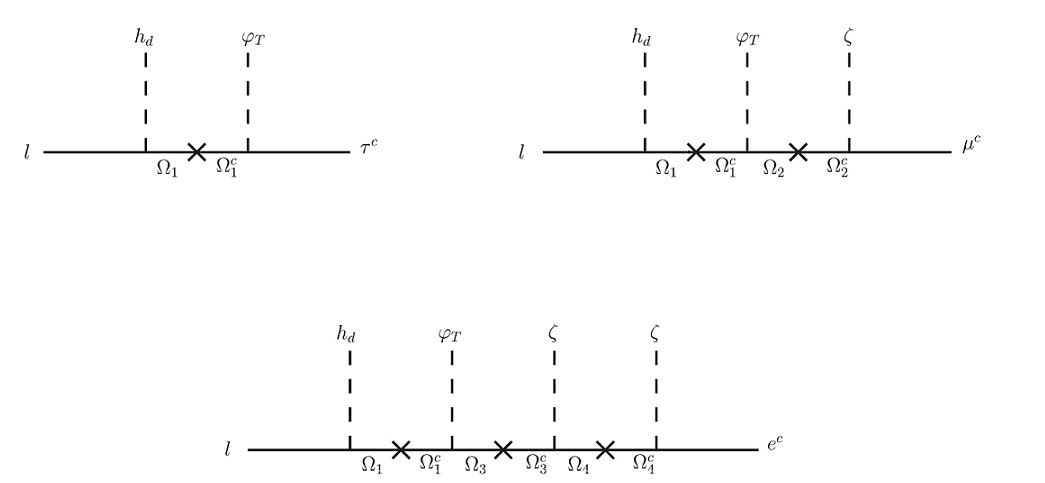}
\caption{\label{fig:charged_renor}The diagrams which generate the effective operators for the charged lepton masses, where crosses indicate the mass insertions for fermions.}
\end{center}
\end{figure}

\begin{figure}[t!]
\begin{center}
\includegraphics[width=0.9\textwidth]{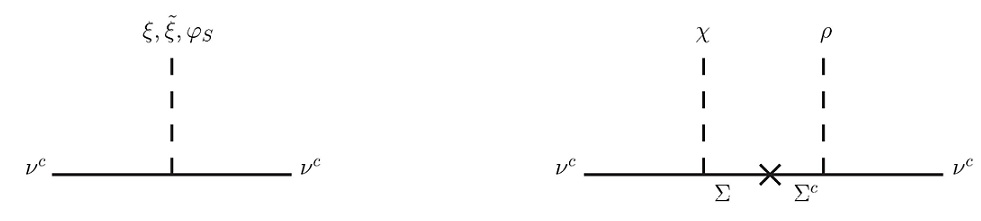}
\caption{\label{fig:neutrino_renor}The diagrams for the RH neutrino masses, where crosses indicate the mass insertions for fermions.}
\end{center}
\end{figure}

To begin, the driving superpotential $w_d$ of Eq.~\eqref{eq:driving} is already renormalisable, and therefore the vacuum alignment given in
Eqs.~(\ref{eq:vev_charged},\ref{eq:vev_neutrino},\ref{eq:vev_rho}) is kept intact. The effective terms for the charged lepton masses in $w_{l}$ of Eq.~\eqref{eq:wl_LO} is non-renormalisable. Thus in order to reproduce these terms through the combination of renormalisable terms, we minimally increase the field content to introduce four pairs of messenger fields $\Omega_i$ and $\Omega^c_i\;(i=1,2,3,4)$. The transformation properties of the all the messenger fields under the family symmetry $A_4\times Z_4\times Z_6$ are listed in Table \ref{tab:UV_transformation}. Notice that these messengers are chiral superfields with non-vanishing hypercharge $+2(-2)$ for $\Omega_i$ ($\Omega^c_i$). We can straightforwardly write down the renormalisable charged lepton superpotential
\begin{eqnarray}
\nonumber
w_l&=&z_1\left(l\Omega_1\right)h_d+z_2\left(\Omega^c_1\varphi_T\right)''\tau^c+z_3\left(\Omega^c_1\varphi_T\right)'\Omega_2+z_4\Omega^c_2\zeta\mu^c+z_5\left(\Omega^c_1\varphi_T\right)\Omega_3
+z_6\Omega^c_3\Omega_4\zeta\\
\nonumber&&+z_7\Omega^c_4\zeta e^c+M_{\Omega_1}\left(\Omega^c_1\Omega_1\right)+M_{\Omega_2}\Omega^c_2\Omega_2+M_{\Omega_3}\Omega^c_3\Omega_3+M_{\Omega_4}\Omega^c_4\Omega_4\,,
\end{eqnarray}
where all the coupling constants $z_i\;(i=1\ldots7)$ and the messenger
masses $M_{\Omega_i}\;(i=1\ldots4)$ are real because of the imposed
generalised CP symmetry. Integrating out the heavy messenger fields
$\Omega_i$ and $\Omega^c_i$, the corresponding Feynman diagrams are shown in
Fig.~\ref{fig:charged_renor}, we obtain the effective superpotential for the
charged lepton masses
\begin{eqnarray}
w^{eff}_l=-\frac{z_1z_2}{M_{\Omega_1}}\left(l\varphi_T\right)''\tau^ch_d+\frac{z_1z_3z_4}{M_{\Omega_1}M_{\Omega_2}}\left(l\varphi_T\right)'\zeta \mu^ch_d-\frac{z_1z_5z_6z_7}{M_{\Omega_1}M_{\Omega_3}M_{\Omega_4}}\left(l\varphi_T\right)\zeta^{2}e^ch_d\,.
\end{eqnarray}
Taking into account the vacuum alignments $\langle\varphi_T\rangle=(0,v_T,0)$ and $\langle\zeta\rangle=v_{\zeta}$ of
Eq.~\eqref{eq:vev_charged}, we obtain a diagonal charged lepton mass matrix
with
\begin{eqnarray}
\hskip-0.2in
m_{e}=-z_1z_5z_6z_7\frac{v_{T}v^2_{\zeta}}{M_{\Omega_1}M_{\Omega_3}M_{\Omega_4}}v_d,\quad
m_{\mu}=z_1z_3z_4\frac{v_Tv_{\zeta}}{M_{\Omega_1}M_{\Omega_2}}v_d,\quad
m_{\tau}=-z_1z_2\frac{v_T}{M_{\Omega_1}}v_d\,.
\end{eqnarray}
For the neutrino sector, we introduce the messenger fields $\Sigma$ and
$\Sigma^c$ which are chiral superfields carrying zero hypercharge. The
renormalisable superpotential relevant to the neutrino masses reads
\begin{equation}
w_{\nu}=w^{LO}_{\nu}+w^{\Sigma}_{\nu}\,,
\end{equation}
with
\begin{eqnarray}
w^{LO}_{\nu}&=&y\left(l\nu^c\right)h_u+y_1\left(\nu^c\nu^c\right)\xi+\tilde{y}_1\left(\nu^c\nu^c\right)\tilde{\xi}+y_3\left(\nu^c\nu^c\varphi_S\right)\,,
\\
w^{\Sigma}_{\nu}&=&x_1\left(\nu^c\Sigma\right)'\chi+x_2\left(\nu^c\Sigma^c\right)\rho+M_{\Sigma}\left(\Sigma^c\Sigma\right)\,,
\end{eqnarray}
where all couplings and the mass $M_{\Sigma}$ are real due to the
generalised CP invariance. The first term of $w^{LO}_{\nu}$ gives rise to
the Dirac neutrino mass matrix
\begin{equation}
\label{eq:mD_UV}m_D=y\left(\begin{array}{ccc}
1  &  0  &  0  \\
0  &  0  &  1  \\
0  &  1  &  0
\end{array}\right)v_u\,.
\end{equation}
The RH neutrino masses receive contributions from both $w^{LO}_{\nu}$ and
$w^{\Sigma}_{\nu}$, as shown in Fig.~\ref{fig:neutrino_renor}. Integrating
out the messenger fields $\Sigma$ and $\Sigma^c$ leads to the NLO effective
operator
\begin{equation}
w^{NLO}_{\nu}=-\frac{x_1x_2}{M_{\Sigma}}\left(\nu^c\nu^c\right)'\chi\rho\,,
\end{equation}
which corresponds to the last term of the NLO corrections $\delta w_{\nu}$
in Eq.~\eqref{eq:wnu_NLO} with $y_4=-x_1x_2\Lambda/M_{\Sigma}$. However, the
corrections from the shifted vacuum of $\tilde{\xi}$ and $\varphi_S$
disappear in the present renormalisable model. The reason is that the
messenger fields introduced do not affect the driving superpotential, and thus the vacuum alignment is preserved. Combining the contributions from both $w^{LO}_{\nu}$ and $w^{NLO}_{\nu}$, the RH neutrino mass matrix $m_M$ is given by
\begin{equation}
m_M=\left(\begin{array}{ccc}
y_1v_{\xi}+2y_3v_S/3  &   -y_3v_S/3-x_1x_2v_{\chi}v_{\rho}/M_{\Sigma}   &
-y_3v_S/3  \\
-y_3v_S/3-x_1x_2v_{\chi}v_{\rho}/M_{\Sigma} & 2y_3v_S/3   &
y_1v_{\xi}-y_3v_S/3  \\
-y_3v_S/3   &   y_1v_{\xi}-y_3v_S/3  &
2y_3v_S/3-x_1x_2v_{\chi}v_{\rho}/M_{\Sigma}
\end{array}\right)\,.
\end{equation}
Being similar to the effective model, the VEVs $v_{\xi}$ and $v_{S}$ should have the same phase up to relative sign otherwise the light neutrino masses will be degenerate at LO. Furthermore, the phase difference between $v_{\chi}v_{\rho}$ and $v_{\xi}$ is $0$, $\pi$ or $\pm\pi/2$, as previously emphasised.

For the former cases, i.e. the phase difference is $0$ or $\pi$, the light neutrino mass matrix is real once the common phase of $v_{\xi}$,
$v_{S}$ and $v_{\chi}v_{\rho}$ is absorbed by field redefinition. The resulting PMNS matrix is of the trimaximal form shown in Eq.~\eqref{eq:pmns_Z2_Z3}. Therefore lepton mixing angles compatible with the experimental data can be achieved, and CP is conserved. For the remaining case in which the phase difference of $v_{\chi}v_{\rho}$ and $v_{\xi}$ is $\pm\pi/2$, $m_M$ can be parametrised as in Eq.~\eqref{eq:mM_NLO} with
\begin{equation}
m_M=y_1v_{\xi}\left[\left(\begin{array}{ccc}
1 & 0 & 0 \\
0 & 0 & 1 \\
0 & 1 & 0
\end{array}\right)+x\left(\begin{array}{ccc}
2/3  &  -1/3   &  -1/3  \\
-1/3   & 2/3   &  -1/3 \\
-1/3   &  -1/3  & 2/3
\end{array}
\right)+iz\left(\begin{array}{ccc}
0 & 1 & 0  \\
1 & 0 & 0  \\
0 & 0 & 1
\end{array}\right)\right]\,,
\end{equation}
where $z=i\frac{x_1x_2}{y_1}\frac{v_{\chi}v_{\rho}}{M_{\Sigma}v_{\xi}}$. We can straightforwardly obtain the light neutrino mass matrix from the seesaw formula\cite{seesaw} and then apply a tri-bimaximal transformation, i.e.
\begin{equation}
m'_{\nu}=-U^{T}_{TB}\left(m_Dm^{-1}_{M}m^{T}_D\right)U_{TB}=m_0\left(\begin{array}{ccc}
\frac{2-2x-iz}{2\left(1-x^2-z^2-iz\right)} &  0  &
\frac{i\sqrt{3}\;z}{2\left(1-x^2-z^2-iz\right)}  \\
0  &  \frac{1}{1+iz}   &  0  \\
\frac{i\sqrt{3}\;z}{2\left(1-x^2-z^2-iz\right)}  &  0   &
\frac{-2-2x+iz}{2\left(1-x^2-z^2-iz\right)}\,,
\end{array}\right)
\end{equation}
with $m_0\equiv-y^2v^2_u/(y_1v_{\xi})$. Notice that the neutrino sector is
described by three real parameters $m_0$, $x$ and $z$ at low energy, and therefore this model is rather predictive. As shown in Appendix \ref{sec:appendix_B}, the mass matrix $m'_{\nu}$ can be diagonalised exactly as
\begin{equation}
U'^{T}_{\nu}m'_{\nu}U'_{\nu}=\text{diag}(m_1,m_2,m_3)\,,
\end{equation}
where the unitary matrix $U'_{\nu}$ is of the form
\begin{equation}
U'_{\nu}=\left(\begin{array}{ccc}
e^{i(\pi/4+\phi_1/2)}\cos\theta   &  0   & e^{i(\pi/4+\phi_3/2)}\sin\theta
\\
0   &    e^{i\phi_2/2}   &   0   \\
-e^{-i(\pi/4-\phi_1/2)}\sin\theta    & 0    &
e^{-i(\pi/4-\phi_3/2)}\cos\theta   \end{array}\right),
\end{equation}
where the angle $\theta$ satisfies
\begin{equation}
\tan2\theta=\frac{\sqrt{3}z}{2x}\,,
\end{equation}
and the phases $\phi_{1,2,3}$ are given by
\begin{eqnarray}
\nonumber&&\phi_1=-\text{arg}\left(\frac{z+i(2-2x\cos2\theta-\sqrt{3}\;z\sin2\theta)}{1-x^2-z^2-iz}\right)\,,\\
\nonumber&&\phi_2=\text{arg}\left(1+iz\right)\,,\\
&&\phi_3=-\text{arg}\left(\frac{z+i(2+2x\cos2\theta+\sqrt{3}\;z\sin2\theta)}{1-x^2-z^2-iz}\right)\,.
\end{eqnarray}
where the overall phase of $m_0$ has been omitted. Therefore the PMNS matrix is of the form
{\small\begin{eqnarray}
\nonumber&&\hskip-0.5in U_{PMNS}=U_{TB}U'_{\nu}\\
&&\hskip-0.4in=\left(\begin{array}{ccc}
\frac{2}{\sqrt{6}}\cos\theta e^{i(\pi/4+\phi_1/2)}   &
\frac{1}{\sqrt{3}}e^{i\phi_2/2}   &   \frac{2}{\sqrt{6}}\sin\theta
e^{i(\pi/4+\phi_3/2)} \\
  &   &  \\ [-0.15in]
\left(-\frac{1}{\sqrt{6}}\cos\theta
-\frac{i}{\sqrt{2}}\sin\theta\right)e^{i(\pi/4+\phi_1/2)} &
\frac{1}{\sqrt{3}}e^{i\phi_2/2}  &  \left(-\frac{1}{\sqrt{6}}\sin\theta
+\frac{i}{\sqrt{2}}\cos\theta\right)e^{i(\pi/4+\phi_3/2)}\\
  &   &  \\ [-0.15in]
\left(-\frac{1}{\sqrt{6}}\cos\theta+\frac{i}{\sqrt{2}}\sin\theta\right)e^{i(\pi/4+\phi_1/2)}
& \frac{1}{\sqrt{3}}e^{i\phi_2/2}  &
\left(-\frac{1}{\sqrt{6}}\sin\theta-\frac{i}{\sqrt{2}}\cos\theta\right)e^{i(\pi/4+\phi_3/2)}
\end{array}
\right).
\end{eqnarray}}
From this, we can immediately extract the lepton mixing angles and CP phases:
\begin{eqnarray}
\nonumber&\sin^2\theta_{13}=\frac{1}{3}\left(1-\cos2\theta\right),\quad
\sin^2\theta_{12}=\frac{1}{2+\cos2\theta}=\frac{1}{3\cos^2\theta_{13}},\quad\sin^2\theta_{23}=\frac{1}{2}\,,\\
&\delta_{\rm{CP}}=\text{sign}(xz)\frac{\pi}{2},\quad
\alpha_{21}=\phi_2-\phi_1-\frac{\pi}{2},\quad
\alpha_{31}=\phi_3-\phi_1+\text{sign}(xz)\pi\,.
\end{eqnarray}
It is remarkable that this model predicts maximal Dirac CP violation
$\delta_{\rm{CP}}=\pm\frac{\pi}{2}$ and maximal atmospheric neutrino mixing in this case. For the measured values $\sin^2\theta_{13}=0.0227$, the solar
mixing angle is predicted to be $\sin^2\theta_{12}\simeq0.341$ which is
compatible with the experimentally allowed regions. Finally, we remark that the light neutrino masses $m_{1,2,3}$ are given by
\begin{eqnarray}
\nonumber&&m_1=|m_0|\sqrt{\frac{1+x^2+z^2-\text{sign}(x\cos2\theta)\sqrt{4x^2+3z^2}}{\left(1-x^2-z^2\right)^2+z^2}}\,,\\
\nonumber&&m_2=\frac{|m_0|}{\sqrt{1+z^2}}\,,\\
&&m_3=|m_0|\sqrt{\frac{1+x^2+z^2+\text{sign}(x\cos2\theta)\sqrt{4x^2+3z^2}}{\left(1-x^2-z^2\right)^2+z^2}}\,.
\end{eqnarray}
As a result, the solar and atmospheric mass-squared splittings are predicted
to be
\begin{eqnarray}
\nonumber&&\Delta
m^2_{sol}=\frac{\left(x^2-3\right)\left(x^2+z^2\right)+\text{sign}(x\cos2\theta)(1+z^2)\sqrt{4x^2+3z^2}}{\left(1+z^2\right)\left[(1-x^2-z^2)^2+z^2\right]}|m_0|^2\,,\\
\nonumber&&\Delta
m^2_{atm}=\frac{2\sqrt{4x^2+3z^2}}{(1-x^2-z^2)^2+z^2}|m_0|^2,\quad\text{for~~NO}\,,\\
&&\Delta
m^2_{atm}=\frac{\left(x^2-3\right)\left(x^2+z^2\right)+(1+z^2)\sqrt{4x^2+3z^2}}{\left(1+z^2\right)\left[(1-x^2-z^2)^2+z^2\right]}|m_0|^2,\quad
\text{for~~IO}\,.
\end{eqnarray}
When we impose the best fit values for the reactor mixing angle
$\sin^2\theta_{13}=0.0227$ and the mass-squared differences $\Delta
m^2_{sol}=7.50\times10^{-5}\text{eV}^2$ and $\Delta
m^2_{atm}=2.473(2.427)\times10^{-3}\text{eV}^2$ for normal (inverted)
ordering, we find six possible solutions to the parameters $x$ and $z$:
\begin{equation}
(x,z)\simeq(0.97, \pm0.44),~~(0.81, \pm0.36),~~ (-2.17, \pm0.98)\,,
\end{equation}
where the first four cases correspond to a normally ordered neutrino mass
spectrum, while latter two correspond to inverted ordering. The
corresponding predictions for the light neutrino masses and the lepton
mixing parameters are presented in Table~\ref{tab:predictions_renorm}.
\begin{table} [t!]
\begin{center}
\begin{tabular}{|c|c|c|c|c|c|c|c|c|}
\hline\hline
 $\left(x,z\right)$ & $~\delta_{\rm{CP}}~$ & $\alpha_{21}$  & $\alpha_{31}$ & $m_1$ &   $m_2$ & $m_3$ &  $|m_{\beta\beta}|$   &   \text{mass order} \\ \hline

$(0.97, 0.44)$  & $\pi/2$   &   $0.17\pi$  &   $0.47\pi$ & \multirow{2}{*}{5.43}  &  \multirow{2}{*}{10.22}  &   \multirow{2}{*}{50.02}   & \multirow{2}{*}{6.37}   &  \multirow{2}{*}{\text{NO}} \\ \cline{1-4}

$(0.97, -0.44)$ &  $3\pi/2$ &  $1.83\pi$   &  $1.53\pi$ &   & &   &  &  \\ \hline

$(0.81, 0.36)$  &  $\pi/2$  &  $0.14\pi$  &   $0.73\pi$  &  \multirow{2}{*}{5.95}   &   \multirow{2}{*}{10.51}  &   \multirow{2}{*}{50.08}  &  \multirow{2}{*}{7.76}   &  \multirow{2}{*}{\text{NO}}   \\ \cline{1-4}

$(0.81,-0.36)$ & $3\pi/2$  &   $1.86\pi$ &  $1.27\pi$   &     &  &   & & \\ \hline

$(-2.17, 0.98)$  &  $3\pi/2$  &   $1.13\pi$   &   $1.84\pi$   &   \multirow{2}{*}{53.46} &  \multirow{2}{*}{54.15}   &   \multirow{2}{*}{22.49}    &  \multirow{2}{*}{18.90}   &  \multirow{2}{*}{\text{IO}} \\ \cline{1-4}

$(-2.17, -0.98)$  &  $\pi/2$   &   $0.87\pi$   &   $0.16\pi$  &  &  &   &  &    \\  \hline\hline

\end{tabular}
\caption{\label{tab:predictions_renorm}The predictions for the leptonic CP phases, the light neutrino masses $m_i(i=1,2,3)$ and the effective mass $|m_{\beta\beta}|$ of the neutrinoless double-beta decay in the UV completion of the effective model, where the unit of mass is meV.}
\end{center}
\end{table}

\section{\label{sec:conclusion}Conclusions}

A promising and attractive approach to the well-known family puzzle is to invoke (spontaneously broken) discrete family symmetry to describe the observed patterns. The lepton mixing angles and CP violating phases can be predicted simultaneously from a family symmetry $G_f$ combined with a generalised CP symmetry $H_{\rm{CP}}$, which is broken  to different remnant symmetries in the neutrino and charged lepton sectors. In this work, we have focused on the most popular $A_4$ family symmetry. For the faithful representation $\mathbf{3}$, we find that the generalised CP symmetry is $S_4$ which is the automorphism group of $A_4$. However, only half of these 24 generalised CP transformations are consistent with the nontrivial singlet representations $\mathbf{1}'$ and $\mathbf{1}''$. We performed a comprehensive study of lepton mixing angles and CP phases which can be produced from the original symmetry $A_4\rtimes H_{\rm{CP}}$ breaking to different remnant symmetries. Of all the possibilities, we find that only the case with $G^{\nu}_{\rm{CP}}=Z_2\times H^{\nu}_{\rm{CP}}$ and $G^{l}_{\rm{CP}}=Z_3\rtimes H^{l}_{\rm{CP}}$ is phenomenologically viable, in which the second column of the corresponding lepton mixing matrix is proportional to $(1,1,1)^{T}$. Furthermore, there is no CP violation in this case, namely $\delta_{CP}=0,\pi$, with Majorana phases $\alpha_{21}=0,\pi$ and $\alpha_{31}=0,\pi$.

Motivated by this general analysis, we have constructed an effective SUSY model for leptons based on the  $A_4\rtimes
H_{\rm{CP}}$ symmetry with  auxiliary $Z_4\times Z_6$ symmetries.   This model reproduces the correct mass hierarchies among the three charged leptons. At LO, the lepton mixing is of the tri-bimaximal form, which is further reduced to trimaximal mixing by the NLO contributions. Consequently the reactor mixing angle arises as a NLO correction, and thus it is of the correct order of magnitude. It is notable that the Dirac phase is predicted to be trivial or approximately maximal, namely $\delta_{CP}=0,\pi$ or $\delta_{CP}=\pm \pi/2$, with Majorana phases
$\alpha_{21}$ and $\alpha_{31}$ being more general. For the case $\delta_{CP}=0,\pi$, the residual symmetry in the neutrino sector is $G^{\nu}_{\rm{CP}}=Z_2\times H^{\nu}_{\rm{CP}}$ with $H^{\nu}_{\rm{CP}}=\left\{\rho_{\mathbf{r}}(1),\rho_{\mathbf{r}}(S)\right\}$.
While for the nearly maximal CP violation case, i.e. $\delta_{\rm{CP}}\simeq\pm\frac{\pi}{2}$, the generalised CP symmetry
is broken completely in the neutrino sector.

Furthermore, we have promoted this effective model to a renormalisable one, where the non-renormalisable terms arise from integrating out heavy messenger fields and some higher dimensional operators included at the effective level are eliminated. As a result, the model becomes rather predictive, and the light neutrino mass matrix depends on only three real parameters which are fixed to reproduce the observed values of $\Delta m^2_{sol}$, $\Delta m^2_{atm}$ and $\theta_{13}$.  Then all the other observables including $\theta_{12}$, $\theta_{23}$, Dirac phase $\delta_{\rm{CP}}$, Majorana phases and the absolute neutrino mass scale are related, leading to the
definite predictions shown in Table~\ref{tab:predictions_renorm}. In particular, both the atmospheric mixing angle $\theta_{23}$ and Dirac phase $\delta_{\rm{CP}}$ are maximal.

\section*{Acknowledgements}

We are grateful to Christoph Luhn for his participation in the early stage of the work. The research was partially supported by  the National Natural Science Foundation of China under Grant Nos. 11275188 and 11179007, the EU ITN grants UNILHC 237920 and INVISIBLES 289442. SK and AJS acknowledge support from the STFC Consolidated ST/J000396/1 grant.

\newpage

\appendix

\section{\label{sec:appendix_A}Group theory of $A_4$}
\cleqn

$A_4$ is the even permutation group of four objects.  As such, it has 12 elements. Geometrically, it is isomorphic to the symmetry group of a regular tetrahedron. The elements of $A_4$ can be generated by two generators $S$ and $T$ satisfying the relation:
\begin{equation}
\label{eq:relation}S^2=T^3=(ST)^3=1\,.
\end{equation}
The 12 elements of $A_4$ are obtained as $1$, $S$, $T$, $ST$, $TS$, $T^2$,
$ST^2$, $STS$, $TST$, $T^2S$, $TST^2$ and $T^2ST$. Without loss of
generality, we can choose
\begin{equation}
S=(14)(23),~~~~~T=(123)\,,
\end{equation}
where the cycle $(123)$ represents the permutation $(1,2,3,4)\rightarrow(2,3,1,4)$ and $(14)(23)$ means $(1,2,3,4)\rightarrow(4,3,2,1)$. The $A_4$ elements belong to 4 conjugacy classes:
\begin{eqnarray}
\nonumber&&1C_1: 1\\
\nonumber&&4C_3: T=(123),~~ST=(134),~~TS=(142),~~STS=(243)\\
\nonumber&&4C_3^2: T^2=(132),~~ST^2=(124),~~T^2S=(143),~~ST^2S=(234)\\
&&3C_2: S=(14)(23),~~T^2ST=(12)(34),~~TST^2=(13)(24)\,.
\end{eqnarray}
In the above, we have adopted Schoenflies notation in which $m C_n^k$ denotes a conjugacy class of $m$ elements of rotations by an angle $\frac{2\pi k
}{n}$. $A_4$ has four inequivalent irreducible representations: three singlet representations $\mathbf{1}$, $\mathbf{1}'$, $\mathbf{1}''$ and one triplet
representation $\mathbf{3}$ which is a faithful representation of $A_4$. The
representation matrices of the generators $S$ and $T$ are listed in
Table~\ref{tab:representation}.
\begin{table}[t!]
\begin{center}
\begin{tabular}{|c|c|c|}\hline\hline
 ~~  &  $S$  &   $T$     \\ \hline
~~~$\mathbf{1}$  ~~~ & 1   &  1  \\ \hline
   &   &      \\ [-0.16in]

~~~$\mathbf{1}'$  ~~~ & 1   & $\omega^2$  \\ \hline
   &   &       \\ [-0.16in]

~~~$\mathbf{1}''$  ~~~ & 1   &  $\omega$  \\ \hline
   &   &      \\ [-0.16in]

$\mathbf{3}$~~ & $\frac{1}{3} \left(\begin{array}{ccc}
    -1& 2  & 2  \\
    2  & -1  & 2 \\
    2 & 2 & -1
    \end{array}\right)$
    & $\left( \begin{array}{ccc}
    1 & 0 & 0 \\
    0 & \omega^{2} & 0 \\
    0 & 0 & \omega
    \end{array}\right) $
   \\[0.22in] \hline\hline
\end{tabular}
\caption{\label{tab:representation}The representation matrices for the $A_4$
generators $S$ and $T$ in different irreducible representations, where
$\omega=e^{2\pi i/3}$ is the cube root of unit.}
\end{center}
\end{table}
The Kronecker products between various irreducible representations are as
follows:
\begin{eqnarray}
\nonumber&&\mathbf{1}\otimes
R=R,~~\mathbf{1}'\otimes\mathbf{1}''=\mathbf{1},~~~\mathbf{1}'\otimes\mathbf{1}'=\mathbf{1}'',~~~\mathbf{1}''\otimes\mathbf{1}''=\mathbf{1}',\\
&&\mathbf{3}\otimes\mathbf{1}'=\mathbf{3},~~\mathbf{3}\otimes\mathbf{1}''=\mathbf{3},~~~\mathbf{3}\otimes\mathbf{3}=\mathbf{1}\oplus\mathbf{1}'\oplus\mathbf{1}''\oplus\mathbf{3}_S\oplus\mathbf{3}_A\,,
\end{eqnarray}
where $R$ denotes any $A_4$ representation, and the subscript $S$ ($A$) denotes symmetric (antisymmetric) combinations. For two $A_4$ triplets
$\alpha=(\alpha_1,\alpha_2,\alpha_3)\sim\mathbf{3}$ and
$\beta=(\beta_1,\beta_2,\beta_3)\sim\mathbf{3}$, the irreducible
representations obtained from their product are:
\begin{eqnarray}
\begin{gathered}
\mathbf{1}\equiv(\alpha\beta)=\alpha_1\beta_1+\alpha_2\beta_3+\alpha_3\beta_2\,,
\\
\mathbf{1}'\equiv(\alpha\beta)'=\alpha_3\beta_3+\alpha_1\beta_2+\alpha_2\beta_1\,,
\\
\mathbf{1}''\equiv(\alpha\beta)''=\alpha_2\beta_2+\alpha_1\beta_3+\alpha_3\beta_1\,,
\\
\mathbf{3}_S\equiv(\alpha\beta)_{3_S}=\frac{1}{3}\left(\begin{array}{c}
2\alpha_1\beta_1-\alpha_2\beta_3-\alpha_3\beta_2 \\
2\alpha_3\beta_3-\alpha_1\beta_2-\alpha_2\beta_1\\
2\alpha_2\beta_2-\alpha_1\beta_3-\alpha_3\beta_1
\end{array}\right),\quad
\mathbf{3}_A\equiv(\alpha\beta)_{3_A}=\frac{1}{2}\left(\begin{array}{c}
\alpha_2\beta_3-\alpha_3\beta_2\\
\alpha_1\beta_2-\alpha_2\beta_1\\
\alpha_3\beta_1-\alpha_1\beta_3
\end{array}\right)\,,
\end{gathered}
\end{eqnarray}
where we have followed the same convention of Ref.~\cite{Altarelli:2005yx}.

Finally $A_4$ has three $Z_2$ subgroups, four $Z_3$ subgroups and one $K_4\cong Z_2\times Z_2$ subgroup, which can be expressed in terms of the generators $S$ and $T$ as follows:
\begin{itemize}
  \item {$Z_2$ subgroups}
  \begin{equation}
  \label{eq:Z2_subg}Z^{S}_2=\left\{1,\;S\right\},\quad
  Z^{T^2ST}_2=\left\{1,\;T^2ST\right\},\quad
  Z^{TST^2}_2=\left\{1,\;TST^2\right\}\,.
  \end{equation}

  \item {$Z_3$ subgroups}
  \begin{eqnarray}
  \nonumber&&Z^{T}_3=\left\{1,T,T^2\right\},\quad
  Z^{ST}_3=\left\{1,ST,T^2S\right\}, \\
  \label{eq:Z3_subg}&&Z^{TS}_3=\left\{1,TS,ST^2\right\}, \quad
  Z^{STS}_3=\left\{1,STS, ST^2S\right\}\,.
  \end{eqnarray}

  \item {$K_4$ subgroup}
  \begin{equation}
  K_4=\left\{1,S,T^2ST,TST^2\right\}\,.
  \end{equation}
\end{itemize}
We note that $K_4$ is the normal subgroup of $A_4$, all $Z_3$ subgroups are conjugate to each other, and all $Z_2$ groups are conjugate to each other as well.

\section{\label{K4}Implication of $G_{\nu}=K_4\cong Z_2\times Z_2$}
\cleqn

We first show that the remnant subgroup $G_{\nu}=K_4$ in the neutrino sector can not lead to phenomenologically acceptable lepton mixing angles even if we only impose the $A_4$ family symmetry. In order to be able to uniquely fix the mixing pattern from the group structure, the residual family symmetry in the charged lepton sector is taken to be $Z_3$ abelian subgroups. Thus, there are four possible choices for the preserved charged lepton subgroup $G_l$ of $A_4$ with $G_{\nu}=K_4$, i.e. $G_l=Z^{T}_3$, $G_l=Z^{ST}_3$,
$G_l=Z^{TS}_3$ or $G_l=Z^{STS}_3$. All four of these combinations lead to the same mixing parameters:
\begin{equation}
\sin^2\theta_{13}=1/3,\quad \sin^2\theta_{12}=\sin^2\theta_{23}=1/2,\quad
|\sin\delta_{\rm{CP}}|=1\,.
\end{equation}
The same results have also been found in Refs.~\cite{Lam:2011ag,deAdelhartToorop:2011re}. Obviously this mixing
pattern is not consistent with the present data. This result confirms that it is impossible to generate tri-bimaximal mixing by preserving the complete Klein symmetry group of $A_4$ in the neutrino sector. In order to produce tri-bimaximal mixing in $A_4$, one should use flavons transforming as $\mathbf{3}$ not $\mathbf{1}'$ or $\mathbf{1}''$
to break the family symmetry such that only the $Z^{S}_2$ subgroup together
with another accidental $Z_2$ $\mu-\tau$ symmetry is preserved in the neutrino sector. Moreover, if we choose $G_l=K_4$, the resulting mixing matrix will be the identity matrix up to permutation of rows and columns.  This case is clearly not viable.

As an academic exercise to further convince the reader that $G_{\nu}$ can not be $K_4$ subgroup when considering $G_f=A_4$, it is insightful to investigate the constraints that the residual CP and family symmetries impose on the mass matrices. Considering the $K_4$ family symmetry first, the following constraints are found on $m_{\nu}$:
\begin{eqnarray}
\nonumber&&\rho^{T}_{\mathbf{3}}(S)m_{\nu}\rho_{\mathbf{3}}(S)=m_{\nu},\\
&&\rho^{T}_{\mathbf{3}}(TST^2)m_{\nu}\rho_{\mathbf{3}}(TST^2)=m_{\nu}\,,
\end{eqnarray}
because $K_4=\left\{1,S,TST^2,T^2ST\right\}$ can be generated by $S$ and $TST^2$. Then, the most general neutrino mass matrix satisfying these equations has the form
\begin{equation}
\label{eq:nu_k4}m_{\nu}=\left(\begin{array}{ccc}
m_{11}  &   m_{12}  &  m_{13}  \\
m_{12}  &  m_{13}   &  m_{11}  \\
m_{13}  &  m_{11}   & m_{12}
\end{array}\right)\,,
\end{equation}
where $m_{11}$, $m_{12}$ and $m_{13}$ are complex parameters. It can be
diagonalised by the unitary transformation
\begin{equation}
\label{eq:um_K4}U_{K}=\frac{1}{\sqrt{3}}\left(\begin{array}{ccc}
1  &  \omega &  \omega^2  \\
1  &  \omega^2  &  \omega \\
1  &   1   &  1
\end{array}\right)\,,
\end{equation}
where $\omega=e^{2\pi i/3}$.  Thus,
\begin{equation}
U^{T}_{K}m_{\nu}U_{K}=\text{diag}(m_1,m_2,m_3)\,,
\end{equation}
where
\begin{eqnarray}
\begin{array}{cccccc}
&m_1=m_{11}+m_{12}+m_{13},&m_2=\omega^2m_{11}+m_{12}+\omega
m_{13},&m_3=\omega m_{11}+m_{12}+\omega^2m_{13}.&~&
\end{array}
\end{eqnarray}
The light neutrino mass matrix $m_{\nu}$ of Eq.~\eqref{eq:nu_k4} is further
constrained by the remnant CP symmetry $H^{\nu}_{CP}$, as shown in
Eq.~\eqref{eq:inv_CP}, and the associated consistency equations are
\begin{equation}
X_{\mathbf{r}\nu}\rho^{*}_{\mathbf{r}}(S)X^{-1}_{\mathbf{r}\nu}=\rho_{\mathbf{r}}(S')
,\quad
X_{\mathbf{r}\nu}\rho^{*}_{\mathbf{r}}(TST^2)X^{-1}_{\mathbf{r}\nu}=\rho_{\mathbf{r}}(g'),\quad
S',g'\in K_4\,.
\end{equation}
By considering all possible values for $S'$ and $g'$, we find that all
twelve CP transformations of $A_4$ in Eq.~\eqref{eq:CP_A4} are acceptable, i.e.
\begin{equation}
\label{eq:CP_K4}X_{\mathbf{r}\nu}=\rho_{\mathbf{r}}(g),\quad g\in A_4\,,
\end{equation}
where $g$ is any group element of $A_4$. We further find that $H^{\nu}_{\rm{CP}}$ can be classified
into three cases:
\begin{itemize}
  \item {
      $X_{\mathbf{r}\nu}=\rho_{\mathbf{r}}(1),\rho_{\mathbf{r}}(S),\rho_{\mathbf{r}}(TST^2),
      \rho_{\mathbf{r}}(T^2ST)$}

      In this case, $m_{11}$, $m_{12}$ and $m_{13}$ are constrained to be
      real, and thus we have the degeneracy $|m_2|^2=|m_3|^2$. The
      mass-squared splittings $\Delta m^2_{sol}\equiv|m_2|^2-|m_1|^2$ and
      $\Delta m^2_{atm}\equiv\big||m_3|^2-|m_1|^2(|m_2|^2)\big|$ have been
      precisely measured to be non-zero\footnote{The atmospheric
      mass-squared difference $\Delta m^2_{atm}\equiv|m_3|^2-|m_1|^2$ for the
      normal ordered neutrino mass spectrum and $\Delta
      m^2_{atm}\equiv|m_2|^2-|m_3|^2$ for the inverted ordering.},
      consequently the three light neutrinos should be of different
      masses. Moreover, partially degenerate light neutrino masses are
      disfavoured by the recent Planck results~\cite{Ade:2013zuv}.
      Therefore this case is not viable.

  \item {$X_{\mathbf{r}\nu}=\rho_{\mathbf{r}}(T), \rho_{\mathbf{r}}(ST),
      \rho_{\mathbf{r}}(TS), \rho_{\mathbf{r}}(STS)$}

       In this case, the parameters $m_{11}$, $\omega m_{12}$ and $\omega^2m_{13}$ are
       required to be real. Therefore, it leads to the degeneracy
       $|m_1|^2=|m_3|^2$, which is not compatible with the experimental
       data.

  \item {$X_{\mathbf{r}\nu}=\rho_{\mathbf{r}}(T^2),
      \rho_{\mathbf{r}}(ST^2), \rho_{\mathbf{r}}(T^2S),
      \rho_{\mathbf{r}}(ST^2S)$}

     The parameters $m_{11}$, $\omega^2m_{12}$ and $\omega m_{13}$ have to be real in this case.  Therefore,
    the degeneracy $|m_1|^2=|m_2|^2$ is produced.  This scenario is also not
    in accordance with three distinct neutrino masses.
\end{itemize}
As a result, if both the $K_4$ subgroup and the associated generalised CP
symmetry are preserved in the neutrino sector, the neutrino mass matrix is
 strongly constrained such that the resulting light neutrino mass spectrum
is partially degenerate, and the PMNS matrix cannot be determined uniquely.  Thus, as determined before from mixing considerations, $G_{\nu}=K_4$ is not phenomenologically viable.

\section{\label{GlZ2}Implication of $G_{l}=Z_2$ }
\cleqn

In this appendix, we consider the possibility that $G_l$ is a $Z_2$ subgroup of $A_4$. It is sufficient to discuss the representative case $G_{l}=Z^{S}_2$. As shown in Eq.~\eqref{eq:CP_Z2}, the CP symmetry $H^{l}_{\rm{CP}}$ consistent with $Z^{S}_2$ is
\begin{equation}
H^{l}_{\rm{CP}}=\left\{\rho_{\mathbf{r}}(1), \rho_{\mathbf{r}}(S),
\rho_{\mathbf{r}}(T^2ST), \rho_{\mathbf{r}}(TST^2)\right\}\,.
\end{equation}
The hermitian combination $m_{l}m^{\dagger}_{l}$ is constrained by the
remnant symmetry $G^{l}_{\rm{CP}}\cong Z^{S}_2\times H^{l}_{\rm{CP}}$ as
\begin{eqnarray}
\nonumber&&\rho^{\dagger}_{\mathbf{3}}(S)m_{l}m^{\dagger}_{l}\rho_{\mathbf{3}}(S)=m_{l}m^{\dagger}_{l},\\
\label{eq:cc_mm_Z2}&&X^{\dagger}_{\mathbf{3}l}m_{l}m^{\dagger}_{l}X_{\mathbf{3}l}=(m_{l}m^{\dagger}_{l})^{*}\,,
\end{eqnarray}
which allows us to straightforwardly reconstruct $m_{l}m^{\dagger}_{l}$.
There are two possible scenarios:
\begin{itemize}
  \item {$X_{\mathbf{r}l}=\rho_{\mathbf{r}}(1), \rho_{\mathbf{r}}(S)$}

  The mass matrix $m_{l}m^{\dagger}_{l}$ fulfilling Eq.~\eqref{eq:cc_mm_Z2}  is of the form

  \begin{equation}
  m_{l}m^{\dagger}_{l}=\tilde{\alpha}\left(\begin{array}{ccc}
2 & -1 & -1 \\
-1 & 2 & -1 \\
-1 & -1 &  2
\end{array}\right)+\tilde{\beta}\left(\begin{array}{ccc}
1 &  0 & 0 \\
0 & 0  &  1  \\
0 &  1 & 0
\end{array}\right)+\tilde{\gamma}\left(\begin{array}{ccc}
0 &  1 & 1 \\
1 &  1 & 0  \\
1 &  0 &  1
\end{array}\right)+\tilde{\epsilon}\left(\begin{array}{ccc}
0 & 1 & -1  \\
1 & -1 &  0 \\
-1 & 0 & 1
\end{array}\right)\,,
  \end{equation}
where $\tilde{\alpha}$, $\tilde{\beta}$, $\tilde{\gamma}$ and
$\tilde{\epsilon}$ are real parameters. After performing a tri-bimaximal
transformation, we have
\begin{equation}
U^{\dagger}_{TB}m_{l}m^{\dagger}_{l}U_{TB}=\left(\begin{array}{ccc}
3\tilde{\alpha}+\tilde{\beta}-\tilde{\gamma}  & 0  &
-\sqrt{3}\;\tilde{\epsilon}  \\
0  &  \tilde{\beta}+2\tilde{\gamma}  & 0  \\
-\sqrt{3}\;\tilde{\epsilon}  & 0  &
3\tilde{\alpha}-\tilde{\beta}+\tilde{\gamma}
\end{array}\right)\,,
\end{equation}
which can be further diagonalised by a (1,3) rotation $R(\vartheta)$,
\begin{equation}
R(\vartheta)=\left(\begin{array}{ccc}
\cos\vartheta  &  0  &  \sin\vartheta  \\
0  &   1   &    0  \\
-\sin\vartheta  &  0   &  \cos\vartheta
\end{array}\right)\,,
\end{equation}
with
$\tan2\vartheta=\sqrt{3}\;\tilde{\epsilon}/(\tilde{\beta}-\tilde{\gamma})$.
The squared charged lepton masses are given by
\begin{eqnarray}
\nonumber&&m^2_e=3\tilde{\alpha}-\sqrt{(\tilde{\beta}-\tilde{\gamma})^2+3\tilde{\epsilon}^{2}},\\
\nonumber&&m^2_{\mu}=\tilde{\beta}+2\tilde{\gamma},\\
&&m^2_{\tau}=3\tilde{\alpha}+\sqrt{(\tilde{\beta}-\tilde{\gamma})^2+3\tilde{\epsilon}^2}\,.
\end{eqnarray}
In order to account for the observed hierarchies among the charged lepton masses $m_e$, $m_{\mu}$ and $m_{\tau}$, a moderate fine-tuning of the
parameters $\tilde{\alpha}$, $\tilde{\beta}$, $\tilde{\gamma}$ and
$\tilde{\epsilon}$ is needed.

  \item {$X_{\mathbf{r}l}=\rho_{\mathbf{r}}(T^2ST),
      \rho_{\mathbf{r}}(TST^2)$}

  In this case, $m_{l}m^{\dagger}_{l}$ is of the form
  \begin{equation}
  m_{l}m^{\dagger}_{l}=\left(\begin{array}{ccc}
  R_{11}  &  R_{12}  &  R_{12}  \\
  R_{12}  & R_{11}   &  R_{12}  \\
  R_{12}  & R_{12}   &  R_{11}
  \end{array}\right)\,,
  \end{equation}
where $R_{11}$ and $R_{12}$ are real. After applying a tri-bimaximal transformation, it becomes
\begin{equation}
U^{\dagger}_{TB}m_{l}m^{\dagger}_{l}U_{TB}=\text{diag}(R_{11}-R_{12},R_{11}+2R_{12},R_{11}-R_{12})\,,
\end{equation}
which implies $m^2_e=m^2_{\tau}$.  This is obviously not viable.
\end{itemize}
In the cases of $G_{l}=Z^{T^2ST}_2$ and $G_{l}=Z^{TST^2}_2$, we can immediately obtain the corresponding consistent CP transformations and the mass matrix with the aid of the relations in
Eqs.~(\ref{eq:CP_conju},\ref{eq:mass_matr_conju}).

Assuming that $G^{\nu}_{\rm{CP}}\cong Z_2\times H^{\nu}_{\rm{CP}}$ (the only viable possibility for the neutrino sector) and $G^{l}_{\rm{CP}}\cong Z_2\times H^{l}_{\rm{CP}}$ (as discussed in this Appendix) then the corresponding PMNS matrix is of the form
\begin{equation}
U_{PMNS}=R^{\dagger}(\vartheta)U^{\dagger}_{TB}\rho^{m}_{\mathbf{3}}(T)U_{TB}R(\theta),\quad
m=0,\pm1\,.
\end{equation}
For $m=0$, which corresponds to the remnant $Z_2$ symmetry in $G^{\nu}_{\rm{CP}}$ and $G^{l}_{\rm{CP}}$ being the same, the lepton mixing angles are
\begin{equation}
\label{eq:mixing_angle1_Z2}\sin^2\theta_{13}=\sin^2(\theta-\vartheta),\quad
\sin^2\theta_{12}=\sin^2\theta_{23}=0\,.
\end{equation}
For the case $m=\pm1$, where the $Z_2$ factors in $G^{\nu}_{\rm{CP}}$ and
$G^{l}_{\rm{CP}}$ are different, the lepton mixing angles are
\begin{equation}
\label{eq:mixing_angle2_Z2}\sin^2\theta_{13}=1/4,\qquad
\sin^2\theta_{12}=\sin^2\theta_{23}=2/3\,.
\end{equation}
Obviously the predictions in both Eq.~\eqref{eq:mixing_angle1_Z2} and
Eq.~\eqref{eq:mixing_angle2_Z2} are disfavoured by experimental data.
Therefore we exclude the possibility that $G_{l}=Z_2$.

\section{\label{GlK4} Implication of $G_{l}=K_4$}
\cleqn

In this appendix, we discuss the last possibility $G_{l}=K_4$, which implies that
\begin{eqnarray}
\nonumber&&\rho^{\dagger}_{\mathbf{3}}(S)m_{l}m^{\dagger}_{l}\rho_{\mathbf{3}}(S)=m_{l}m^{\dagger}_{l},\\
&&\rho^{\dagger}_{\mathbf{3}}(TST^2)m_{l}m^{\dagger}_{l}\rho_{\mathbf{3}}(TST^2)=m_{l}m^{\dagger}_{l}\,.
\end{eqnarray}
Then, the mass matrix $m_{l}m^{\dagger}_{l}$ is determined to be of the form
\begin{equation}
\label{eq:cc_mm_K4}m_{l}m^{\dagger}_{l}=\left(\begin{array}{ccc}
\tilde{m}_{11}  &  \tilde{m}_{12}  &   \tilde{m}^{*}_{12}  \\
\tilde{m}^{*}_{12}  &  \tilde{m}_{11}  &  \tilde{m}_{12}  \\
\tilde{m}_{12}   &  \tilde{m}^{*}_{12}  &  \tilde{m}_{11}
\end{array}\right)\,,
\end{equation}
where $\tilde{m}_{11}$ is real and $\tilde{m}_{12}$ is complex. It is
diagonalised by the unitary transformation $U_K$ of Eq.~\eqref{eq:um_K4},
\begin{equation}
U^{\dagger}_Km_{l}m^{\dagger}_{l}U_K=\text{diag}(m^2_e, m^2_{\mu},
m^2_{\tau})\,,
\end{equation}
with
\begin{eqnarray}
\nonumber&&m^2_e=\tilde{m}_{11}+\tilde{m}_{12}+\tilde{m}^{*}_{12},\\
\nonumber&&m^2_{\mu}=\tilde{m}_{11}+\omega\tilde{m}_{12}+\omega^2\tilde{m}^{*}_{12},\\
\label{eq:ch_ma_K4}&&m^2_{\tau}=\tilde{m}_{11}+\omega^2\tilde{m}_{12}+\omega\tilde{m}^{*}_{12}\,.
\end{eqnarray}
The hermitian combination $m_{l}m^{\dagger}_{l}$ also respects the CP symmetry $H^{l}_{\rm{CP}}$. As shown in Eq.~\eqref{eq:CP_K4}, all twelve CP
transformations are consistent with the $K_4$ subgroup, i.e.,
\begin{equation}
X_{\mathbf{r}l}=\rho_{\mathbf{r}}(g),\quad g\in A_4\,,
\end{equation}
where $X_{\mathbf{r}l}$ is the element of $H^{l}_{\rm{CP}}$. It is clear that invariance under the action of $H^{l}_{\rm{CP}}$ yields
\begin{equation}
X^{\dagger}_{\mathbf{3}l}m_{l}m^{\dagger}_{l}X_{\mathbf{3}l}=(m_{l}m^{\dagger}_{l})^{*}\,,
\end{equation}
which further constrains the parameter $m_{12}$ of Eq.~\eqref{eq:cc_mm_K4} in various ways for different preserved CP subgroups as follows:
\begin{itemize}
  \item
      {$X_{\mathbf{r}l}=\rho_{\mathbf{r}}(1),\rho_{\mathbf{r}}(S),\rho_{\mathbf{r}}(TST^2),
      \rho_{\mathbf{r}}(T^2ST)$}

      In this case, the parameter $m_{12}$ is real, which leads to
      $m_{\mu}=m_{\tau}$.

  \item {$X_{\mathbf{r}l}=\rho_{\mathbf{r}}(T), \rho_{\mathbf{r}}(ST),
      \rho_{\mathbf{r}}(TS), \rho_{\mathbf{r}}(STS)$}

      $\omega m_{12}$ is constrained to be real, and thus the degeneracy
      $m_e=m_{\tau}$ arises.

  \item {$X_{\mathbf{r}l}=\rho_{\mathbf{r}}(T^2), \rho_{\mathbf{r}}(ST^2),
      \rho_{\mathbf{r}}(T^2S), \rho_{\mathbf{r}}(ST^2S)$ }

      $\omega^2m_{12}$ is real in this case, and the relation
      $m_{e}=m_{\mu}$ follows immediately.

\end{itemize}
Therefore the symmetry breaking $G^{l}_{\rm{CP}}\cong K_4\rtimes H^{l}_{\rm{CP}}$ leads to partial degeneracy among the charged lepton masses. Hence, it is not viable.

\section{\label{sec:appendix_B}Diagonalisation of a $2\times2$ symmetric
complex matrix }
\cleqn

If neutrinos are Majorana particles, their mass matrix is symmetric and
generally complex. In the following, we present the result for
the diagonalisation of a general $2\times2$ symmetric complex matrix, which
is of the form
\begin{equation}
\mathcal{M}=\left(
\begin{array}{cc}
a_{11}e^{i\phi_{11}}  &  a_{12}e^{i\phi_{12}} \\
a_{12}e^{i\phi_{12}}  &  a_{22}e^{i\phi_{22}}
\end{array}
\right)\,,
\end{equation}
where $a_{ij}$ and $\phi_{ij}$ $(i,j=,1,2)$ are real. It can be diagonalised by a unitary matrix $U$ via
\begin{equation}
U^{T}\mathcal{M}U=\text{diag}(\lambda_1,\lambda_2)\,,
\end{equation}
where the unitary matrix $U$ can be written as
\begin{equation}
U=\left(\begin{array}{cc}
\cos\theta e^{i(\phi+\varrho)/2}   &  \sin\theta e^{i(\phi+\sigma)/2}   \\
-\sin\theta e^{i(-\phi+\varrho)/2}    &    \cos\theta e^{i(-\phi+\sigma)/2}
\end{array}\right)\,,
\end{equation}
with the rotation angle $\theta$ satisfying
\begin{eqnarray}
&&\tan2\theta=\frac{2a_{12}\sqrt{a^2_{11}+a^2_{22}+2a_{11}a_{22}\cos(\phi_{11}+\phi_{22}-2\phi_{12})}}{a^2_{22}-a^2_{11}}\,.
\end{eqnarray}
The eigenvalues $\lambda_1$ and $\lambda_2$ can always set to be positive
with
\begin{eqnarray}
\nonumber&&\lambda^2_1=\frac{1}{2}\left\{a^2_{11}+a^2_{22}+2a^2_{12}-\mathcal{S}\sqrt{(a^2_{22}-a^2_{11})^2+4a^2_{12}\left[a^2_{11}+a^2_{22}+2a_{11}a_{22}\cos(\phi_{11}+\phi_{22}-2\phi_{12})\right]}\right\}\,,\\
\nonumber&&\lambda^2_2=\frac{1}{2}\left\{a^2_{11}+a^2_{22}+2a^2_{12}+\mathcal{S}\sqrt{(a^2_{22}-a^2_{11})^2+4a^2_{12}\left[a^2_{11}+a^2_{22}+2a_{11}a_{22}\cos(\phi_{11}+\phi_{22}-2\phi_{12})\right]}\right\}\,,
\end{eqnarray}
where
$\mathcal{S}=\text{sign}\big(\left(a^2_{22}-a^2_{11}\right)\cos2\theta\big)$.
Finally the phases $\phi$, $\varrho$ and $\sigma$ are given by
\begin{eqnarray}
\nonumber&&\sin\phi=\frac{-a_{11}\sin(\phi_{11}-\phi_{12})+a_{22}\sin(\phi_{22}-\phi_{12})}{\sqrt{a^2_{11}+a^2_{22}+2a_{11}a_{22}\cos(\phi_{11}+\phi_{22}-2\phi_{12})}}=\frac{Im\left(\mathcal{M}_{11}^{*}\mathcal{M}_{12}+\mathcal{M}_{22}\mathcal{M}_{12}^{*}\right)}{\left|\mathcal{M}_{11}^{*}\mathcal{M}_{12}+\mathcal{M}_{22}\mathcal{M}_{12}^{*}\right|},\\
\nonumber&&\cos\phi=\frac{a_{11}\cos(\phi_{11}-\phi_{12})+a_{22}\cos(\phi_{22}-\phi_{12})}{\sqrt{a^2_{11}+a^2_{22}+2a_{11}a_{22}\cos(\phi_{11}+\phi_{22}-2\phi_{12})}}=\frac{Re\left(\mathcal{M}_{11}^{*}\mathcal{M}_{12}+\mathcal{M}_{22}\mathcal{M}_{12}^{*}\right)}{\left|\mathcal{M}_{11}^{*}\mathcal{M}_{12}+\mathcal{M}_{22}\mathcal{M}_{12}^{*}\right|}\,,\\
\nonumber&&\sin\varrho=-\frac{\left(\lambda^2_1-a^2_{12}\right)\sin\phi_{12}+a_{11}a_{22}\sin(\phi_{11}+\phi_{22}-\phi_{12})}{\lambda_1\sqrt{a^2_{11}+a^2_{22}+2a_{11}a_{22}\cos(\phi_{11}+\phi_{22}-2\phi_{12})}}\,,\\
\nonumber&&\cos\varrho=\frac{\left(\lambda^2_1-a^2_{12}\right)\cos\phi_{12}+a_{11}a_{22}\cos(\phi_{11}+\phi_{22}-\phi_{12})}{\lambda_1\sqrt{a^2_{11}+a^2_{22}+2a_{11}a_{22}\cos(\phi_{11}+\phi_{22}-2\phi_{12})}}\,,\\
\nonumber&&\sin\sigma=-\frac{\left(\lambda^2_2-a^2_{12}\right)\sin\phi_{12}+a_{11}a_{22}\sin(\phi_{11}+\phi_{22}-\phi_{12})}{\lambda_2\sqrt{a^2_{11}+a^2_{22}+2a_{11}a_{22}\cos(\phi_{11}+\phi_{22}-2\phi_{12})}}\,,\\
&&\cos\sigma=\frac{\left(\lambda^2_2-a^2_{12}\right)\cos\phi_{12}+a_{11}a_{22}\cos(\phi_{11}+\phi_{22}-\phi_{12})}{\lambda_2\sqrt{a^2_{11}+a^2_{22}+2a_{11}a_{22}\cos(\phi_{11}+\phi_{22}-2\phi_{12})}}\,.
\end{eqnarray}

\end{document}